\documentclass[fleqn,usenatbib,useAMS]{mnras}

\usepackage{graphicx}	
\usepackage{amsmath}	
\usepackage{amssymb}	
\usepackage{multicol}        
\usepackage{multirow}
\usepackage{bm}		
\usepackage{pdflscape}	

\makeatletter
\newlength{\abovecaptionskip}%
\makeatother
\usepackage{threeparttable}

\newcommand{\codename}[1]{\texttt{#1}}
\newcommand{\nodata}{ ~$\cdots$~ }

\usepackage[T1]{fontenc}
\usepackage{ae,aecompl}

\usepackage{newtxtext,newtxmath}

\title[Star Cluster Identification in PHANGS-HST]{PHANGS-HST: New Methods for Star Cluster Identification in Nearby Galaxies}

\author[Thilker et al.]
{David~A.~Thilker,$^{1}$\thanks{Contact e-mail: \href{mailto:dthilker@jhu.edu}{dthilker@jhu.edu}}
Bradley C. Whitmore,$^{2}$
Janice C. Lee,$^{3,4}$
Sinan~Deger,$^{5}$
\newauthor
Rupali~Chandar,$^{6}$ 
Kirsten~L.~Larson,$^{4}$
Stephen Hannon,$^{4,7}$
Leonardo Ubeda,$^{2}$
\newauthor
Daniel A. Dale,$^{8}$
Simon C. O. Glover,$^{9}$
Kathryn~Grasha,$^{10}$
Ralf S.\ Klessen,$^{9,11}$
\newauthor
J. M. Diederik Kruijssen,$^{12}$
Erik Rosolowsky, $^{13}$
Andreas Schruba, $^{14}$
\newauthor
Richard L. White, $^{2}$
and Thomas G. Williams $^{15}$
\\
\\
$^{1}$Department of Physics and Astronomy, The Johns Hopkins University, Baltimore, MD, USA\\
$^{2}$Space Telescope Science Institute, 3700 San Martin Drive, Baltimore, MD, USA\\
$^{3}$Gemini Observatory/NSF’s NOIRLab, 950 N. Cherry Avenue, Tucson, AZ, USA\\
$^{4}$Caltech/IPAC, California Institute of Technology, Pasadena, CA, USA\\
$^{5}$TAPIR, California Institute of Technology, Pasadena, CA 91125\\
$^{6}$Department of Physics and Astronomy, University of Toledo, Toledo, OH, USA\\
$^{7}$Department of Physics and Astronomy, University of California, Riverside, CA, USA\\
$^{8}$Department of Physics and Astronomy, University of Wyoming, Laramie, WY 82071, USA\\
$^{9}$Universit\"{a}t Heidelberg, Zentrum f\"{u}r Astronomie, Institut f\"{u}r Theoretische Astrophysik, Albert-Ueberle-Str 2, D-69120 Heidelberg, Germany\\
$^{10}$Research School of Astronomy and Astrophysics, Australian National University, Canberra, ACT 2611, Australia\\
$^{11}$Universit\"{a}t Heidelberg, Interdisziplin\"{a}res Zentrum f\"{u}r Wissenschaftliches Rechnen, Im Neuenheimer Feld 205, D-69120 Heidelberg, Germany\\
$^{12}$Astronomisches Rechen-Institut, Zentrum f\"{u}r Astronomie der Universit\"{a}t Heidelberg, M\"{o}nchhofstra\ss e 12-14, D-69120 Heidelberg, Germany\\
$^{13}$Department of Physics, University of Alberta, Edmonton, AB T6G 2E1, Canada\\
$^{14}$Max-Planck-Institut f\"{u}r extraterrestrische Physik, Giessenbachstra{\ss}e 1, D-85748 Garching, Germany\\
$^{15}$Max-Planck-Institut f\"{u}r Astronomie, K\"{o}nigstuhl 17, D-69117, Heidelberg, Germany}

\date{Last updated \today}

\pubyear{2021}

\begin{document}
\label{firstpage}
\pagerange{\pageref{firstpage}--\pageref{lastpage}}
\maketitle

\begin{abstract}
We present an innovative and widely applicable approach for the
detection and classification of stellar clusters, developed for the
PHANGS-HST Treasury Program, an $NUV$-to-$I$ band imaging
campaign of 38 spiral galaxies.  Our pipeline first generates a
unified master source list for stars and candidate clusters, to enable
a self-consistent inventory of all star formation products.  To
distinguish cluster candidates from stars, we introduce the Multiple
Concentration Index (MCI) parameter, and measure inner and outer MCIs
to probe morphology in more detail than with a single, standard
concentration index (CI). We improve upon cluster candidate selection,
jointly basing our criteria on expectations for MCI derived from
synthetic cluster populations and existing cluster catalogues,
yielding model and semi-empirical selection regions (respectively).
Selection purity (confirmed clusters versus candidates, assessed via
human-based classification) is high (up to 70\%) for
moderately luminous sources in the semi-empirical selection region, and
somewhat lower overall (outside the region or fainter). The number of
candidates rises steeply with decreasing luminosity, but
pipeline-integrated Machine Learning (ML) classification prevents this
from being problematic.  We quantify the performance of our PHANGS-HST
methods in comparison to LEGUS for a sample of four galaxies in common
to both surveys, finding overall agreement with 50--75\% of human
verified star clusters appearing in both catalogues, but also subtle
differences attributable to specific choices adopted by each
project.  The PHANGS-HST ML-classified Class 1 or 2 catalogues reach
$\sim1$ magnitude fainter, $\sim2\times$ lower stellar mass, and are
$2{-}5\times$ larger in number, than attained in the human classified
samples.

\end{abstract}




\section{Introduction}
\label{sec:intro}

Just as resolved stellar populations are routinely used to measure the evolutionary history of a galaxy, revealing information otherwise hidden in plain view, young star clusters yield unique insight into the physical conditions supporting star formation.  They are critically useful as `clocks' that allow for estimating timescales within the galactic ecosystem.  The ensemble distribution of cluster mass and age for a galaxy informs us regarding the processes that limit unbridled star formation \citep[e.g. feedback mechanisms in][]{Krumholz2019review} and those that encourage cluster dissolution (e.g. natal gas expulsion, \citealt{Baumgardt2007}; stellar dynamics, \citealt{Lamers2010,ReinaCampos2018}) or more active disruption (e.g. due to tidal shocks, \citealt{Gieles2016,Webb2019}).  In clusters that can be resolved into individual constituent stars and thereby age dated precisely, our understanding of stellar evolution was originally nurtured, and continues to be refined for rare phases, binary stars, and models incorporating rotation.  These (and other) beneficial aspects of star cluster science have encouraged the production of a `legacy' of studies, both observational \citep[e.g.,][plus innumerable studies in the Local Group, Magellanic Clouds, and Milky Way]{Whitmore1993,Whitmore1995,Whitmore1999,LarsenRichtler1999,Larsen1999,LarsenRichtler2000,LarsenBrodie2000,Larsen2001,Larsen2002,Larsen2004,Bastian2005,Bastian2005hierarchy,Whitmore2007,Bastian2008,Larsen2009,Adamo2010,chandar10,Adamo2011,Whitmore2011,Johnson2012,Adamo2015,AdamoBastian2015,Johnson2015, chandar16,Adamo2017,Ryon2017,Johnson2017,Grasha2017,Grasha2018,Messa2018,Messa2018b,Grasha2019,Hannon2019,Elmegreen2020,Whitmore2020,Adamo2020SSRv,adamo2020} and theoretically focused \citep[e.g. ][]{Hunter2003, Bastian2005, Lamers2005a, Lamers2005b, Gieles2006, Larsen2009, Kruijssen2011, Kruijssen2012, Pfeffer2019}.  

All of these works have led toward an emphasis on understanding the complex, cyclic relation between the interstellar medium (ISM) and star formation products.   The goal is a thorough assessment spanning the complete range of physical conditions and spatiotemporal scales on which the process of star formation, and hence galaxy evolution, operates.  In the Galaxy, observers study compact (very young) pre-stellar cores occupying giant molecular clouds (GMCs), whereas in nearby extragalactic targets we are able to more clearly delineate larger, progressively evolved, superstructures.  Unfortunately, these two perspectives rarely overlap. Only recently, this situation has begun to change \citep[with a handful of exceptions, e.g. ][describing observations of molecular cores in the Magellanic Clouds]{Cores1,Cores2,Cores3,Indebetouw2020}.  Further afield, ALMA ({\it Atacama Large Millimeter Array}) and $HST$ ({\it Hubble Space Telescope}), respectively, now allow us to image the high resolution ($1\arcsec \approx 50~{\rm pc}$ at 10~Mpc) distribution of the molecular ISM at GMC scales across entire galaxies and (in the same targets) allow us to inventory clusters and luminous stars an order of magnitude in size smaller yet.   

The PHANGS-HST\footnote{\url{https://phangs.stsci.edu}} survey \citep[][]{PHANGSHSTsurvey} was designed to capitalise on this opportunity in a large sample of galaxies.  Specifically, we obtained five-band near-ultraviolet through red imaging for selected targets from the primary ALMA survey of nearby galaxies (PHANGS\footnote{\url{https://www.phangs.org}}, Physics at High Angular resolution in Nearby GalaxieS; \citealt{PHANGSSurvey}) for which sensitive \mbox{CO(2--1)} maps were available.  About half of our $HST$ targets are also being studied using integral field spectroscopy at the VLT ({\it Very Large Telescope}) with MUSE ({\it Multi Unit Spectroscopic Explorer}) in a companion survey called PHANGS-MUSE \citep[][]{Emsellem2021}.

We build upon the effort and results of LEGUS (Legacy ExtraGalactic UV Survey, \citealt[][]{LEGUSsurvey}) in a very direct manner, through co-aligned scientific goals (with the invaluable addition of ALMA data for PHANGS-HST targets), further development in measurement and analysis techniques, and via use of the public LEGUS catalogues for baseline testing of our new methodologies.   For instance, while we have adopted the cluster classification system of LEGUS (Class~1: Symmetric, single-peaked cluster; Class~2: Asymmetric, single-peaked cluster; Class~3: Multi-peaked compact association; see \citealt{Adamo2017}, new details in \citealt{Whitmore2021}), the current paper describes our evolved approach to cluster detection and selection of likely candidates. We note that the PHANGS-HST sample significantly improves LEGUS coverage of the star-forming galaxy main sequence at higher stellar mass and SFRs \citep[see Fig. 1 of][]{PHANGSHSTsurvey}, but does not contain systems classified as a starburst or a dwarf starburst (that are well represented in LEGUS). PHANGS-HST galaxies are generally at larger distances than the LEGUS sample.

Common extragalactic cluster detection methods (e.g.\ used by the studies above) vary significantly in terms of source identification schemes, but the final catalogues almost always rely on concentration index\footnote{Defined here as the change in magnitude for a source when measured in two concentric circular apertures of different radius, typically $r=1$ and 3 pixels.} (CI)-based selection followed by confirmation via human inspection/classification.   This is necessary because galaxies in the Local Volume and just beyond show both resolved individual stars and clusters when observed with $HST$, imparting significant confusion to the cluster identification process.  Unfortunately, analysis of resolved stellar populations and compact clusters generally tend to be decoupled from the start, with one detection method used for stars (e.g.\ PSF-fitting photometry) and another independent process for clusters (e.g.\ SExtractor, Bertin \& Arnouts 1996).  However, this begets inconsistency in the eventual outcomes, and a degree of double counting in the overall census of star formation activity.  For this reason it would be ideal if our PHANGS-HST pipeline were able to identify them both using a unified approach.  

This paper describes such a process, dedicated to jointly recovering clusters and individual stars with a single, versatile detection method.  Working on both populations at once does increase the importance of accurately distinguishing clusters from stars, and accordingly we use more information (beyond a single CI value) from our high-resolution imaging than is typical for cluster candidate selection. Specifically, to better quantify the radial profile of each detected source, we introduce the concept of a Multiple Concentration Index (MCI), averaging multiple, strategically re-normalised CI.  We then use a pair of independent MCI values, MCI$_{\mathrm{in}}$ and MCI$_{\mathrm{out}}$, focused on inner and outer source morphology, respectively, to define an `MCI plane'.  We ultimately base our cluster candidate selection on cuts in these morphological metrics which have been informed using extensive synthetic cluster observations and MCI values measured for previously confirmed cluster catalogues.  Besides guiding selection, such synthetic clusters will soon also be used to quantify completeness as a function of cluster properties (e.g.\ age, mass, plus critically also morphology and environment)\footnote{Detailed completeness analysis of this sort will be included in a forthcoming publication.}, and to gain insight into the physical properties of cluster ensembles.

This paper is part of a series which documents the major components of the overall PHANGS-HST data products pipeline.  Details are provided regarding: survey design and implementation \citep[][]{PHANGSHSTsurvey}; aperture correction methods (\citealt{Deger2021}); source detection, selection, and aperture photometry of compact star cluster candidates (this work); star cluster candidate classification (\citealt{Whitmore2021}); neural network classification proof-of-concept demonstration \citep[][]{Wei2020}; stellar association and hierarchical structure determination, photometry, and determination of physical properties \citep{Larson2021}; spectral energy distribution (SED) fitting with \codename{CIGALE} \citep[][]{Turner2021}; and distance determination using PHANGS-HST parallel observations for Tip of the Red Giant Branch (TRGB) analysis \citep[][]{Anand2021}.

Section~\ref{sec:detect} provides details of source detection and photometry in our unified stellar and cluster context.  Section~\ref{sec:synthetic} describes creation of synthetic cluster models.  In Section~\ref{sec:selection} the criteria for cluster candidate selection are given.  A large portion of  Section~\ref{sec:selection} is devoted to introducing the MCI plane and describing how we employ synthetic clusters and observed stars to guide selection in this multi-variate, morphologically-sensitive context. Our classification of cluster candidates, both via human inspection and Machine Learning (ML), is described in Section~\ref{sec:classification}.  The familiar face-on disk galaxy NGC~628 (M~74) is used for illustration of our methods in Secs.~\ref{sec:detect}--\ref{sec:classification}. Section~\ref{sec:results} provides a detailed description of our observational data and the resultant cluster catalogues for a collection of four PHANGS-HST galaxies and compares them to existing products from LEGUS.   We discuss the implications of our work and specific scientific benefits of our method in Section~\ref{sec:discussion}. Finally, in Section~\ref{sec:conclusions} we end with a brief summary, including possible future improvements.   

\section{Source Detection and Photometry}
\label{sec:detect}

The completeness and fidelity of any stellar cluster catalogue is ultimately limited by its initial construction, specifically by the detection of sources that may be candidate clusters.  This Section explains our approach to this initial step, and describes the 
aperture photometry performed on all detected sources.

Preferred extragalactic star cluster detection methods vary considerably depending upon the physical spatial resolution of the data.
When clusters are resolved or partially-resolved into individual stars, well-known spatial clustering methods (e.g.\ friends-of-friends, nearest-neighbours, Minimum Spanning Tree (MST) separation, \codename{DBSCAN}, \codename{OPTICS}; \citealt{Battinelli1991}, \citealt{Wilson1991}, \citealt{Oey2004}, \citealt{Schmeja2011}, \citealt{SansFuentes1017}, \citealt{Zari2019})
applied to point source (stellar) catalogues are typically preferred.
At somewhat lower resolution, when only a handful of individual stars can be identified in any given cluster but not enough to support the use of clustering algorithms, a matched filter, cross-correlation approach, with kernels tuned to emphasise extended, cluster-like objects, are preferred (e.g. star cluster catalogues based on Source Extractor \citep{BertinArnouts1996} with filtering enabled and a kernel having FWHM broader than the PSF \citep[][]{Forbes2014,LimLee2015}.)
The PHANGS-HST data used here approaches the limit of resolution where this latter method begins to be disfavored.
Indeed, the vast majority of clusters in our $HST$ observations appear quasi-point-like, and are easily detected with PSF-fitting photometry codes which were originally designed for resolved stellar population work.
Note that at the median 16~Mpc distance to a PHANGS-HST galaxy, the $0.08\arcsec$ PSF FWHM (2-pixel) resolution of the {\em WFC3/UVIS} camera is $\sim6$~pc. A typical cluster has an intrinsic effective radius of $r_{\text{eff}}\sim 3$~pc, which after convolution by the PSF, yields a source with apparent FWHM $\sim\sqrt{2}\times$ broader than a point source.

\subsection{Source detection using \codename{DOLPHOT}}
\label{sec:dolphot}

PHANGS-HST has adopted the \codename{DOLPHOT} photometry package \citep[v2.0, ][]{Dolphin2000} as the principal source detection code in our pipeline.  Specifically, we use the {\em WFC3} and {\em ACS} modules,
and main distribution tarball downloaded\footnote{\url{http://americano.dolphinsim.com/dolphot/}} on 4~Dec 2019.  This choice
was motivated by a preference to have a single, unified catalogue
for selection of both candidate clusters and stellar sources \citep[which
are used to define multi-scale associations; see ][]{PHANGSHSTsurvey,Larson2021}.
PSF-fitting source detection also takes full advantage of our
high-resolution $HST$ imaging, de-blending closely neighbouring sources in an optimal manner.

Source detection using \codename{DOLPHOT} is governed by an extensive set of parameters. 
A detailed list of the parameters used is given in Appendix~B (Table~\ref{tab:dolphotparam}).
Briefly, we detect sources to the $3.5\sigma$
level
using simultaneous PSF-fitting on the individual exposure (*flc.fits) images in all filters.  The drizzled $F555W$ ($V$) band image is used as the positional reference for each target, meaning that exposure level detections in $F275W$, $F336W$, $F438W$,
$F814W$ are joined into a unique source list with sky positions based on the drizzled $F555W$ world coordinate solution (WCS).  Each detected source 
is photometered in all the bands, with non-detections flagged.
We adopt the Tiny Tim PSF library (default within \codename{DOLPHOT}) 
since the \codename{DOLPHOT}-specific implementation of
Anderson PSFs \citep[][]{Anderson2016,Anderson2018} leads to systematic differences across the field (A.~Dolphin priv.~comm.). 
\codename{DOLPHOT} makes refinements to the PSF and implements aperture
corrections based on a subset of well-detected, isolated stars in each chip.  Magnitudes are calibrated onto the Vega system using
STScI-supplied zeropoints in the image headers.  The photometric uncertainties reported by \codename{DOLPHOT} tend to be 
underestimated, as they do not account for the noise
contributed by neighbouring sources in crowded fields \cite[see][where it is shown that artificial star
testing can be used to constrain the true
uncertainties]{Williams2014}. 

\codename{DOLPHOT} is run using a set of python wrappers originally developed
for the LEGUS project (by PHANGS-HST member L.~Ubeda), and later
adapted for PHANGS-HST (by D.~Thilker).  These wrappers organize the
data prior to photometry, generate a complete parameter file specific
to the data set, and manage execution of the \codename{DOLPHOT} package codes.


We adopt the $V$-band \codename{DOLPHOT} source list as our principal inventory of astrophysical sources (clusters and stars) in each observed target.  Testing during pipeline development showed that occasional high-confidence clusters were not included in the \codename{DOLPHOT} catalogue.  In order to assess the degree to which \codename{DOLPHOT} is effective at recovering clusters for our PHANGS-HST distance range, we match the \codename{DOLPHOT} catalogue against the LEGUS Class~1 and~2 cluster population for NGC~628 \citep[][]{Adamo2017}.  We find that $\sim98\%$ of LEGUS clusters have a \codename{DOLPHOT} source within 2 pixels, and for $\sim80\%$ the closest match is within 1 pixel.  Given the methodological difference in detection algorithms (LEGUS used SExtractor; PHANGS-HST uses DOLPHOT), it is unsurprising that small positional shifts in the source positions are encountered at this level.

Because clusters with large angular sizes can be missed by \codename{DOLPHOT}, we also run the \codename{DAOStarFinder} code (in \codename{astropy}/\linebreak[0]{}\codename{photutils}), which is better suited to detecting extended objects.  \codename{DAOStarFinder} employs a convolution-based source identification method, whereas \codename{DOLPHOT} requires explicit local maxima.  We adopt a \codename{DAOStarFinder} kernel with a FWHM of 2.5 pixels ($0.1\arcsec$,  e.g. slightly broader than the PSF, so as to emphasise extended objects), and only add distinct new sources, plus those where the summed \codename{DOLPHOT} catalogue flux within 2 pixels is $>2.5\times$ fainter than measured by \codename{DAOStarFinder}.  Typically only $\lesssim1$\% of all eventual candidate clusters originate from \codename{DAOStarFinder}.
The \codename{DOLPHOT} catalogue augmented by \codename{DAOStarFinder} sources is referred to as the PHANGS-HST ``all-source'' detection list.

\subsection{Aperture photometry using \codename{photutils}}
\label{sec:aperphot} 

\codename{DOLPHOT} is used {\em only for source detection} in our cluster pipeline\footnote{The \codename{DOLPHOT} PSF-fitted magnitudes are used in a parallel pipeline devoted to point sources.}.  Once the all-source detection list (\codename{DOLPHOT}+\codename{DAOStarFinder}) is ready, {\em circular aperture photometry is performed using the \codename{photutils} python package}.

Because most clusters appear round in $HST$ images, we measure photometry in a series of circular apertures with radii of 1.0, 1.5, 2.0, 2.5, 3.0, 4.0, 5.0 pixels.  The background level is measured in a circular annulus 7--8 pixels away from the centre of each source, then subtracted off.  This is a fairly standard background annulus when working in crowded fields imaged with $HST$ (e.g., \citealt{Sabbi2018}), and balances the need to measure the background level close to each source while being sufficiently distant ($\approx 14-29$~pc for typical galaxies in our sample) to minimise light from the source itself within the background annulus.
We estimate the sky value as a robust median, using the {\tt sigma\_clipped\_stats} function to mitigate the influence of neighbours.  The sigma-clipped mean and standard deviation are also recorded by our pipeline, to allow for investigation of inevitable systematic and random errors in the adopted background.   Calculations of the photometric uncertainty follow the prescription of F.~Masci\footnote{Eq.~13 in \url{http://wise2.ipac.caltech.edu/staff/fmasci/ApPhotUncert\_corr.pdf}} with specific allowance for spatially-correlated noise in the pixels of the source aperture and sky annulus.





For sources selected as candidate star clusters (see Section~\ref{sec:selection}), the total magnitude in each filter is determined by measuring the flux in a 4 pixel radius ($\approx 12$~pc at the median PHANGS-HST galaxy distance) aperture, which captures $\sim$50\% of the total flux, and then applying an aperture correction \citep[e.g.,][]{Adamo2017, Cook2019}.  

A detailed description of how PHANGS-HST aperture corrections are determined is given in \citet{Deger2021}.  Briefly, 
human inspection-based selection of an aperture correction sample (consisting of highly confident clusters, compact associations, and stars) in each target enables measurement of the correction for comparatively isolated sources. 
One of us (Bradley C. Whitmore, hereafter BCW) inspects the drizzled image products before our pipeline run and manually identifies 10 sources of each cluster Class~(1,~2,~3).  
For each source, we compute aperture corrections to convert the $r=4$~pixel ($7{-}8$~pixel background annulus) $V$-band photometry to an estimated total (sky-subtracted at $r=20{-}21$~pixel) magnitude.  Rather than letting the relative aperture correction between bands float on a galaxy-by-galaxy basis, which would add noise, we normalise our average aperture corrections as offsets with respect to the measured $F555W$ aperture correction in each target.  These normalised corrections versus band are then combined into a mean survey-wide aperture correction offset table, after outlier rejection.  For each galaxy we then adopt the final aperture correction as the galaxy-specific $F555W$ value plus the survey-wide offsets for other bands.  The offsets are $-0.19$, $-0.12$, $-0.03$, $0.00$, $-0.12$~mag in $F275W$, $F336W$, $F438W$, $F555W$, and $F814W$, respectively.  For the galaxies processed at this time, the $F555W$ aperture corrections range from $0.6$ to $0.85$~mag. 

We refer to the complete photometric database resulting from these measurements as the 'all-source' catalogue, to which morphological information (e.g. MCI) is later added (Sec.~\ref{sec:MCIdefined}).


\section{Creation of Synthetic Cluster Models}
\label{sec:synthetic}

The {\em selection} of candidate star clusters from a larger source catalogue that includes individual stars, chance super-positions, background galaxies, etc. is critical for the study of cluster populations in nearby galaxies.
Much of our new approach to cluster selection (Sec.~\ref{sec:selection}) uses morphological metrics determined from both synthetic clusters and previously identified (real) clusters.
Synthetic clusters also provide an estimate of the true photometric uncertainties and any biases in the measurements.
This Section describes a database of synthetic cluster models which are used for these purposes, and will be helpful for future studies.

\subsection{Overview of the synthetic cluster database}

Synthetic cluster sources are added to the $V$-band image for each target galaxy, since these images are the ones used for source detection and cluster selection.  The procedure is generalized sufficiently that we can easily add artificial sources to any PHANGS-HST band, although this is not currently implemented in the pipeline.
However, we envision future work along these lines for the purpose of machine learning cluster classification training (e.g.\ \citealt{Wei2020}, \citealt{Whitmore2021}).  

Random positions which are uniformly distributed are generated for the synthetic clusters. We considered using positions weighted to the observed source locations, but decided against it so that this step can be run independently from the rest of the pipeline.  Note that if positionally-weighted properties are desired for the synthetic clusters, these can be selected post-facto.  The location of a cluster influences detectability and photometric scatter/bias (with objects in complex and/or high background regions of a galaxy being harder to identify and measure than those in uncrowded and/or faint environments).  Thus for future catalogue completeness analysis (beyond the scope of the current paper) we expect to impose such post-facto positional weighting. 

Our fundamental goal in this portion of the pipeline is to generate a database of synthetic sources which adequately covers the plausible range of parameters expected for clusters in the PHANGS-HST imaging.  This implicitly means we need to sample the age--mass plane, range of $A_V$ values and all plausible morphologies, and add enough realisations of clusters with a given set of parameters to fully represent the galactic environment and quantify the measured scatter in morphological metrics.  To the degree that the entire cluster parameter space is represented in a target's source population, the locus traced by morphological metrics most frequently seen in our synthetic database should resemble the observed distribution of confirmed clusters.  Even so, we stress that our aim in the current paper is not to forward-model predict the expected distribution of metrics for a hypothetical cluster population, but rather to delineate the bounds of the distribution containing objects of interest.     

We assign physical characteristics (age, mass, $A_V$, and morphology) to each synthetic cluster, to enable completeness studies (as a function of cluster type) and build a foundation for the forward-modelling framework alluded to above.  
We permit ages log($t$) = [6.0, 6.5, 7.0, 7.5, 8.0, 8.5, 9.0] yr and extinction values of $A_V$ = [0.0, 0.25, 0.5] mag (in addition to Milky Way foreground), plus a wide range of Moffat profiles which sample intrinsic size and radial shape; see Sec.~\ref{sec:Moffat} for details. The chosen limits of our model parameter space cover the vast majority of clusters found in nearby galaxies, with the particular exception of the very youngest, still embedded population.  These particularly elusive, dust-enshrouded clusters \citep[e.g.][]{LadaLada2003} will be one focus of our approved JWST program (GO-2107) targeting half of our PHANGS-HST sample.  Their expected numbers are uncertain and highly dependent on environment \citep{Romita2016}.  For each combination of age, $A_V$, and morphology, we generate 1000 cluster realisations from a flat mass distribution between $10^3$ to $10^5~M_\odot$,  sampling every  $10^2~M_\odot$.
Fixed step sizes in age, mass, $A_V$ are used for convenience. If we eventually further explore forward-modelling applications (as mentioned above), we would then likely switch to randomized sampling to eliminate any possibility of unwanted biases. Population synthesis models within \codename{CIGALE} \citep[][]{Boquien2019} are used to predict the $V$-band magnitude for each synthetic cluster, based on its assigned age, mass, and $A_V$.
In this way, we establish consistency between our synthetic clusters and the SED-fitting \citep[][]{Turner2021} of the PHANGS-HST pipeline.   
In total, we inserted some 4.5+ million synthetic clusters in each PHANGS-HST target, 200 objects at a time, simulating diverse cluster morphologies accounting for the PSF and incorporating Poisson noise on a pixel-by-pixel basis as described in Sec.~\ref{sec:Moffat}~and~\ref{sec:modelrealisation}.   The same procedure used to determine photometry of real sources is applied to the synthetic ones.

\subsection{Moffat model}
\label{sec:Moffat}

We adopt Moffat profiles to represent our synthetic clusters, since these give a particularly good fit to the measured profiles of young clusters.  This choice is well-supported by observational studies \citep[e.g.][]{Elson1987,Larsen1999,McLaughlin2005}, although there are other reasonable options like King models \citep[][]{King1962,King1966}.   To reduce computational burden, our Moffat models are circularly symmetric, though in future work we may include asymmetric synthetic cluster models as they would be a superior representation of Class 2 clusters found observationally.  Our model based selection (Sec.~\ref{sec:modelregions}) is `loose' enough that we do not expect to be biasing against Class 2 objects in the present work.  However, if we later attempt forward-modelling of the cluster population, this complexity should be included.

The Moffat surface brightness profile is given by:
\[\mu(r) = \mu_0 \left(1 + r^2 / a^2\right)^{-\eta}\;,\]
where $\mu_0$ is the central surface brightness, $a$ is a characteristic radius, and $\eta$ is the power-law exponent of the profile wings. Note that
$\eta$ is equal to $\gamma/2$ in equation~(1) of \citet{Elson1987}.   As $\eta$ approaches infinity the Moffat profile becomes a Gaussian.  The characteristic radius can be expressed as a function of full-width-half-max (FWHM) and~$\eta$:
\[a = \frac{\mathrm{FWHM}}{2}{\left(2^{1/\eta}-1\right)}^{-1/2}\;.\]

Studies of clusters in nearby galaxies typically measure the effective radius $r_{\text{eff}}$, which is defined to be the
radius of a circle containing half the integrated light of the model. For a Moffat profile,  $r_{\text{eff}}$ is given by:
\[r_{\rm{eff}} = a {\left((1/2)^{\frac{1}{1-\eta}}-1\right)}^{1/2}\;,\]
which is only valid for $\eta > 1.0$.  Although the effective radius is undefined for $\eta \le 1.0$, and the integral of the profile infinite, the Moffat parameterisation can still be evaluated (and matching cluster morphologies observed) for this regime. In the real-world case it simply means that the cluster profile is truncated or lost to the background at some large radius.   

The parameter space of circular Moffat clusters spans core size (via $a$ or FWHM) and halo power-law slope (via~$\eta$).  We adopted 8 values of $\eta$ (= 0.75, 0.875, 1.0, 1.25, 1.5, 2.0, 3.0, 4.0) and pre-PSF-convolution FWHM ranging from 0.5 to 7.0~pix, with step sizes of 0.25~pix.  In total, taking all permutations across this parameter space, we allow for 216 (8$\times$27) distinct cluster morphologies.  Models with high $\eta$ values (e.g.\ 3,~4) have minimal halo structure, and are dominated by their progressively Gaussian core.  Conversely, low $\eta$ values correspond to very `fluffy' cluster morphologies. Guiding the choice of minimum cluster core FWHM for our models, parametric fitting studies of marginally resolved stellar clusters have shown that reliable structural parameters can be obtained for effective radii approaching down to $10{-}20$\% of the PSF FWHM, dependent on signal-to-noise and cluster morphology \cite[e.g.][]{Larsen1999,Ryon2017,Brown2021}.  We expect to have a morphological limit of $r_{\text{eff}}\sim1$~pc at 16~Mpc (median PHANGS-HST distance, \citet{Anand2021}) with {\em WFC3/UVIS}.
\begin{figure*}
\includegraphics[width=7in]{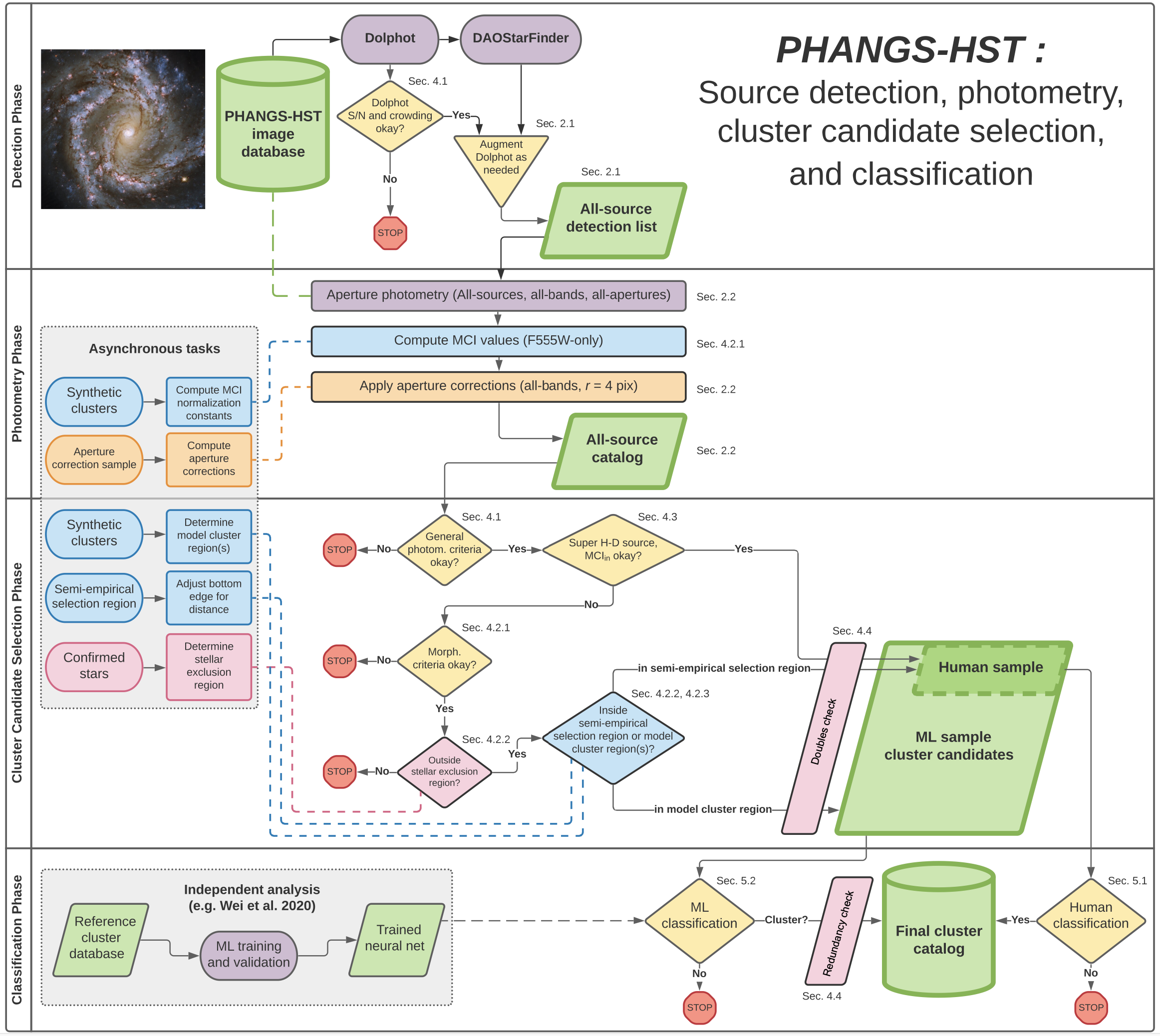}
\caption{Flowchart summarising the PHANGS-HST source detection, photometry, cluster candidate selection, and classification process. }
\label{fig:flowchart}
\end{figure*}

\subsection{Model realisation}
\label{sec:modelrealisation}

Regardless of our particular choice for parameterising the structure of model clusters, we require a method to generate and realistically insert them into the observations in a way that mirrors the observational strategy (e.g.\ multiple exposures, dithering) and instrumental characteristics (e.g.\ resolution, noise).  It is particularly important that the intrinsic cluster profile be convolved with the PSF.  The assumed PSF is adopted from the empirically determined PSF products of J.~Anderson, using a PSF representation appropriate to the centre of the {\em WFC3/UVIS} camera field of view\footnote{\url{https://www.stsci.edu/~jayander/STDPSFs/}} \footnote{\url{https://www.stsci.edu/hst/instrumentation/wfc3/data-analysis/psf}}. This choice allows the synthetic cluster database to be generated in advance of cluster detection/\linebreak[0]{}selection pipeline runs.\footnote{In later versions of our cluster catalogue we may adopt the PSF determined by \codename{DOLPHOT} instead of generic representations.}

Although it was developed to study galaxy morphology via parametric model fitting, \codename{IMFIT} \citep[][]{IMFITpaper} is well suited for our purpose of generating PSF-convolved models, especially as the package includes Moffat model capabilities.  It is adopted for the initial steps of our cluster model realisation process (the stages of generating the intrinsic structure and convolving with the PSF), then a custom python script is used to do the remaining tasks, including adding noise, reprojecting exposure-level models to the drizzled grid, coadding, and finally adding the synthetic clusters to the PHANGS-HST images.

Images of real clusters may or may not be centred at the centre of a pixel.
To account for this effect, which can become important for compact clusters, for each `batch' of clusters that is inserted into an image, we adopt one of five different centres offset from the pixel centre by (0.0,~0.0), (0.25,~0.25), (0.5,~0.5), (0.25,~0.0), and (0.5,~0.0), to allow the centres to range from the centre to a pixel edge.  Provided our step size is sufficiently fine, these offsets are sufficient to represent any actual cluster centre, as all other offset choices are equivalent via reflection and rotation. 


For each batch of 200 synthetic clusters, of fixed age, $A_V$, FWHM, $\eta$, and subpixel location, we use \codename{IMFIT} to generate a noise-free Moffat model using factor of 4 oversampling.  The Anderson PSF appropriate to the filter and camera of the synthetic run (in our case $WFC3/UVIS/F555W$ almost always) is convolved with the model to produce a noise-free template synthetic cluster (still oversampled by a factor of~4) for each of our dither positions (generally three).  Next we use these to generate native-scale ($0.03962\arcsec/\mathrm{pix}$) models (again one per dither position / $HST$ exposure) which were subsequently scaled to the integrated intensity corresponding to the adopted mass [magnitude, see beginning of this Section], independently per cluster.  At this point we compute a Poisson variate for every significant pixel in the penultimate flc-exposure-level-appropriate model realisations, in a manner that accounted for the local area specifically around each cluster position as recorded in the unmodified image.  Finally, we co-add the noisy models (per cluster) and add the result to the drizzled $V$-band image.

\section{Cluster Candidate Selection Criteria}
\label{sec:selection}

Our cluster candidate selection method was developed with the goal of maximising completeness and minimising contamination.
Our criteria include cuts on: (1) data-quality and photometric measures (Sec.~\ref{sec:generalcriteria}, e.g.\ number of bands with acceptable photometric error, $V$-band $\mathrm{S/N}$, absolute $V$-band magnitude, and crowding) and (2) novel morphological metrics, MCI$_{\mathrm{in}}$, MCI$_{\mathrm{out}}$ (Sec.~\ref{sec:MCI}),
as guided by synthetic cluster modelling (Sec.~\ref{sec:modelregions}) and published cluster catalogues (Sec.~\ref{sec:EmpiricalSelectionReg}).  We also impose selection rules to: (3) guarantee candidate inclusion of sources too luminous to be ordinary stars (Sec.~\ref{sec:maglimits}), and (4) to eliminate double counting associated with instances of overly aggressive source deblending (Sec.~\ref{sec:doubles}). 

In Fig.~\ref{fig:flowchart}, we present an overview of the cluster identification pipeline, beginning with the source detection and photometry phases already described in Sec.~\ref{sec:detect},  continuing to the subsequent cluster candidate selection phase which is the topic here, and ending with the classification phase that yields our final cluster catalogue.  We position this figure prior to describing details of selection, so that it is easier for the reader to obtain an understanding of the overall method and specifically how the pieces fit together.  We direct the reader to a fully comprehensive flow chart for the entire PHANGS-HST survey (thus also including multi-scale asssociations) in Fig.~5 of \citet{PHANGSHSTsurvey}.  Our Fig.~\ref{fig:flowchart} represents a more detailed view at the data flow and logical path required for the cluster-specific aspects of the overall project.

\subsection{General candidate selection criteria}
\label{sec:generalcriteria}

We begin with the set of sources detected by \codename{DOLPHOT}, and impose a limit of $\mathrm{S/N} \ge 5$ and crowding $\le 0.667$ mag (rationale on this in Sec.~\ref{sec:doubles}) for a source to enter cluster candidate processing.  \codename{DAOStarFinder} then adds in a very limited number of objects missed by \codename{DOLPHOT}, meeting the conditions for augmentation described at the end of Section~\ref{sec:dolphot}.  Imposition of the DOLPHOT $\mathrm{S/N} \ge 5$ cut (even though our cluster candidate sample selection later demands $\mathrm{S/N} \ge 10$) excludes potentially unreliable compact sources from the determination of whether a given DAOStarFinder detection should be added to the all-source detection list, or if it was already preferentially well split into two or more DOLPHOT detections of reasonable quality. Next, aperture photometry and MCI computation is conducted for each source.
From the resulting combined ``all-source catalogue'' (= bright+uncrowded \codename{DOLPHOT} plus \codename{DAOStarFinder}), we cull out poorly measured sources by requiring detections in $\ge 3$ bands with photometric error $\le 0.3$ mag.  We further require that V is one of these bands.   These general criteria reduce the all-source catalogue to only $\sim 4{-}9\%$ of its initial entries.  As described 
below, we further down-select this subset by considering two populations: (1) typical sources which meet a full set of morphological criteria (see Sec.~\ref{sec:MCI}) and (2) those that are so bright they are surely clusters or artefacts \citep[brighter than the Humphreys--Davidson (H-D) limit,][]{Humphreys1979},  for which we impose a more lenient set of criteria (see Sec.~\ref{sec:maglimits}).  

\codename{DOLPHOT}'s goodness of \mbox{(PSF-)}fit metric ($\chi^2$) and sharpness parameter
are both limited to the inner portion of sources ($\le 3$ pixels for $\chi^2$, and the central pixel alone for sharpness), which is not ideal for selecting sources broader than the PSF.
For this reason we do not use these quantities for candidate  cluster selection.

\subsection{Morphology criteria: Multiple Concentration Index (MCI)}
\label{sec:MCI}

One of the primary aims of our study is to improve on the cluster selection methodology used in previous studies, which has mostly relied on a single concentration index
\citep[CI, e.g.][]{Holtzman1992, Whitmore1993, Adamo2017, Cook2019}. 
High-resolution $HST$ images of clusters in galaxies at distances between $\sim5{-}25$~Mpc allow for constructing maximally-informative (though still azimuthmally-averaged) constraints on cluster morphology, enabling more robust discrimination of point-like or cluster-like appearance than has been done to date.

\begin{figure*}
 \includegraphics[width=7in]{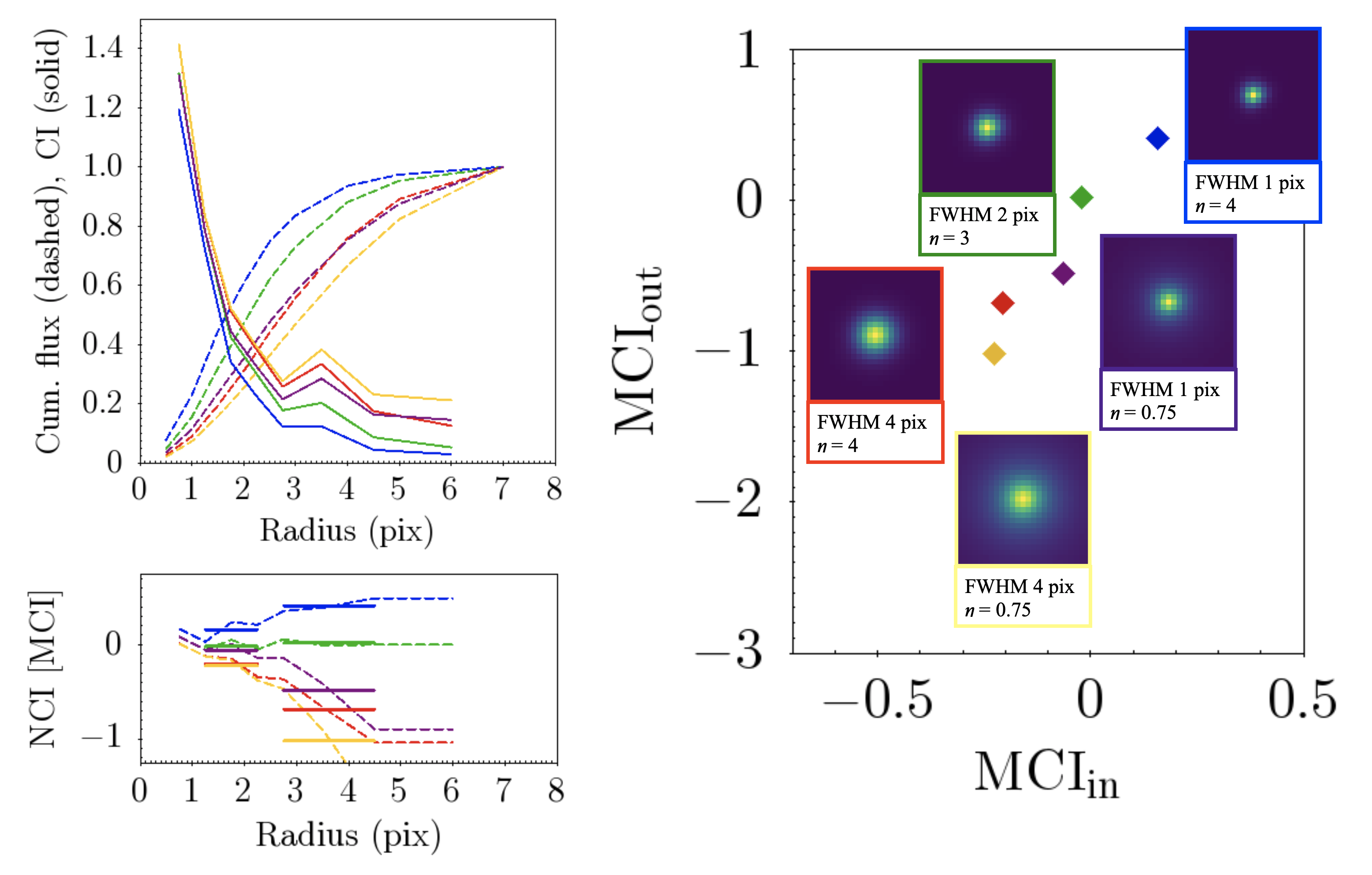}
 \caption{The inter-relation between cluster appearance, profile shape, curve of growth, NCI values, MCI, and ultimately position of different characteristic morphologies in the MCI plane. Line and marker colours in each panel are linked to the colour of the box outlying each synthetic cluster image shown. In order to obtain MCI$_{\mathrm{in}}$ and MCI$_{\mathrm{out}}$ for any source, (top left panel) we first measure curve of growth photometry and then convert these measurements to CI values. (bottom left) We re-normalise the CI values with respect to the adopted fiducial cluster to obtain NCI and subsequently average NCI values for adjacent positions within the radial profile to obtain MCI metrics. (right) The MCI plane position for a source is thereby linked to how locally steep or shallow the profile appears with respect to the fiducial morphology.  See the text for a complete description.  Synthetic clusters used {\it for this figure only} are idealized in the sense that they are noise-free and were not inserted into our $HST$ images. Note that the images of model clusters are scaled from zero to their maximum value, so as to enable relative morphological comparison.}
 \label{fig:MCIexample1}
\end{figure*}

We compute Multiple Concentration Index (MCI) metrics to probe radial profile shape over a wide range of distance from a source.  These metrics allow us to compare the shape to that of a fiducial cluster morphology (via the specific choice of normalisation constants).  In the subsections below, we formally define the MCI$_{\mathrm{in}}$ and MCI$_{\mathrm{out}}$ parameters used in our cluster candidate selection method, give examples of how cluster morphology varies with location in the MCI plane, and provide motivation for the area(s) of the MCI plane in which candidates are selected.

\subsubsection{Definitions and examples}
\label{sec:MCIdefined}

Our MCI metrics are based on the concentration index concept (which by itself is a reformulation of curve-of-growth analysis), although we take advantage of many independent measures of CI (each based on a different pair of radii), rather than a single CI measure.  To simplify the interpretation, we re-normalise the CI with respect to a fiducial cluster (of fixed angular size), so that the {\em normalised CI (NCI) indicates whether the source surface brightness profile is steeper or flatter than the fiducial one at the measured radii.}  Specifically, for radii $i$ and~$j$ we define: 

\[\mathrm{NCI}(i,j) = 1 - \frac{\mathrm{CI}(i,j)}{\mathrm{CI}(i,j)_{\mathrm{fiducial}}}\]

We adopt a mildly extended cluster morphology as our fiducial cluster model, with pre-PSF-convolution FWHM of 2 pixels and $\eta = 3$.  The choice is arbitrary, as a different morphology for the fiducial cluster would lead to systematically different extent and scatter in the resulting observed NCI (eventually MCI) distribution, but these changes would also correspondingly change the MCI distribution of our entire synthetic cluster database (used below to define the selection region) and the observed source distribution against which it is compared.  
{\em It is important to note that the sense of compactness used here is reversed when compared with the traditional measure of CI (1~pixel to 3~pixel), with higher NCI values indicating a more concentrated (star-like) object.}

In order to allow readers to use our methods without generating synthetic fiducial clusters, in Table~\ref{tab:normalisationconstants} we provide representative normalisation constants required to compute NCI values from observed concentration indices.  In practice within our pipeline, we compute constants specifically for each galaxy (observed field)  using 21,000 fiducial morphology clusters fully sampling all permutations of age, mass, $A_V$ from the synthetic cluster database described in Sec.~\ref{sec:synthetic}.  The constants differ very little amongst galaxies, except for those targets with data (in $V$-band) from {\em ACS/WFC} rather than {\em WFC3/UVIS}.

\begin{table}
    \caption{Representative normalisation constants for transforming CI to NCI}
	\begin{center}
    \begin{tabular}{lccl}
 \hline
    Quantity & $i$ & $j$ & $\mathrm{CI}(i,j)_{\mathrm{fiducial}}$ value\\ 
 & (pix) & (pix) & \\
    \hline    
$\mathrm{CI(1.0,1.5)}$ & 1.0 & 1.5 & 0.746 \\
$\mathrm{CI(1.5,2.0)}$ & 1.5 & 2.0 & 0.447 \\
$\mathrm{CI(2.0,2.5)}$ & 2.0 & 2.5 & 0.288 \\
$\mathrm{CI(2.5,3.0)}$ & 2.5 & 3.0 & 0.186 \\
$\mathrm{CI(3.0,4.0)}$ & 3.0 & 4.0 & 0.201 \\
$\mathrm{CI(4.0,5.0)}$ & 4.0 & 5.0 & 0.0858 \\

    \hline    
    \label{tab:normalisationconstants}
    \end{tabular}
	\end{center}
    \vspace{-10pt}
    \begin{tablenotes}
	\small
\item[$a$] Note: Listed values ($\mathrm{CI}(i,j)_{\mathrm{fiducial}}$) are concentration indices appropriate for the fiducial cluster morphology (pre-PSF-convolution FWHM of 2 pixels and $\eta = 3$).  $i$ and $j$ are the aperture radii for the small and large (respectively) circular apertures used for each CI. We insert 21,000 realisations of the fiducial cluster spanning our synthetic parameter space of age, mass, $A_V$ into all galaxies of our sample and then take the median of the measured CI values to produce this table. For reference, the standard CI(1.0,3.0) typically used by previous studies has a value of 1.67 for the fiducial cluster morphology. 
    \end{tablenotes}
\end{table}

Because the NCI values are normalised, we can combine multiple NCI values into a summary metric to reduce the `noise' in the measured profiles, and thereby mitigate the impact that an image artefact or nearby object could have on any single NCI measure.  
We define the Multiple Concentration Index (MCI), for radius pairs $(a,b), (b,c), (c,d),$ as:

\[\mathrm{MCI} = \overline{[\mathrm{NCI}(a,b), \mathrm{NCI}(b,c), \mathrm{NCI}(c,d)]}\;,\]
where the overline notation indicates averaging.

In Section~\ref{sec:aperphot}, we described our circular aperture photometry employing a series of aperture radii (in particular, 1.0, 1.5, 2.0, 2.5, 3.0, 4.0, 5.0 pixels).  We use these measurements to construct two independent MCI metrics, MCI$_{\mathrm{in}}$ and MCI$_{\mathrm{out}}$.  For MCI$_{\mathrm{in}}$ we take radii 1.0, 1.5, 2.0, 2.5 as $a, b, c, d$ in the expression above.  MCI$_{\mathrm{out}}$ is based on photometry with aperture radii of 2.5, 3.0, 4.0, 5.0 pixels.  {\em As such, MCI$_{\mathrm{in}}$ probes the inner portion of a source (out to $r \approx 0.1\arcsec$ or 1 resolution element for $HST$), dominated by the cluster core at the distances of our target galaxies, whereas MCI$_{\mathrm{out}}$ traces the faint outer, low surface brightness structure of a source.}  

The plots in Fig.~\ref{fig:MCIexample1} illustrate how steep the {\em local} radial profile of a source is compared to the fiducial cluster model, based on the MCI metrics.  Our aperture photometry is first used to create curves of cumulative flux versus aperture radius (interior to the sky annulus), as shown by the dashed curves in the upper left panel.  Synthetic models for five clusters are compared, including the fiducial model (green), two with pre-PSF-convolution FWHM 1~pix (blue with $\eta=4$, purple with $\eta=0.75$), and two with FWHM of 4~pix (red with $\eta=4$, yellow with $\eta=0.75$).  Images of these cluster realisations are shown in the right panel.  The blue boxed model is barely resolved, and is the closest to a point source, but nevertheless can be distinguished as a cluster given sufficient signal-to-noise. For this demonstration no Poisson noise was incorporated, and the images have been scaled linearly from 0.0 to their respective maximum values.  In the upper left panel, we further show the photometry after conversion into a series of CI values (again as a function of radius) as the solid lines.  The `bump' at $r=3.5$ pix is due to the first aperture radii pair with separation of 1.0~pix rather than 0.5~pix, as used near the centre.  In the lower left panel, the CI are first normalised against the fiducial cluster, introducing the sense of local slope with respect to this mildly extended source.  Positive values of NCI (MCI also) indicate progressively steeper profiles, whereas negative NCI values happen whenever the source profile is shallow compared to the fiducial.  To obtain MCI$_{\mathrm{in}}$ and MCI$_{\mathrm{out}}$ we average the three adjacent NCI values (each) as specified above.  MCI in the lower left plot is marked only for the radial range probed by each of the inner and outer metrics.  In the right panel, we show the MCI plane, marking the position of each model in this morphologically indicative space.  Note the fiducial (green) is at ($0,0$), by definition.  The position of a cluster in the MCI plane is largely dependent on the size of the source, here represented by the pre-PSF-convolution FWHM.  {\em Large sources are in the lower left and small in the upper right}.  However, the assumed halo slope in the outer portion of a cluster can have a significant influence on the MCI plane position for fixed FWHM, most significantly for intrinsically small sources in which the halo is able to skew the MCI$_{\mathrm{out}}$ and MCI$_{\mathrm{in}}$ values.  Note the large shift in MCI$_{\mathrm{in}}$ and MCI$_{\mathrm{out}}$ for the purple cluster in comparison to the blue one (with a steep, i.e.\ insignificant, halo).  By the time a source is quite extended even in its core, changes in halo slope only induce shifts in MCI$_{\mathrm{out}}$ (see red versus yellow cluster).  

The illustration of Fig.~\ref{fig:MCIexample1} is idealised. In real observed clusters, in addition to the influence of noise, both source crowding and a variable diffuse environment can perturb the MCI measurements away from expectations for a cluster of given intrinsic morphology.  As our circular aperture photometry does not (yet) include masking of neighbour sources, or a sloped sky estimate, we account for this by inserting our synthetic clusters of varied size and morphology -- not just the fiducial chosen for MCI normalization -- in the actual $HST$ $V$-band images of each sample galaxy.  The outcome of such analysis is discussed in detail in Sec.~\ref{sec:modelregions}, but we first conclude the introduction to MCI with a discussion of the typical source distribution in the observed MCI plane.

\begin{figure*}
    \centering
    \includegraphics[width=\columnwidth]{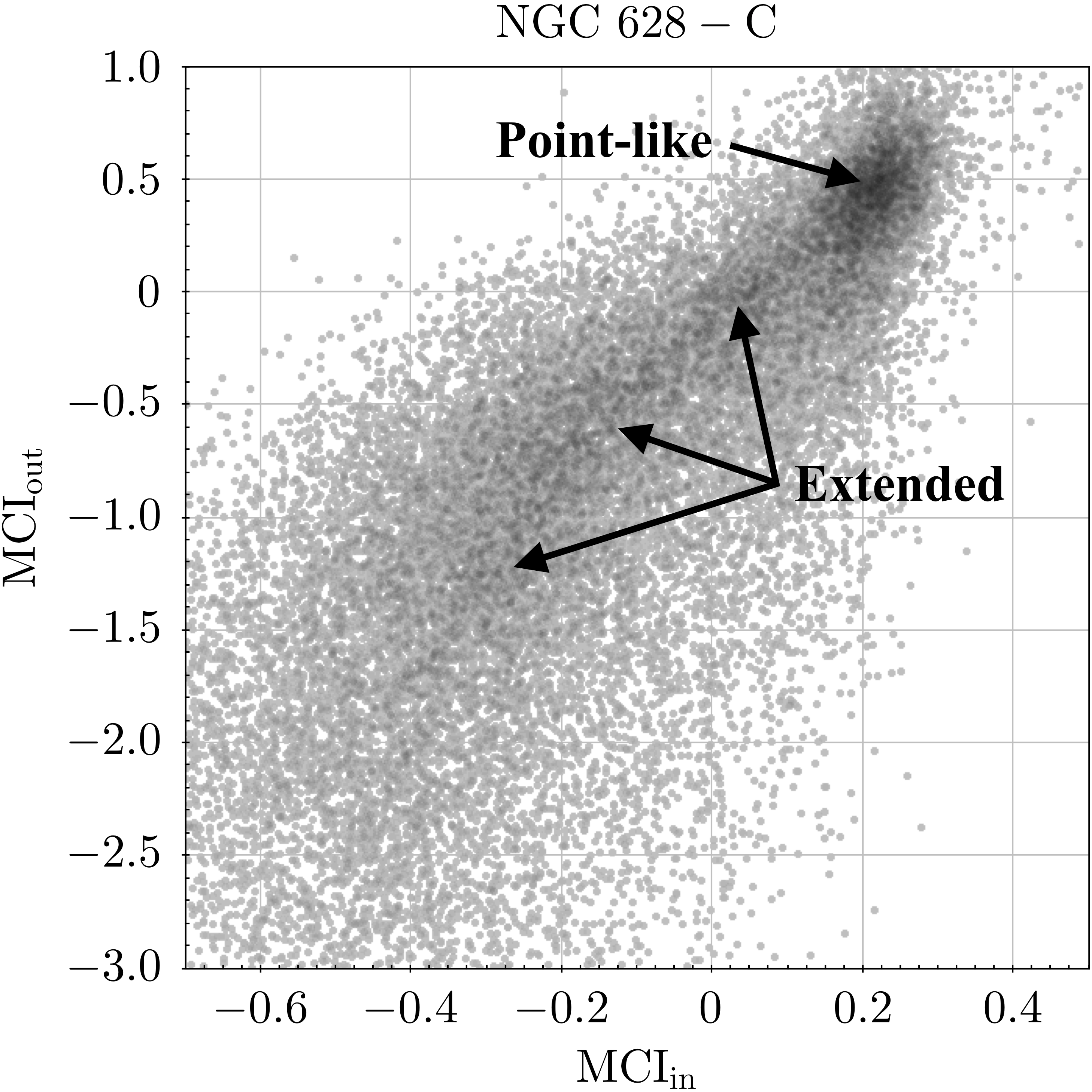}
    \includegraphics[width=\columnwidth]{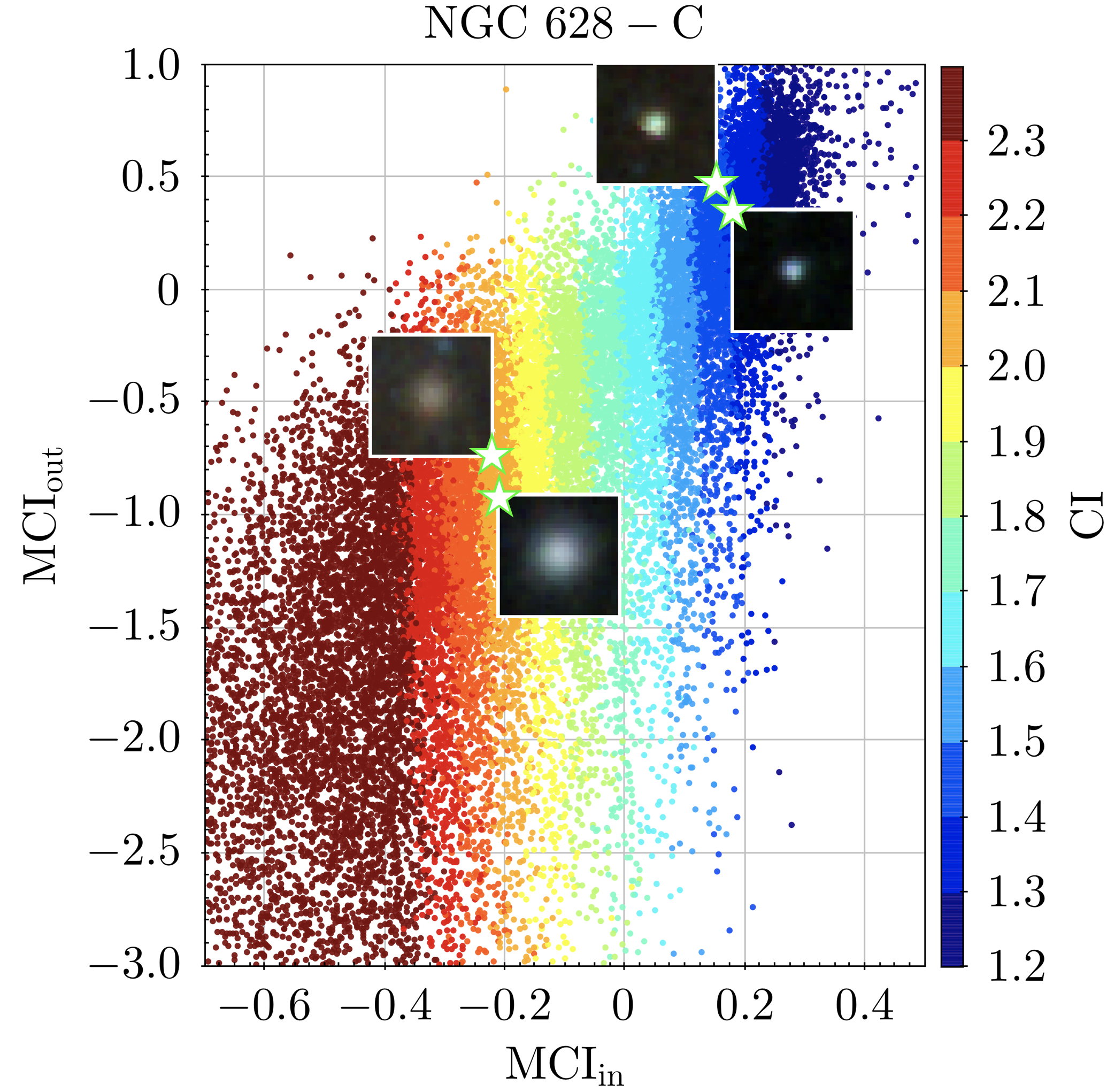}
    \caption{(Left) The MCI plane distribution of detections in our all-source catalogue for the NGC~628-C field with $V$-band $\mathrm{S/N} \ge 10$.  The distribution reflects the population of stars, clusters, and contaminants.  Stars are largely separated to the upper right, but the contribution of clusters is not as tightly defined. (Right) The same plot on the left except coloured by CI.  Cutout images of {\em actual} NGC~628-C clusters linked to their MCI plane location (with star symbols) have been overlaid on the right panel.  As opposed to Fig.~\ref{fig:MCIexample1}, here the {\em actual} clusters are all displayed as observed, not rescaled according to their maxima (such as was done for the synthetic clusters in the previous plot).}
    \label{fig:MCIplane}
\end{figure*}

In the left panel of Fig.~\ref{fig:MCIplane} we show the distribution of measured MCI$_{\mathrm{in}}$ versus MCI$_{\mathrm{out}}$ values for all sources with $\mathrm{S/N} \ge 10$ in our NGC~628-C field (see Sec.~\ref{sec:results} for a description of these data). We also identify the expected location of point-like sources (towards the upper right) and extended sources in this plane. 
Note the largely bimodal distribution, with a strong density of points with positive MCI values (corresponding to stars) and a broad distribution of points dominating the remainder of the plot (including, but not limited to, clusters).  Within this broader distribution, a ridge extending diagonally down and to the left from the location of point-sources is often evident.
Note the distribution of scattered sources around the ridge of extended points.  This scatter is found in all of our fields,  and is due to contaminants, such as close pairs of stars, diffraction spike artefacts, and local maxima in diffuse regions. 
The right panel in Figure~\ref{fig:MCIplane} shows the same data as in the left panel, but now colour-coded by the traditional measure of CI, which has a range of $\sim1.2$ (dark blue) to 2.3 (reddish-brown) in our $HST$ images.  The LEGUS project typically used a value of CI $= 1.3$ to separate cluster candidates ($>1.3$) from stars (CI $\leq 1.3$).  In our plot, this CI cut corresponds to the dark blue colour.  
We show the images of four hand-selected clusters from NGC~628 in the right hand panel of Fig.~\ref{fig:MCIplane} and mark their position with star symbols, to illustrate the systematic change in cluster morphology as a function of MCI plane position.

The set of sources plotted in Fig.~\ref{fig:MCIplane} is intended to be limited to those for which MCI$_{\mathrm{in}}$ and MCI$_{\mathrm{out}}$ can be well-measured (e.g.\ $\le 10$\% error).  This is generally achieved for sources with $V$-band $\mathrm{S/N} \ge 10$, so we opted to display those data points in this figure.   Nevertheless, {\em for our morphology-based cluster candidate selection we further impose limits on MCI statistical error (MCI$_{\mathrm{in, err}}$ $\le 0.3$, MCI$_{\mathrm{out, err}}$ $\le 1.0$) alongside the $\mathrm{S/N} \ge 10$ cut.}\footnote{These MCI error and $\mathrm{S/N} \ge 10$ cuts do not apply to the super H-D sources described in Sec.~\ref{sec:maglimits}.} \footnote{We note that MCI error and signal-to-noise are highly correlated.  In future works we may consider only using a limit on $\mathrm{S/N}$ rather than in combination with limits on MCI$_{\mathrm{in}}$ and MCI$_{\mathrm{out}}$ error.  As it turns out, our present MCI error limits tend to preclude potential candidates even before $\mathrm{S/N} = 10$ is reached.  When assessing whether to make a change in our selection criteria, we will also need to consider that MCI error can be inflated at fixed cluster magnitude due to a crowded, complex environment.} 

\subsubsection{MCI selection: guidance from synthetic clusters} 
\label{sec:modelregions}

\begin{figure*}
 \includegraphics[width=2.2in]{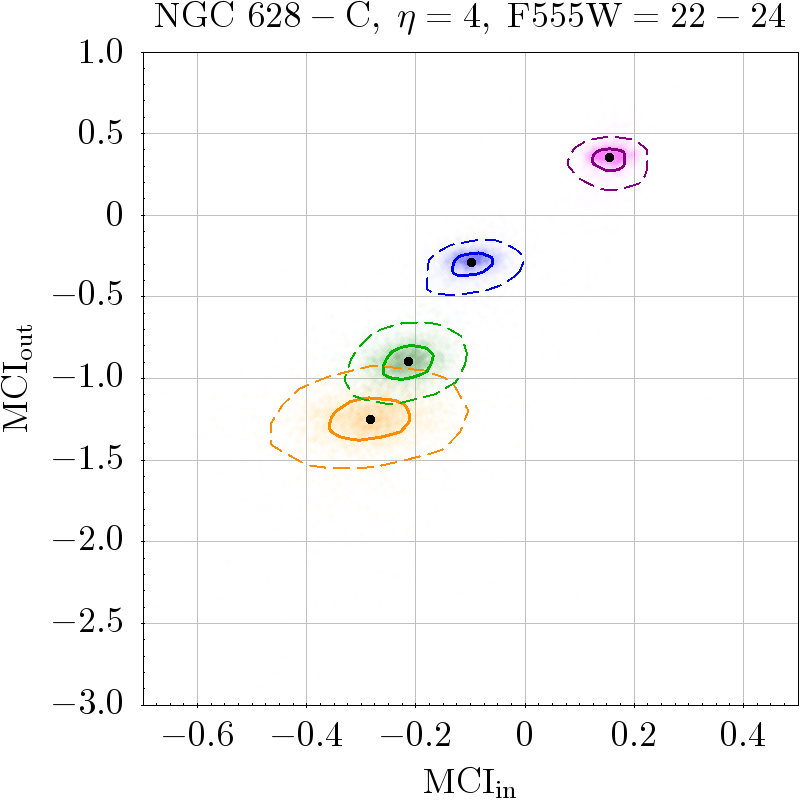}
 \includegraphics[width=2.2in]{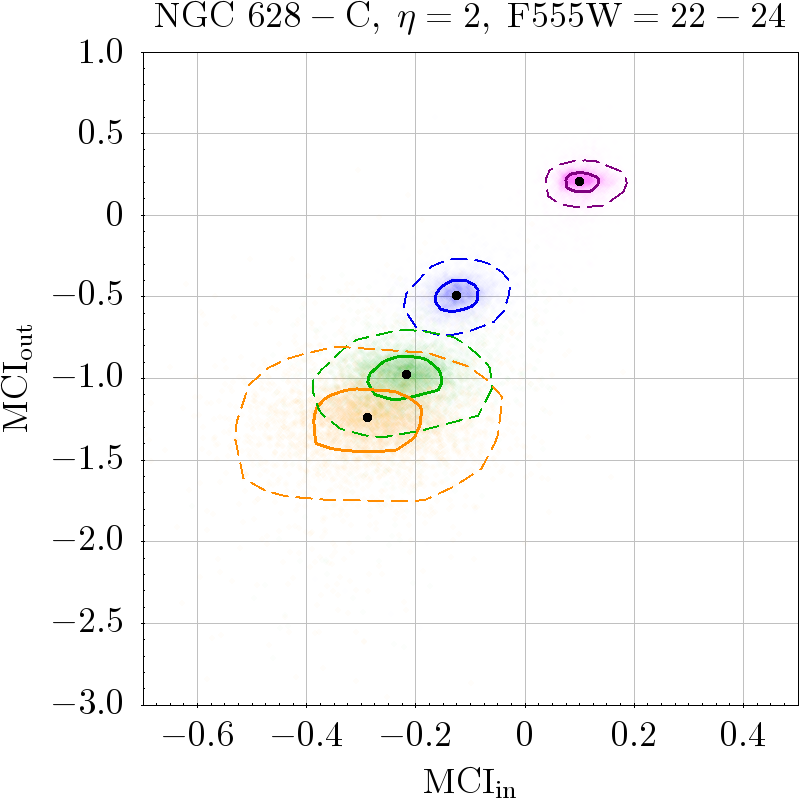}
 \includegraphics[width=2.2in]{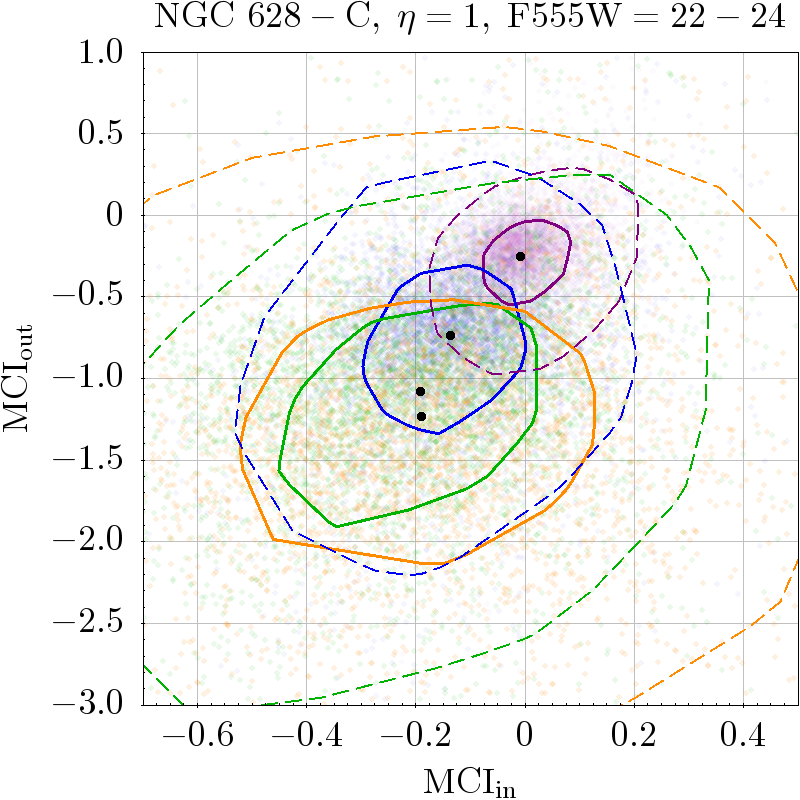}
 \caption{Bag plots illustrating the location of synthetic clusters in the MCI plane as a function of pre-PSF-convolution FWHM and halo slope ($\eta$).  From left to right, panels show models with $\eta = 4.0, 2.0, 1.0$.  In each panel, the transparent points are coloured according to FWHM, ranging from FWHM $= 1, 3, 5, 7$ pixels (purple, blue, green, orange).  This figure only shows synthetic clusters with $m(V) = 22{-}24$. The inner polygon drawn for each (FWHM,~$\eta$) combination is the `bag' containing $\sim50\%$ of the clusters, whereas the outer dashed `loop' indicates the region beyond which only outlier clusters are found. To the degree that bag plots are the bivariate equivalent of box plots, outliers can be thought to have coordinates more than 1.5$\times$ the inter-quartile range outside of the 25th-to-75th percentile zone.  For a univariate normal distribution, this corresponds to the $\sim2.7\sigma$ level.  The black dots mark the centre of each distribution.}
 \label{fig:MCIplanemodelsets}
\end{figure*}

In this Section, we investigate the location of realistic, synthetic clusters in the MCI plane.  This is done in order to guide our selection of actual cluster candidates using the MCI metrics. 
In Fig.~\ref{fig:MCIplanemodelsets} we show the measured values of MCI$_{\rm in}$ and MCI$_{\rm out}$ for thousands of realisations of different Moffat parameter permutations, spanning a range of FWHM and power-law halo slope.  The set of plotted synthetic clusters has been confined to the magnitude range of $m(V)=22-24$, in order to probe the faint end range of cluster magnitudes (thus the most typical population) in PHANGS-HST galaxies. The MCI plane position depends systematically on FWHM and~$\eta$, while scatter, for a constant distribution of masses, increases significantly for decreasing values of~$\eta$ (flatter halo slopes) at fixed FWHM and less so for increasing FWHM at fixed~$\eta$.  For relatively steep halo slopes ($\eta \ge 1.5$), the scatter in this magnitude range is acceptably small to distinguish between cluster morphologies.  However, it is clear (in the right hand panel) that for shallow slopes and large core size clusters the MCI plane scatter eventually renders the metrics less useful.  This is due to two reasons: (1) for this set of synthetic clusters the surface brightness above the background becomes very low, and (2) the range of cluster-centric radii probed by MCI$_{\rm in}$ and MCI$_{\rm out}$ fails to sample an informative portion of the radial surface profile.  In brighter magnitude ranges, such shallow slopes and large core size clusters do become well characterised in the MCI plane. This observation suggested to us that the set of synthetic clusters used to inform cluster candidate selection criteria should be pared down from the complete grid that was originally computed.

Specifically, we sought to delineate those regions in the MCI plane in which clusters could: (1) plausibly exist and (2) likely be detected by our observations.  This knowledge would then most effectively guide our selection of cluster candidates.  We emphasise that a realistic prediction of the cluster population (linked to expectations for the mass function, age distribution and size--mass relationship), subsequently used to establish selection regions in the MCI plane, is a disfavored method because this would create a selection bias tending to produce an observed sample in agreement with our prior notions.

\begin{figure*}
 \includegraphics[width=2.3in]{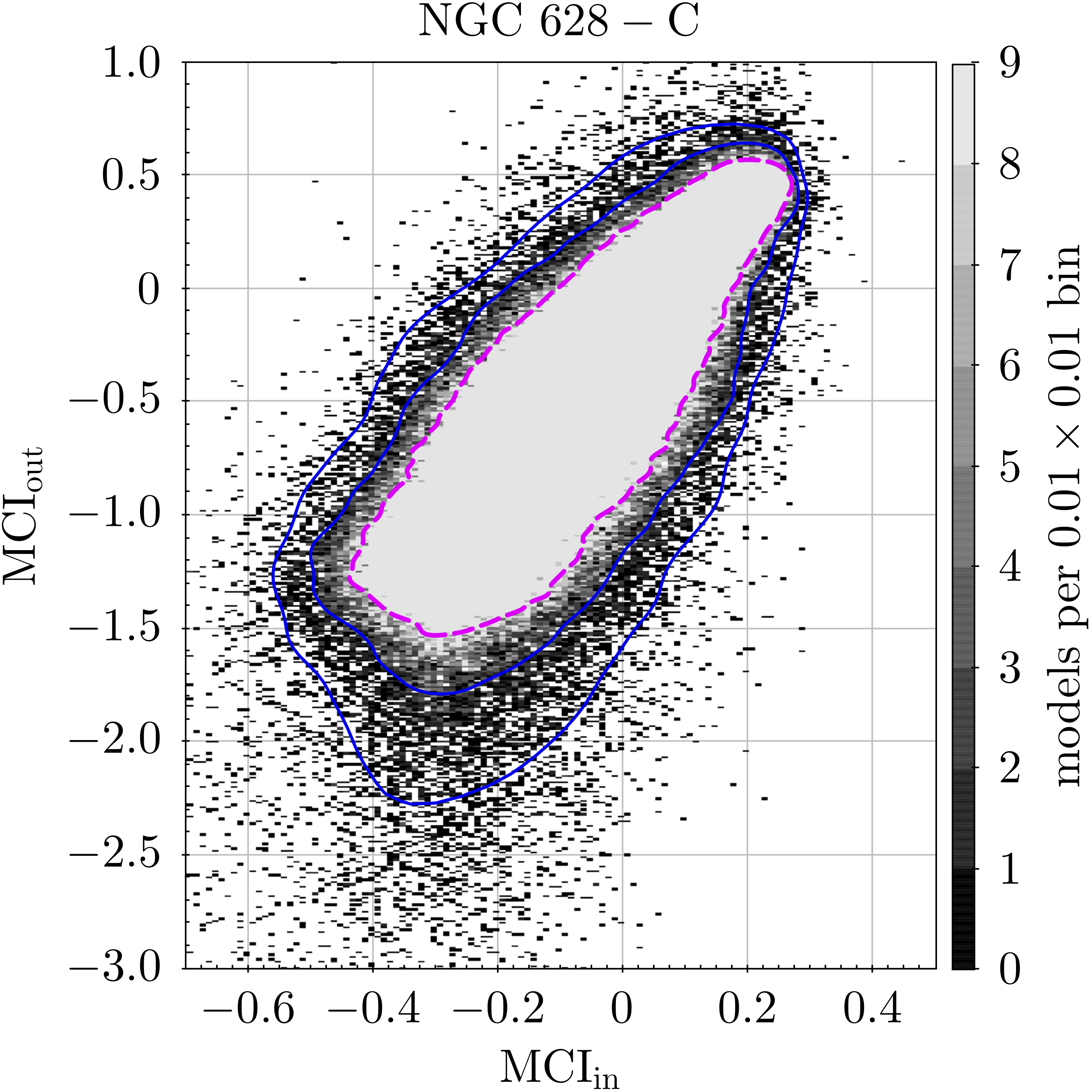}
 \includegraphics[width=2.3in]{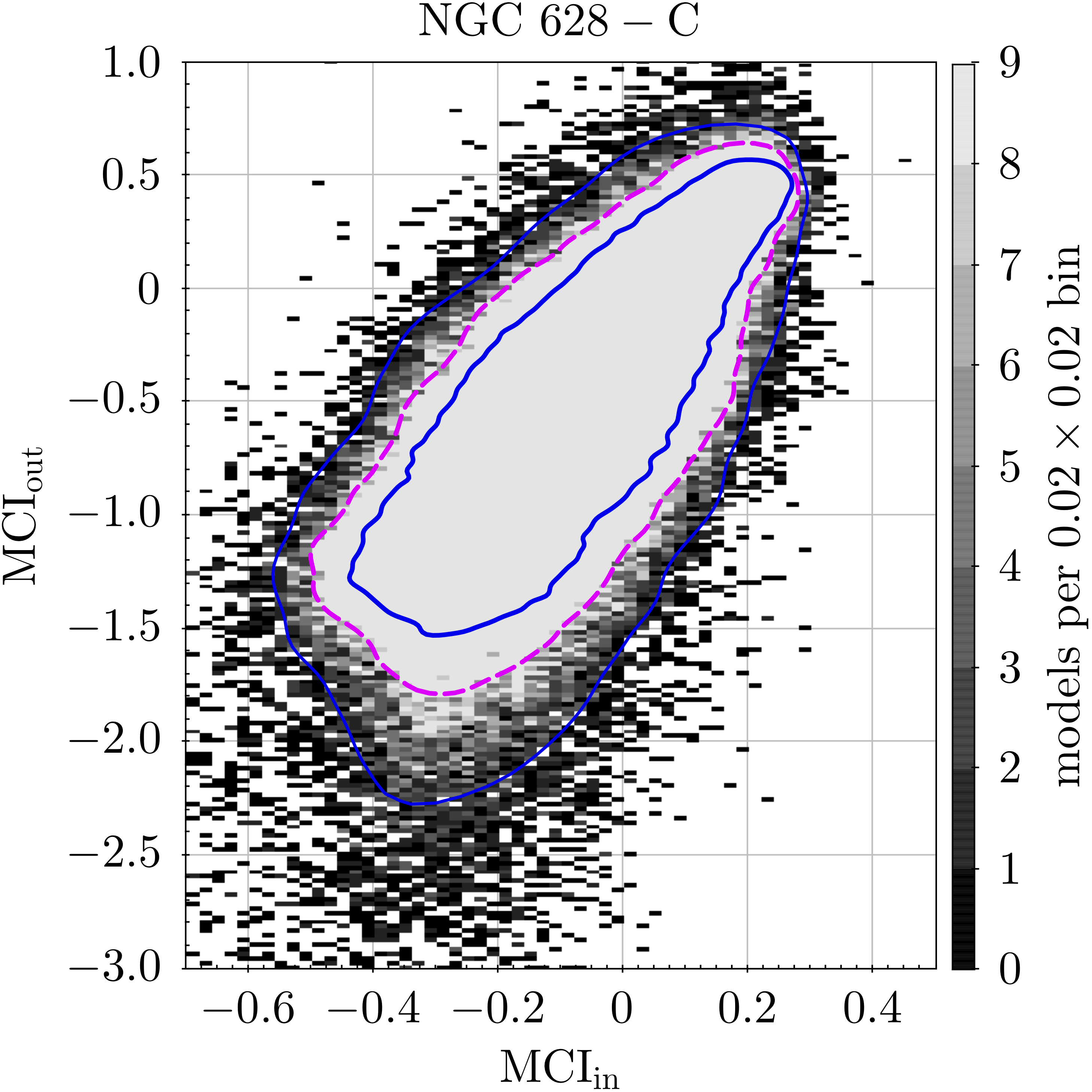}
 \includegraphics[width=2.3in]{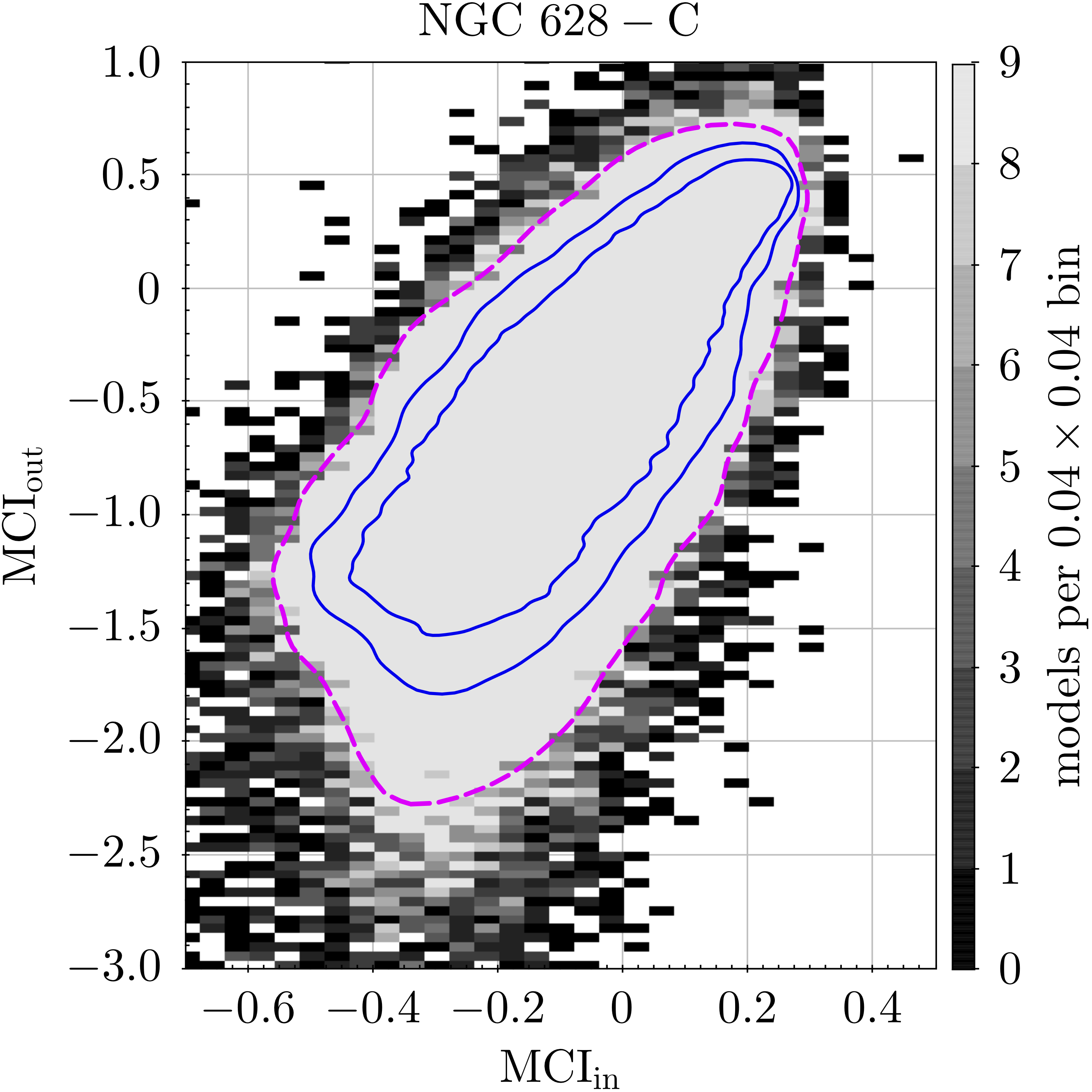}
 \caption{Hit maps computed for synthetic clusters in NGC~628-C, binned at different resolutions 0.01, 0.02, 0.04, and used for the determination of model cluster regions (curves) in the MCI plane for this particular target. Saturated light grey indicates a position registering a `hit' (at least 9 synthetic clusters that are both expected to exist and be detectable) for the plotted binning size. In each panel the model cluster region linked to the current binning resolution is drawn with a dashed magenta curve, whereas the regions at other resolutions are shown with the solid blue curves for comparison. See the text for existence and detection criteria assumed.} 
 \label{fig:hitmaps}
\end{figure*}

We use what we call ``hit maps,'' generated at a series of binning resolutions (0.01, 0.02, 0.04, in both MCI$_{\mathrm{in}}$ and MCI$_{\mathrm{out}}$), to identify the MCI plane regions in which to accept cluster candidates.  Displayed in Fig.~\ref{fig:hitmaps}, these hit maps are essentially histograms of the number of model clusters in the entire synthetic object database meeting {\em existence and detection criteria}, which are then processed to produce region boundaries at each MCI plane resolution.  During processing these histograms first become a binary mask -- we demand at least 9 models per bin to register as a hit (mask value of 1 rather than~0) at the associated bin position.  This guards against noisy boundaries, since Poisson event counting makes such bins significant at the 3$\sigma$ level. Such bins are saturated light grey in the plots of Fig.~\ref{fig:hitmaps}. To further mitigate noise, we then erode and dilate the hit map before contouring to obtain the bin-resolution-dependent model cluster region boundaries.  This process is completed separately for each target, using the specifics of the observations.  

The existence and detection criteria\footnote{Our formulation of the criteria is written using logical operators \& (`AND') and $|$ (`OR').  When interpreting the expression, please be aware of this notation.} are: $(r_{{\rm eff}} \le 10~{\rm pc})$ $\&$ $(\Pi \ge \text{3 | log(age) } \le 7)$ $\&$ (average background-subtracted surface brightness inside a radius of 2.5~pixels exceeding 0.04~e$^{-}$/sec).   The first two criteria ensure we only sample physically plausible clusters.  In the expression, ``boundedness'' (in a statistical sense, not necessarily for a specific object) is defined with the $\Pi$ ratio (= age / crossing time, see \citealt{GielesPortegiesZwart2011} where crossing time is estimated by assuming virial equilibrium) . We take along cluster models of all ages, though for ages greater than 10 Myr we require additionally that $\Pi \ge 3$ (in order to keep synthetic clusters in this post-natal age range which have an appearance/structure and age suggesting they are likely bound). For models younger than 10 Myr we make no such cut, since in this regime it is more difficult to distinguish bound from unbound objects without high-resolution spectroscopic observations.    The last criterion we check (minimum allowed surface brightness) is based on inspection of the distribution of this same quantity in all LEGUS cluster catalogues, and meeting this cut suggests detectability in our imaging since the data from both surveys are comparable in depth.

The end result of the steps we just described is a set of {\bf `model cluster regions'} in the MCI plane, generated specifically for each target/\linebreak[0]{}observation and at three incrementally decreasing resolutions.  The decreasing resolutions (increasing bin size) create nested model cluster regions (blue and magenta curves in Fig.~\ref{fig:hitmaps}), each of which adds a peripheral MCI zone of lower likelihood for viable cluster detections (though to zeroth order only, as we generate them intentionally neglecting prior knowledge about typical cluster populations).  Though not illustrated here, since we show only the NGC~628 central field (NGC~628-C), these regions vary systematically with distance to the target.  We return to this issue in Sec.~\ref{sec:EmpiricalSelectionReg}, while discussing semi-empirically determined selection regions.


Of key importance, we clarify that the model cluster regions were generated using synthetic clusters that approach a point-like appearance (minimum allowed pre-PSF-convolution FWHM of 0.5 pixel), and due to measurement error / environmental confusion, they contain the portion of the MCI plane in which point-sources dominate.  We allow such models to establish realistic expectations for which comparatively small clusters remain adequately distinguishable from stars.  However, it also means that we must take measures to exclude stars from our selection.  We initially experimented with \codename{IMFIT}-generated point source models inserted into the drizzled mosaic and generating model star regions, analogous to our methods for the synthetic Moffat clusters.  However, it was found that slight differences in the actual PSF versus the Anderson PSF, combined with a overly simplistic parameterised form for the magnitude distribution of inserted point sources, sometimes led to slight mismatch between the clear stellar peak in the observed MCI plane and model star regions.  

\begin{figure}
 \includegraphics[width=\columnwidth]{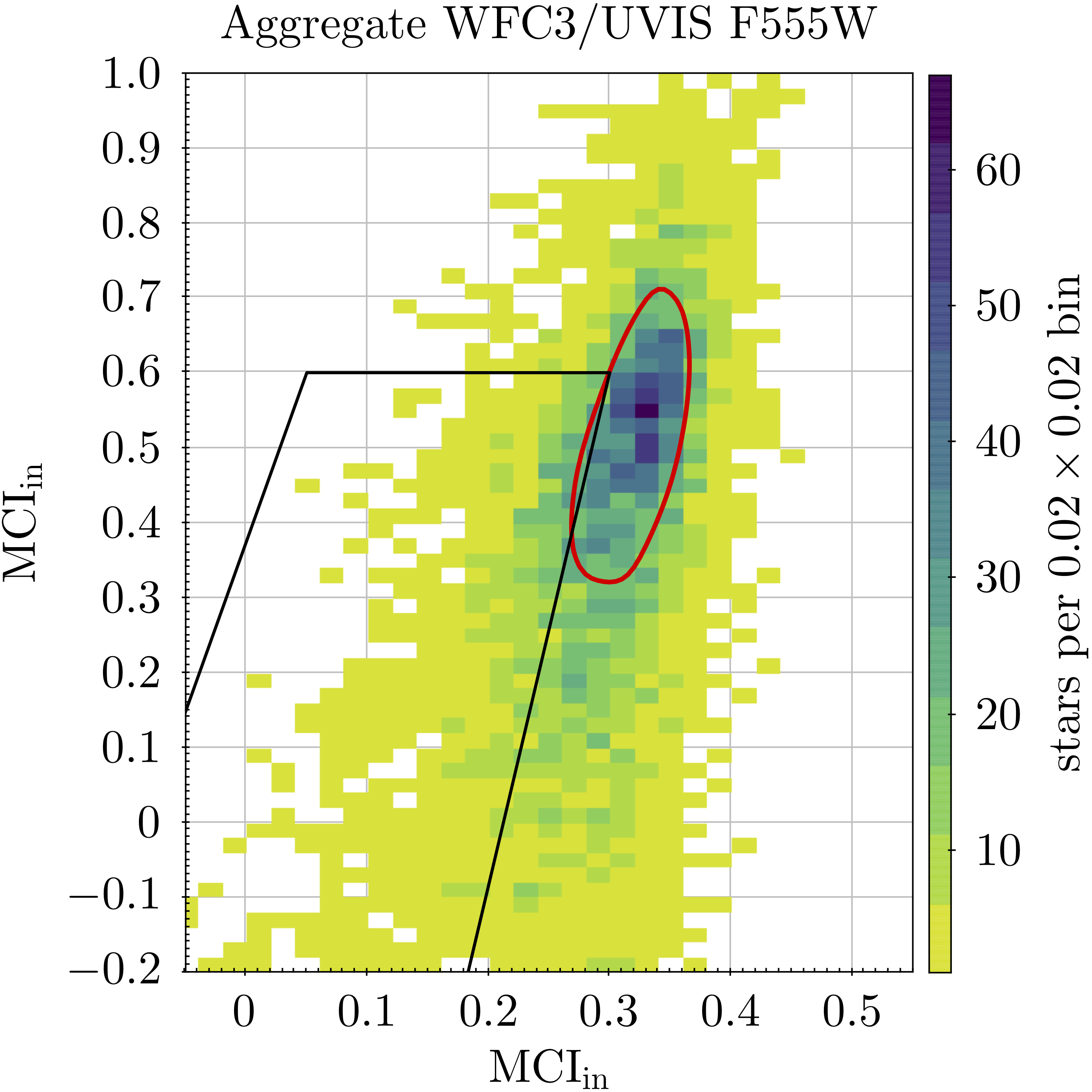}
 \caption{Aggregate stellar exclusion region (red curve) generated for {\em WFC3/UVIS} and $F555W$. Note that this plot is more zoomed in than other depictions of the MCI plane, to show detail. The background histogram shows a histogram of the qualified (see text) point-like sources aggregated over all PHANGS-HST targets in this instrumental configuration processed to date.  The polygonal shape is the upper right portion of the semi-empirical selection region (Sec.~\ref{sec:EmpiricalSelectionReg}), which is drawn for reference. Although not shown here, the equivalent plot for {\em ACS/WFC} (and $F555W$) indicates that point sources occupy a peak at significantly lower MCI$_{\mathrm{in}}$ ($\sim 0.21$, c.f. Fig.~\ref{fig:MCIplane} left panel and Figs.~\ref{fig:PHANGScandMCI}-\ref{fig:SyntheticCIwithallregions}) but comparable MCI$_{\mathrm{out}}$, owing to the lower resolution of the {\em ACS/WFC} PSF compared to {\em WFC3/UVIS}.  Note that we could have specially modified the MCI (NCI) normalisation constants used for {\em ACS/WFC} to eliminate such offset of point sources but chose not to do so, since only three of 38 PHANGS-HST targets (NGC~628, NGC~1300, NGC~3621) rely on archival ({\em ACS/WFC}) $V$-band data for cluster candidate detection/selection, and because our selection/exclusion regions already compensate by construction.}  
 \label{fig:aggregatestellarregion}
\end{figure}

Therefore, we adopt a simpler, yet more empirical method, for defining what we call a {\bf `stellar exclusion region'}.  From the \codename{DOLPHOT} catalogue of each target, we cull a set of actual, high significance stars (\codename{DOLPHOT} $\mathrm{S/N} \ge 40$, sharpness consistent with a point source, and $\chi^2$ less than a tunable limit based on inspection) and use this set to generate a KDE-smoothed distribution of stars in the MCI plane.  Our final adopted stellar exclusion region is defined by a contour at 50\% of the peak in this distribution.  We generate this product for each galaxy individually.  In some cases, we need to increase the degree of KDE smoothing due to low numbers of qualified point sources, or in extreme cases substitute a camera/\linebreak[0]{}band-specific `aggregate' stellar exclusion region for the galaxy-tuned determination.  This latter product is generated from the union of qualified point sources in all targets.  An example of the aggregate product for {\em WFC3/UVIS} datasets is shown in Fig.~\ref{fig:aggregatestellarregion} (red curve).  The stellar exclusion region shifts substantially for {\em ACS/WFC} versus {\em WFC3/UVIS}, due to camera-related differences in effective PSF. The galaxy-specific stellar exclusion regions will be published in our entire-sample cluster catalogue release paper, but the {\em ACS/WFC}-based region for NGC~628-C is plotted later in the current paper (as part of Figs.~\ref{fig:PHANGScandMCI}, \ref{fig:MCImodelregions}, and~\ref{fig:SyntheticCIwithallregions}).  

The stellar exclusion region successfully removes the majority of stars, with limited inevitable exceptions due to low $\mathrm{S/N}$ or confusion. Conversely, the stellar exclusion region is expected to remove only a very small percentage of detectable clusters given the typical mass ($\sim10^{4}$~M$_{\odot}$) of sources recovered in our survey, coupled with the cluster mass-radius relation (e.g. \citealt{Krumholz2019review} with supporting data on Milky Way and M31 clusters from \citealt{Johnson2012,Fouesneau2014,Kharchenko2013}). At a mass of $10^{4}$~M$_{\odot}$ a representative cluster effective radius is $\sim 3$~pc ($\sim 1$~pixel at 15~Mpc), following the observed relation presented in  \citet{Krumholz2019review}, and implying that such a cluster would have MCI values well outside the stellar exclusion region.  Of course, the typical cluster mass we use to support this argument is likely itself modified by selection.  Forthcoming cluster completeness testing as a function of mass, age, and morphology will quantitatively account for candidate loss due to the stellar exclusion region, as well as other cuts of our selection method.

To summarise our MCI-based, synthetic cluster guided selection pathway: for the subset of sources fainter than the H-D limit \citep[][]{Humphreys1979}, we select as a cluster candidate any object meeting the general conditions outlined at the start of Sec.~\ref{sec:selection}, having $V$-band $\mathrm{S/N} \ge 10$ and MCI$_{\mathrm{in, err}}$ $\le 0.3$ plus MCI$_{\mathrm{out, err}}$ $\le 1.0$, and falling within one of the model cluster regions (Fig.~\ref{fig:hitmaps}) but outside the empirically-defined stellar exclusion region (Fig.~\ref{fig:aggregatestellarregion}). After adding possible objects brighter than the H-D limit (Sec.~\ref{sec:maglimits}), this set of candidates is known as our {\bf `ML sample'} (machine learning). Additional illustration of galaxy-specific MCI plane selection model cluster regions and stellar exclusion regions is included in the PHANGS-HST survey paper \citep[][]{PHANGSHSTsurvey}.



\subsubsection{MCI selection: semi-empirical selection region}
\label{sec:EmpiricalSelectionReg}

The synthetic model cluster regions presented above provide an optimally tuned selection of cluster candidates, specific to a given galaxy in terms of its internal structure/\linebreak[0]{}confusion-level, distance, and also the quality of the observation.  However, the construction/\linebreak[0]{}use of such regions requires substantially detailed analysis of synthetic clusters (Sec.~\ref{sec:synthetic}) and high-significance stars in the field of interest.  General users of our MCI-based method may not wish to do this.  For this reason, and also to quickly generate a standardised candidate set of manageable size for human classification, we designed a polygonal {\bf `semi-empirical selection region'} (for cluster candidates) in the MCI plane.  To exclude stars from this selection, in the case of general users, we suggest adopting the aggregate stellar exclusion region of Fig.~\ref{fig:aggregatestellarregion} (though not used by our pipeline, except when stars are too rare), or more simply a lower limit on allowed traditional CI for candidate clusters ($\sim 1.3-1.4$) determined by the user for their own particular dataset. Stellar exclusion is barely even indicated for {\em WFC3/UVIS} data sets but is very much necessary for {\em ACS/WFC} data sets.

\begin{figure*}
 \includegraphics[width=\columnwidth]{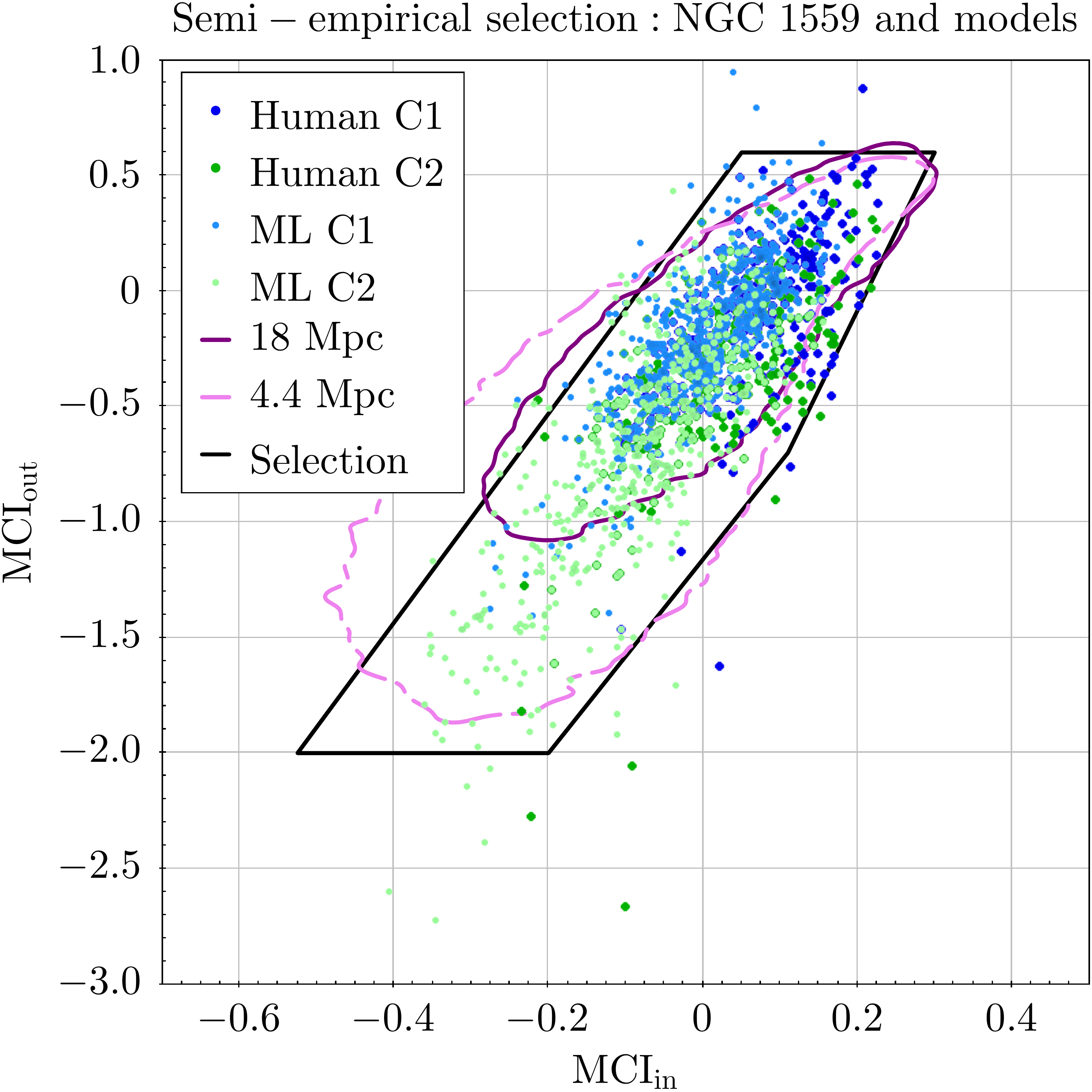}
 \includegraphics[width=\columnwidth]{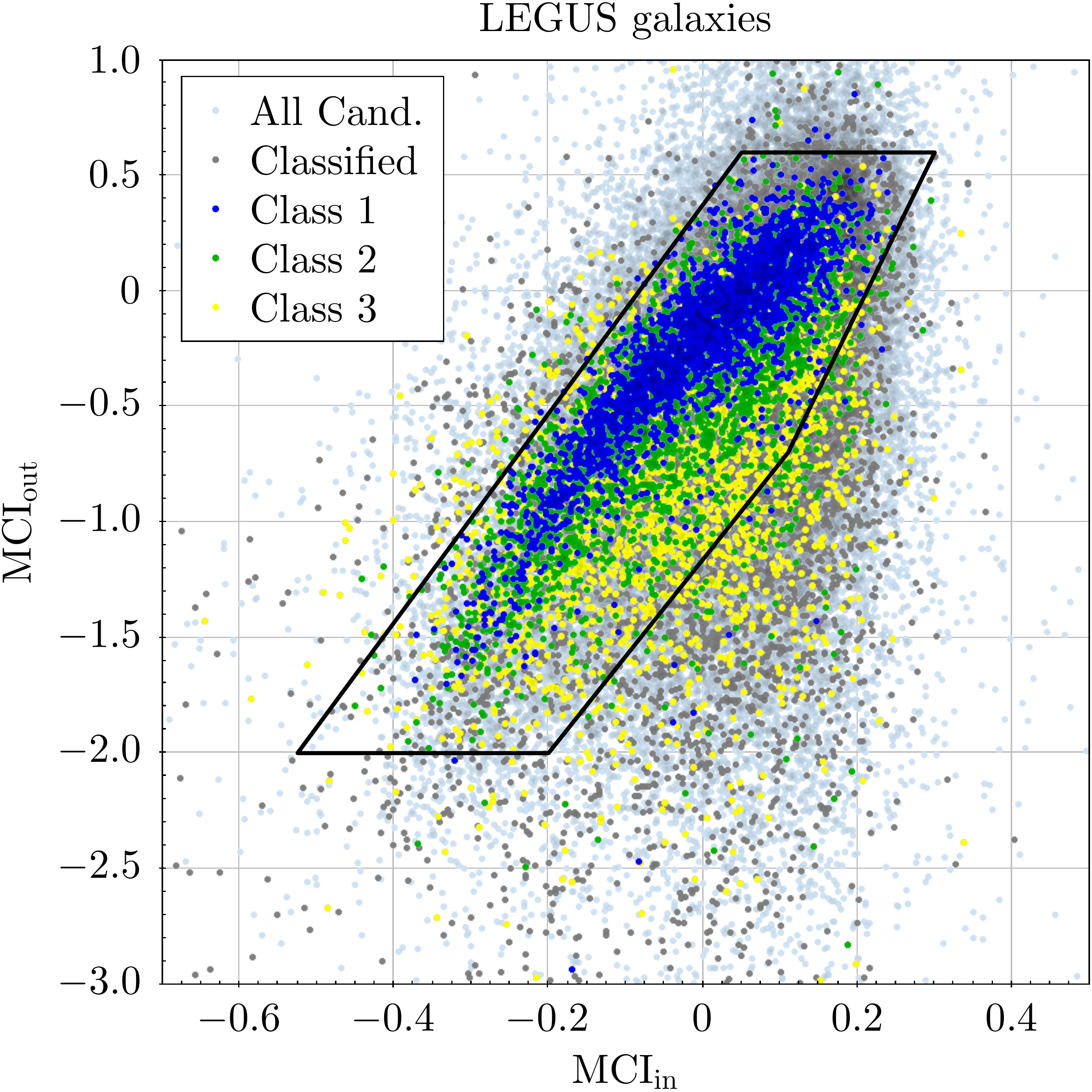}
 \caption{Development of the semi-empirical selection region in the MCI plane. (left) The distribution of clusters in early PHANGS-HST target NGC~1559 (at a distance of 18 Mpc) in comparison to high-resolution model selection regions computed from synthetic clusters (18 Mpc, purple, near the median PHANGS-HST distance; 4.4 Mpc, violet, more suitable for LEGUS galaxies and nearby PHANGS-HST target NGC~4826). Based on this guidance we define the semi-empirical selection region (black polygon), shown here without distance-based cropping at the bottom edge. Symbol colour and size differentiate cluster type (Class 1 and 2) and classification method (human and ML) as indicated in the legend. (right) The MCI plane and the semi-empirical selection region, showing all publicly released LEGUS cluster candidates aggregated across 34 galaxies (light blue, and human classified: grey), with Class~1 clusters marked in dark blue, Class~2 in green, and Class~3 (compact associations) in yellow.  }
 \label{fig:LEGUScandallgalMCI}
\end{figure*}

\begin{figure}
 \includegraphics[width=\columnwidth]{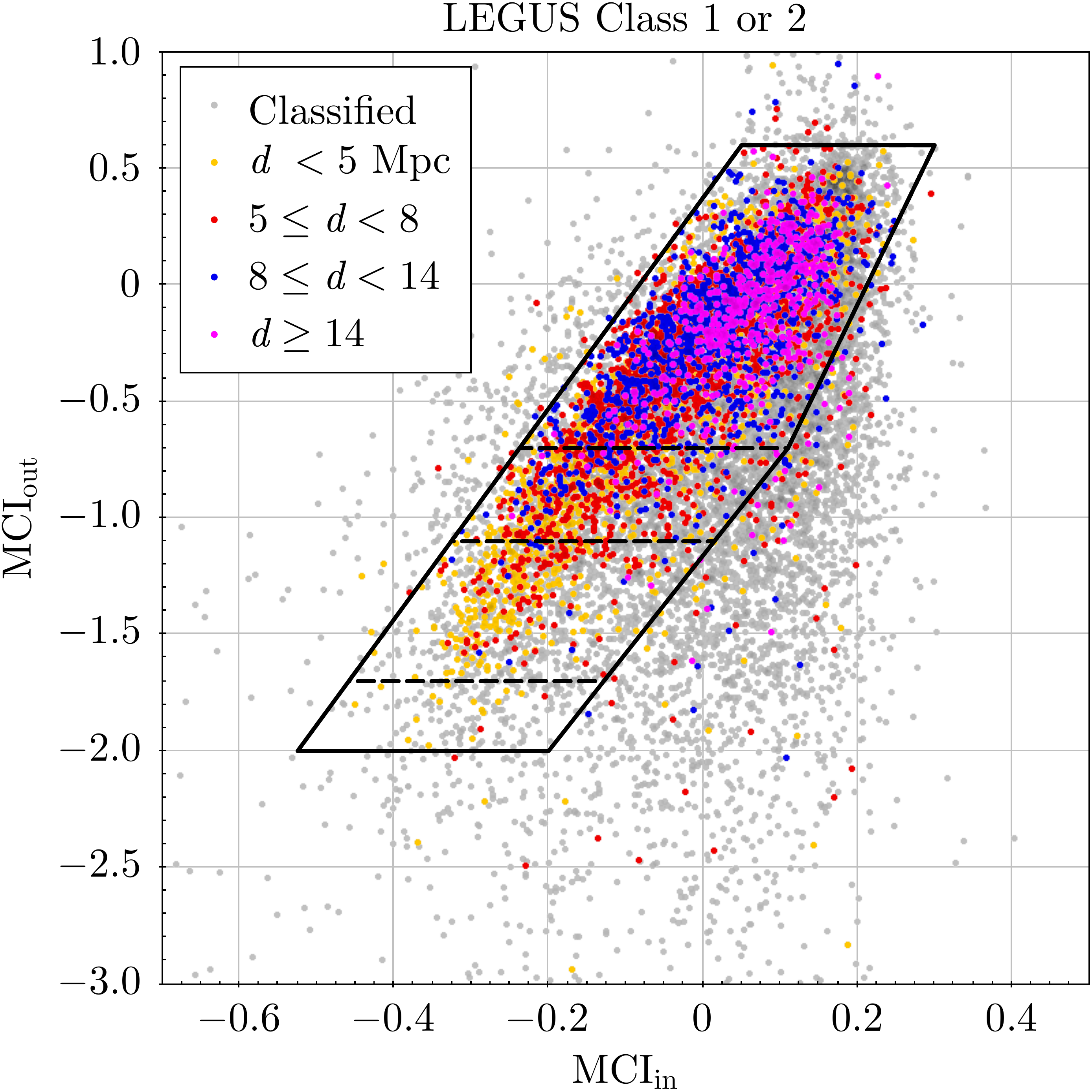}
 \caption{The semi-empirical selection region (thick polygon), with variable lower limits in MCI$_{\mathrm{out}}$ (thin dashed lines).  We show LEGUS confirmed Class~1 or~2 clusters as points colour coded according to the distance range which they occupy.  Note the typical agreement between the observed lower end of the cluster distribution with the progressively lowered MCI$_{\mathrm{out}}$ semi-empirical selection region limit, where the $-1.7$ limit is for $d \le 8$~Mpc, $-1.1$ for $8 < d < 14$~Mpc, $-0.7$ for $d \ge 14$~Mpc.}
 \label{fig:MCIEmpiricalSelectionRegEdge}
\end{figure}

\begin{figure}
 \includegraphics[width=\columnwidth]{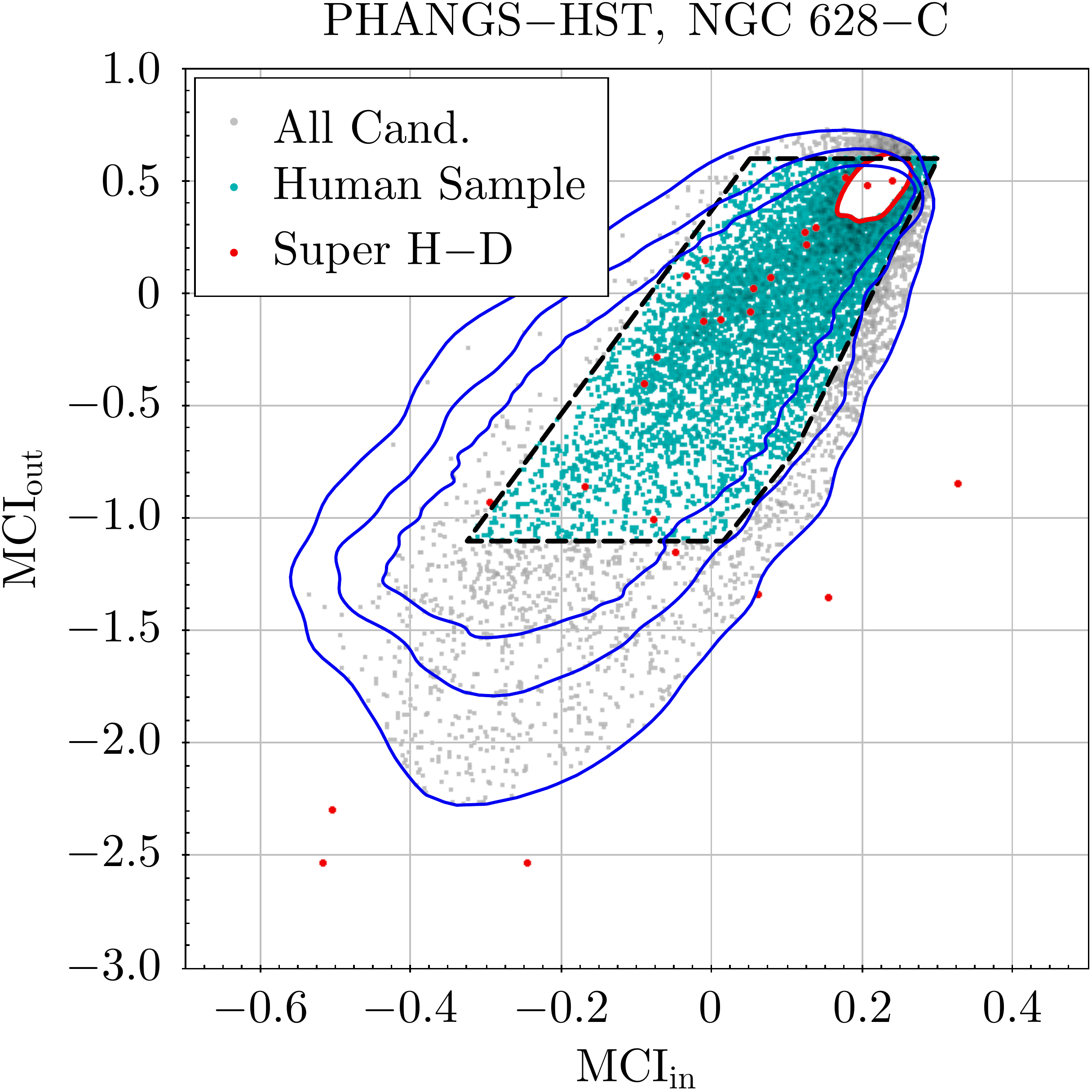}
 \caption{PHANGS-HST NGC~628-C cluster candidates (grey) and the subset of semi-empirical region selected (cyan) or super H-D sources (red).  Solid blue curves are our model cluster regions for 0.01, 0.02, and 0.04 bin sizes.  The dashed black polygon marks the semi-empirical selection region (for this distance), whereas the solid red curve is the stellar exclusion region. Note that the stellar exclusion region intrudes significantly into the selection area, only due to this target being observed with {\em ACS/WFC}.   The majority of the super H-D sources fall along the locus of highest cluster candidate density (within both the semi-empirical selection region and the 0.01 bin size model cluster region) and confirmed by inspection to be luminous clusters. The galaxy nucleus is the super H-D source near MCI$_{\mathrm{out}}$ = -0.3.  We have inspected the super H-D sources outside the model cluster region, finding all but two to be artefacts associated with bright foreground stars. The two super H-D sources in the far bottom left corner of the figure are off-centre detections of the galaxy nucleus. Two super H-D sources fall within the stellar exclusion region.  According to the Gaia GR2 distances of \citet{BailerJones2018}, they are both foreground stars within 2 kpc of the Sun.}
 \label{fig:PHANGScandMCI}
\end{figure}

\begin{figure}
 \includegraphics[width=\columnwidth]{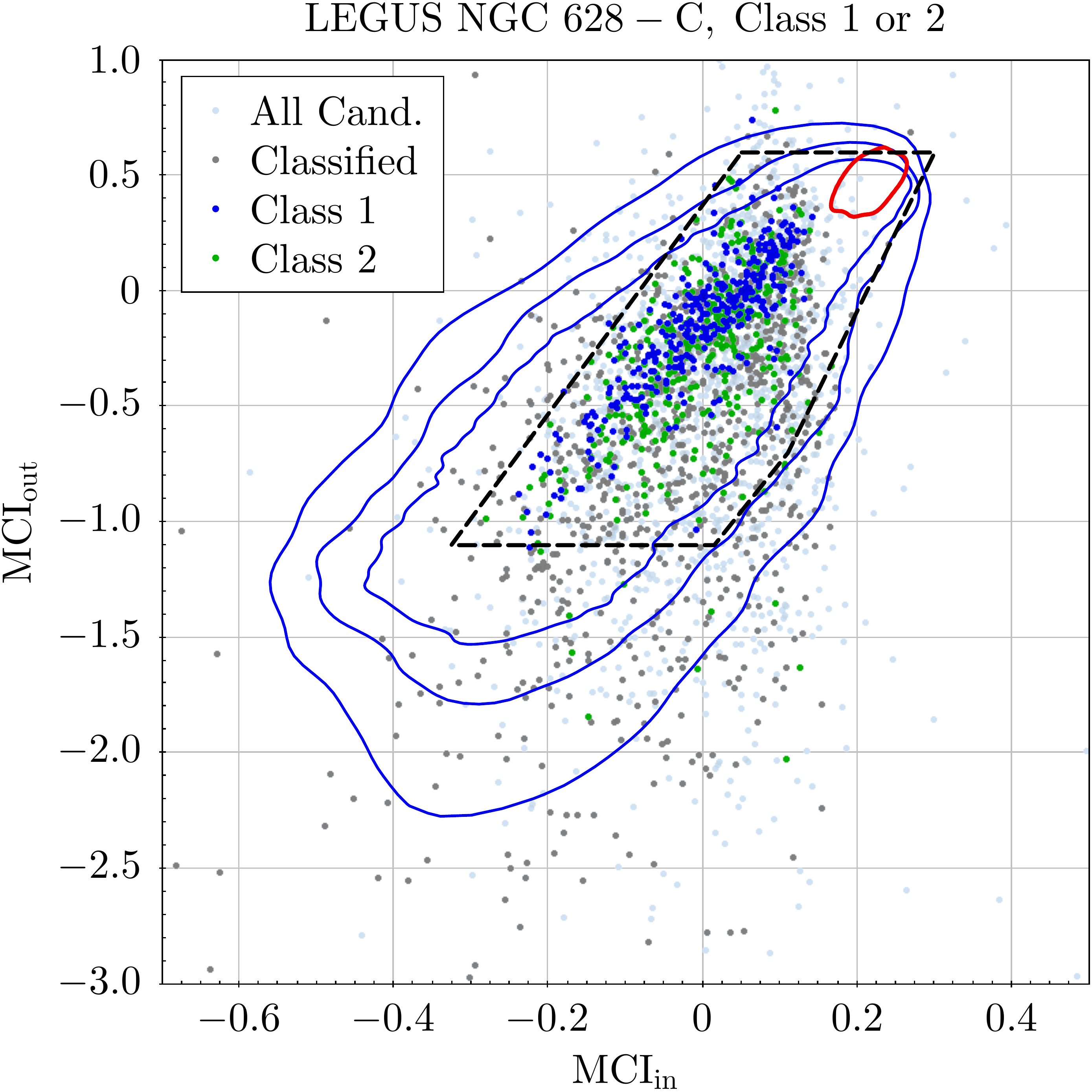}
 \caption{This plot, for NGC~628-C only,  shows LEGUS candidates (light blue, and human classified: grey) and LEGUS human confirmed clusters (of Class~1,~2 -- blue, green, respectively).  Solid blue curves are our model cluster regions for 0.01, 0.02, and 0.04 bin sizes.  The dashed black line marks the semi-empirical selection region (for this distance) and the red curve is the stellar exclusion region for NGC~628-C.   Note that the stellar exclusion region for NGC~628-C intrudes significantly into the selection area, only due to this target being observed with {\em ACS/WFC}.  For {\em WFC3/UVIS} targets, the stellar exclusion region is generally adjacent to, or only slightly overlapping, the semi-empirical selection region.}
 \label{fig:MCImodelregions}
\end{figure}

\begin{figure}
 \includegraphics[width=\columnwidth]{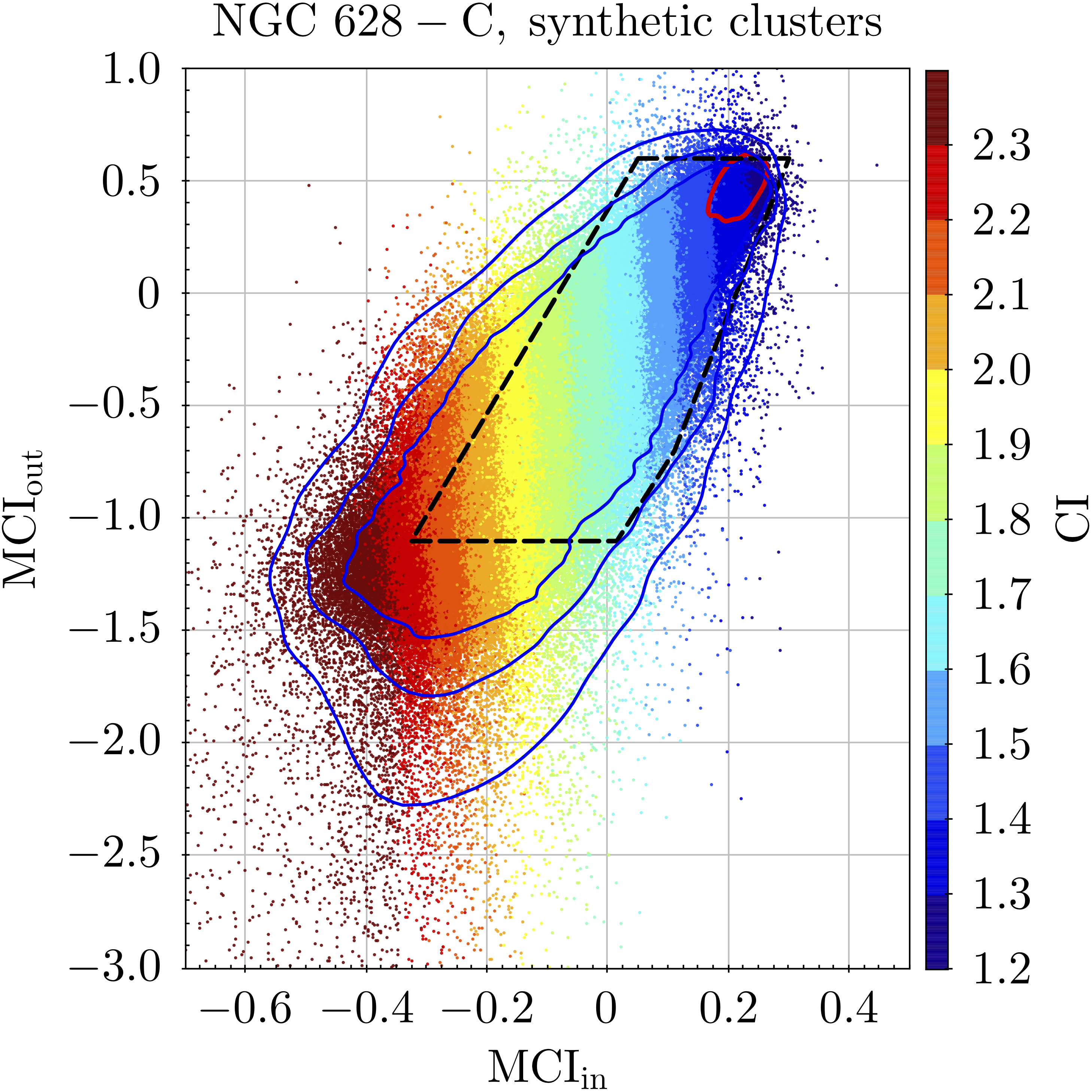}
 \caption{The set of synthetic clusters computed for {\em ACS/WFC} observations of NGC~628-C which could (1) plausibly exist and (2) likely be detected by our observation, colour-coded by CI. We have overplotted all selection/\linebreak[0]{}exclusion regions, with a lower limit appropriate to the distance of NGC~628 in the case of the semi-empirical selection region.}
 \label{fig:SyntheticCIwithallregions}
\end{figure}

In order to decide on the form of the polygon adopted, we jointly used information from actual clusters and synthetic models, with actual clusters  in our analysis coming from the first few PHANGS-HST galaxies studied (e.g. NGC~1559, \citet{Wei2020}; NGC~3351, \citet{Turner2021}; NGC~4548).  At this early stage of our $HST$ program, sources were inspected over a wide swath of the MCI plane (essentially spanning the entire distribution).  Then, the distribution of human verified clusters in combination with the (tightest) model cluster region evaluated for several targets spanning the distance range of our sample was used to semi-empirically define the edges of the adopted polygon.  In Fig.~\ref{fig:LEGUScandallgalMCI} the left hand panel shows the verified clusters of NGC~1559 together with representative model cluster regions, illustrating how such guidance was used to motivate the semi-empirical selection region.  The adopted polygon is shown in various ways in Figs.~\ref{fig:aggregatestellarregion}--\ref{fig:SyntheticCIwithallregions}.  Some slack is allowed in the upper right hand corner of the semi-empirical selection region, in order to allow the smallest recoverable {\em WFC3/UVIS} detected clusters to the maximum degree allowed by our stellar exclusion regions.  We elected to shave off some from a simple quadrilateral in the lower right portion of the selection region, aiming to reduce contamination mildly.  The top edge is fixed at MCI$_{\mathrm{out}}$ = 0.6.

As a validation check (and possible tertiary influence) on the semi-empirical selection region, we examined the MCI plane distribution of LEGUS Class~1 and~2 clusters, and Class~3 compact associations.  The right hand panel of Fig.~\ref{fig:LEGUScandallgalMCI} shows this data set.  Note the tightly confined locus of Class~1 clusters, with the Class~2 clusters scattering to somewhat more negative (extended) MCI$_{\mathrm{out}}$ at fixed MCI$_{\mathrm{in}}$.  Compact associations are even more scattered.  If we ignore Class~3, which are being treated in a more robust manner by the PHANGS-HST pipeline \citep[see the multi-scale associations of][]{Larson2021}, this figure makes it clear that our semi-empirical selection region captures the vast majority of compact clusters.

Because the MCI (and NCI) are normalised to a fiducial cluster of fixed angular size, we expected distance-dependent variations in the MCI plane distribution.  This is illustrated in Fig.~\ref{fig:MCIEmpiricalSelectionRegEdge}.  We adopt a variable lower limit on the bottom edge of the semi-empirical selection region.  Specifically, the region is cropped at MCI$_{\mathrm{out}} = -1.7$ for galaxies with $d \le 8$~Mpc, at MCI$_{\mathrm{out}} = -1.1$ for $8 < d < 14$~Mpc, and at MCI$_{\mathrm{out}} -0.7$ for $d \ge 14$~Mpc.  Future PHANGS-HST papers dealing with cluster catalogues may use a revised ladder of semi-empirical selection region edges, but our goal for the present work was to be conservative, especially in the case of targets for which significant distance uncertainty persists.  Note that for targets considerably closer than 8 Mpc, such as many of the LEGUS galaxies, a limit of MCI$_{\mathrm{out}} = -2$ is appropriate since several LEGUS confirmed Class 2 clusters and even a few Class 1 clusters appear in this regime as can be seen in the right panel of Fig.~\ref{fig:LEGUScandallgalMCI}.  The model region plotted for PHANGS-HST target NGC~4826 (at 4.4 Mpc) in the left panel of Fig.~\ref{fig:LEGUScandallgalMCI} also supports this conclusion.

To enable readers to utilise our MCI-based selection method without running synthetic cluster models of their own, we provide the vertices of the semi-empirical selection region in Table~\ref{tab:empiricalselectionregion} before adjusting the bottom edge according to galaxy distance.

\begin{table}
    \caption{Vertices of the semi-empirical selection region}
	\begin{center}
    \begin{tabular}{rr}
 \hline
    MCI$_{\mathrm{in}}$ & MCI$_{\mathrm{out}}$ \\ 
    \hline    
    0.050 & 0.600 \\
    0.300 & 0.600 \\
    0.110 & -0.700 \\
    -0.200 & -2.000 \\
    -0.525 & -2.000 \\
    0.050 & 0.600 \\
    \hline    
    \label{tab:empiricalselectionregion}
    \end{tabular}
	\end{center}
    \vspace{-10pt}
    \begin{tablenotes}
	\small
\item[$a$] Note: The bottom edge used to crop the semi-empirical selection region on a galaxy-by-galaxy basis depends on the distance to the target (e.g. Fig.~\ref{fig:MCIEmpiricalSelectionRegEdge}), in the sense of the selection region shrinking at the high MCI$_{\mathrm{out}}$ end with increased distance.  Specific recommendations are given in the text. 
    \end{tablenotes}
\end{table}

Recall that in addition to providing for a simple cut in the MCI plane that would work without running synthetic cluster models, the basic motivation for the semi-empirical selection region was to generate a smaller subset of the overall candidate list to be considered for human classification.  Because our semi-empirical selection region (after distance-based cropping) is generally comparable or smaller in area to the highest resolution model cluster region (0.01 bin size), we expect the purity of the selection ($\equiv$~\#~of true clusters / \#~cluster candidates) to be similarly elevated in this subset,  though potentially at the cost of missing cluster candidates of unusually large size or in complex/crowded regions that perturb MCI measurements away from the nominal locus. 

Summarising our semi-empirical selection pathway, we select as a candidate in this {\bf `human classification sample'} any source (meeting the general criteria outlined at the start of Sec.~\ref{sec:selection}, i.e.\ having $V$-band $\mathrm{S/N} \ge 10$, and being inside the semi-empirical selection region but outside the stellar exclusion region defined earlier), \emph{or} brighter than the H-D limit (Sec.~\ref{sec:maglimits}).  We reiterate though that two selection pathways are implemented, and that a broader sample of candidates based on model cluster expectations alone is ultimately passed to ML classification.  In Fig.~\ref{fig:PHANGScandMCI}, we indicate for one target (NGC~628-C) the human subset (cyan) of the entire candidate cluster set (grey).  Both the semi-empirical and model cluster selection regions are plotted, along with the stellar exclusion region, so they can be compared.  Here one can see the {\em ACS/WFC}-specific need to exclude stars from the semi-empirical selection region, not significant for {\em WFC3/UVIS} data sets (see Fig~\ref{fig:aggregatestellarregion}).

As a final double check that our method works as expected, we compared the selection areas for each galaxy in our paper's sample to the MCI plane distribution of LEGUS clusters in that same target.  Figure~\ref{fig:MCImodelregions} shows such a plot for NGC~628-C, and confirms that only a handful of clusters confirmed by LEGUS lie outside our semi-empirical selection region, and only two are missed by our model cluster regions (others may possibly fail general conditions of $\mathrm{S/N}$, \# of low error bands, etc.).

Lastly, in Fig.~\ref{fig:SyntheticCIwithallregions} we compare the MCI plane distribution of our synthetic cluster models for NGC~628-C to standard CI.  This figure is similar to the plot in the right panel of Fig.~\ref{fig:MCIplane}, except that here we show models rather than detected sources.  We also indicate the model cluster regions, stellar exclusion region, and semi-empirical selection region so their connection to the traditional CI metric can be ascertained.

\subsection{Bright object criteria: super H-D cluster candidates}
\label{sec:maglimits}

As noted in Sec.~\ref{sec:generalcriteria}, we treat the brightest sources in our augmented \codename{DOLPHOT}+\codename{DAOStarFinder} catalogue slightly differently than the vast majority of (fainter) possible cluster candidates. The luminosity distribution for stars is strongly limited at the bright end.  The Humphreys--Davidson limit \citep[][a.k.a. H-D limit]{Humphreys1979} effectively delineates a demarcation in the Hertzsprung--Russell diagram above which stars can only briefly exist, in an unstable state.  This limit can be cast into an equivalent boundary in the ($V{-}I$,~$I$) CMD.  There is colour-dependence, but for our purposes we adopt $M_V = -10$ as the H-D limit.  Clusters exist both fainter and brighter than this magnitude, but given the rarity of stars brighter than the H-D limit, it is reasonable to assume that nearly all sources brighter than the H-D limit are clusters.  For this reason, we accept as a cluster candidate any source surviving the cuts outlined as general criteria in Sec.~\ref{sec:generalcriteria} and having $M_V \le -10$ if it also meets the very lax condition of having $-0.55 \le \mathrm{MCI}_{\mathrm{in}} \le 0.45$ (since they can sometimes be [nearly-]saturated or have close neighbours within 5 pixels, leading to corrupted MCI values).  We call these objects {\bf `super H-D candidates'}. They are marked as red symbols in Fig.~\ref{fig:PHANGScandMCI}. In practice, super H-D candidates are a mix of luminous clusters and artefacts.  See the caption of Fig.~\ref{fig:PHANGScandMCI} for more information, but note that contaminating foreground stars (being misinterpreted at the distance of the target galaxy) can also sometimes make it into this subset, and are later discarded to the best of our ability during classification. Objects of this sort should appear within or near the stellar exclusion region, however genuine luminous clusters having intrinsically compact morphology \citep[e.g.][]{Smith2020} would also manifest as super H-D objects found in the exclusion region -- particularly for the more distant galaxies in the PHANGS-HST sample. 

Point-like super H-D cluster candidates present a particular challenge during classification.  We can utilize Gaia Bayesian distance estimates \citep[][]{BailerJones2018} to weed out foreground stars, but such parallax-based ancillary information is not always conclusive since the method depends on sky location with respect to the Galactic model and because parallax uncertainty is excessive for some super H-D sources (e.g. our assumed H-D limit of $M_V = -10$ implies $m_V = 20$ for a galaxy at 10 Mpc).  The appearance of an Airy ring or sharp diffraction spikes can also help to weed out foreground stars.  When neither Gaia nor PSF-wing source morphology helps, we rely on contextual hints from where in the galaxy image a point-like super H-D source is found.  For instance, if it lies in a complex of active star formation or in a comparatively luminous region, and the angular surface density of other point-like super H-D sources nearby in the celestial sphere is low enough that chance projection on such areas is unlikely, then we generally accept the cluster candidate as being a bona fide cluster.  A dedicated, follow-up study of super-HD clusters will explore this issue in more detail.  Finally, one might wonder how complete our catalogues are for super H-D clusters.  \citet{Whitmore2021} looks at this issue for NGC~628, concluding that our catalogues approach 90\% completeness in this bright regime, despite the complexities of presented by point-like super H-D candidates.

\subsection{Eliminating duplicate and/or redundant cluster candidates}
\label{sec:doubles}

As it is originally intended to be a point-source finding and PSF-fitting photometry routine, \codename{DOLPHOT} sometimes chooses to represent an extended source as two or more very closely spaced point sources, rather than passing through only one dominant component. Although well motivated for resolved stellar photometry of crowded fields, and good for close pairs of stellar sources (as in the context of our PHANGS-HST individual star photometry and association analysis), this is a drawback in our case of cluster detection.  We included a general condition on \codename{DOLPHOT} crowding in our candidate selection to mitigate this situation from the start of our cluster-specific work.  See Sec.~\ref{sec:generalcriteria} for specific implementation, but we only allow \codename{DOLPHOT} sources with crowding $\le 0.667$ mag.   
This crowding cut eliminates a majority of cases of duplicative detection, but at the very end of our cluster candidate selection we also explicitly impose a `doubles radius' cut.  After the preliminary candidate list is established, we sort candidates by count rate and work through the list in order of decreasing count rate, disqualifying any fainter \codename{DOLPHOT} artefact (= {\bf `duplicate'}) neighbouring candidate(s) at separations  $\le 2.5$ pixels (hence quasi-unresolved by $HST$).   We cannot make the doubles radius any larger without starting to lose bona fide double objects (not artefacts) of which the fainter source could conceivably be a cluster.  

\codename{DOLPHOT} deblending artefacts that do squeak through this cut are later eliminated by human inspection, or, in the specific case of machine learning classified populations, via a secondary proximity cut to remove redundant links to the same physical object.  The additional need for this secondary cut is best illustrated by Class~3 sources, which are operationally defined as groups of~4 or more point-like sources within 5 pixels radius (further discussion in \citealt{Whitmore2021}).  Our detection code will generally return each point source in such groupings as a candidate, and subsequently most of these candidates (except perhaps those on the edge of a group) will be confirmed as Class~3 objects by machine learning.  Even so, there is only one actual grouping.  Therefore, after ML classification is completed (Sec.~\ref{sec:MLclassification}), we sort Class~3 by decreasing count rate and eliminate {\bf `redundant'} objects within a rejection radius of 5 pixels.  Interactive inspection of the results showed that the method works as expected, keeping the brightest peak within each Class~3 grouping.  The same operation is conducted for Class~1+2 classified candidates jointly -- as a purely conservative step at this time -- even though any pair of true clusters at separations $<5$ pixels would lose its fainter member.  We expect to re-evaluate the Class~1+2 redundancy cut as the PHANGS-HST analysis progresses further.  We do wish to clarify that Class~3 and Class~1+2 objects are not allowed to eliminate cross-class.  That is, a Class~1 or~2 cluster can be allowed within the $2.5{-}5.0$ pixel range from a Class~3 compact association.

\subsection{Summary of cluster candidate selection}
\label{sec:selectionsummary}

The simplest distillation of our cluster candidate selection method is to say that we evaluate each entry in our all-source catalogue (Sec.~\ref{sec:dolphot}), first requiring that photometric and morphological quality assurance conditions are met (Sec.~\ref{sec:generalcriteria}), then allow high quality sources to qualify as clusters either owing to their position in the MCI plane (Secs.~\ref{sec:modelregions} for the maximally inclusive, model-guided {\em ML sample} and~\ref{sec:EmpiricalSelectionReg} for the smaller, semi-empirically selected {\em human sample}), or because they are more luminous than the H-D limit (Sec.~\ref{sec:maglimits}, {\em included in both samples}).  Final checks to prevent any double counting are also implemented as just described. To aid the reader in understanding this complex selection method, we refer them again to the overview flowchart at the start of this Section (Fig.~\ref{fig:flowchart}).  The path in the flowchart from source detection to science-ready cluster catalogues ends with cluster candidate classification, which is the topic of the following Section.

\section{Cluster candidate classification}
\label{sec:classification}

We adopted the cluster classification system introduced by LEGUS, accomplishing the task through a mix of human inspection \citep[][]{Whitmore2021} and deep transfer learning \citep[][]{Wei2020}.  Integrated over the four galaxies included in this paper, approximately 18\% (4246) of the nearly $24{,}000$ candidate clusters were classified by a human for our study (recall these four galaxies also have independent catalogues from LEGUS).   These classifications act as a reference allowing us to judge the performance of our ML method and improve it in the future (\citealt{Whitmore2021}).

\subsection{Human, inspection-based classification}
\label{sec:humaninspection}

Via interactive inspection of the $V$-band ($F555W$) image and also a colour composite made from $B$, $V$, $I$ bands, one of us (BCW) classified objects in the human sample.  Inspection consisted of radial profile analysis and comparison of morphology with known stars during variation of the image intensity transfer function. A detailed description of the procedure is given in \citet{Whitmore2021}.  The large number of candidates in some targets prevented classification of all objects in this sample, and candidates were inspected down to a magnitude limit that included $\sim1000{-}1500$ objects per target. The limit for the four galaxies in this paper ranged from $m(V)$ = 23.0 to 24.1 (see notes on Table~\ref{tab:classification_stats} for details). 

\subsection{Machine learning classification}
\label{sec:MLclassification}

We applied the deep transfer learning ResNet-18 (18 layer residual, \citealt{RESNET}, hereafter ResNet) and VGG-19\_BN (Visual Geometry Group 19 layer with batch normalisation, \citealt{VGG}, hereafter VGG) convolutional neural network models of \citet{Wei2020} to {\em all} of our candidate clusters, even those not included in the human sample.  Specifically, we adopted the Wei et al. models trained using LEGUS-BCW human classifications for ten galaxies. As ML for cluster classification is an actively emerging field, future successful development will benefit from testing various approaches. We direct the interested reader to \citet{Grasha2019} and \citet{Perez2021}.  We note that the method of \citet{Grasha2019} was also used for cluster classification required by the analysis of \citet{Messa2018}.  

Our classification processing was accomplished on Amazon Web Services (AWS) cloud computing hardware, using GPU instances launched on demand.  Running on a single GPU (NVIDIA  Telsa V100-SXM2 16~Gb) we were able to attain a classification rate of $\sim$0.7s per candidate.\footnote{A step-by-step tutorial of our ML procedure is given in \citet{Whitmore2021}, and linked at \url{http://www.stsci.edu/hlsp/phangs-hst}.} For each of ResNet and VGG we evaluated ten independent models.  From this set of results we obtained the mean, median, mode classification, plus standard deviation, for each network.  The mode was adopted as the final single network classification.  In Sec.~\ref{sec:MLresults}, we experiment with various ways to attain a joint classification based on ResNet and VGG together.  It is worth stressing that we do expect our ML classification accuracy to improve in the future, as we are starting further training experiments based on a representative set of BCW human classifications for PHANGS-HST candidates and from synthetic cluster populations, rather than relying on LEGUS classifications for typically more nearby galaxies than in our sample.  As such, we do not consider the current ML classifications to be finalized.

\begin{table*}
    \caption{Galaxy properties}
	\begin{center}
    \begin{tabular}{l|c c c c c c c}
 \hline
    Galaxy Name & RA & Dec & Distance & Stellar Mass & SFR & SFR$_{HST}$ & Morphology\\
        &  [deg] & [deg] & [Mpc] & [$10^{10}$ M$_{\odot}$] & [M$_{\odot}$yr$^{-1}$] & [M$_{\odot}$yr$^{-1}$] & \\
    \hline    
    NGC~628 &	24.1739 &	15.7836 & 9.9& 2.2 ($\pm$0.6)& 1.8 ($\pm$0.5)& 0.93 & Sc \\
    NGC~1433 &	55.5062 &	$-$47.2219 & 8.3& 1.5 ($\pm$0.4)& 0.2 ($\pm$0.06)& 0.11 & SBa \\
    NGC~1566	& 65.0016 & $-$54.9380 & 18 & 6.3	($\pm$1.6) & 4.7 ($\pm$1.2)& 3.3 &  SABb\\
    NGC~3351 &	160.991 &	11.7037 & 10& 2.3 ($\pm$0.6)& 1.3 ($\pm$0.3)& 0.87 & Sb \\
    \hline    
    \label{TAB:gal_properties}
    \end{tabular}
	\end{center}
    \vspace{-10pt}
    \begin{tablenotes}
	\small
\item[$a$] Note: Properties of the PHANGS-HST galaxies analysed in this paper. They were selected for this study because each has a published cluster catalogue from LEGUS and a catalogue from our work, enabling comparison of the new cluster pipeline output with established results.  Despite having updated PHANGS-HST distances in \citet{Anand2021}, we adopt the LEGUS distances here.  This only significantly impacts NGC~1433, which has a much larger revised distance (18.63 versus 8.3 Mpc).  Nominal stellar masses and star formation rates (SFR) are for the entire galaxy and have been scaled for the adopted distance.  They are computed following \citet{Leroy2019} ($z$0MGS), using galaxy integrated {\em GALEX FUV} + {\em WISE W4} for SFR and {\em WISE W1} for stellar mass. However, we also provide the SFR integrated within only the $HST$ footprint, combining multiple fields for NGC~628 and NGC~3351 (see \ref{TAB:exptime}).
    \end{tablenotes}
\end{table*}

\begin{table*}
    \caption{HST Exposure Times}
	\begin{center}
    \begin{tabular}{l|c c c c c c}
 \hline
    Field Name & $F275W$ ($NUV$) & $F336W$ ($U$)& $F438W$ ($B$) & $F555W$ ($V$) & $F814W$ ($I$) & PID \\
        & [s]   &  [s]  &  [s]  &  [s]  &  [s] &  \\
    \hline  
    NGC~628-C & 2481 & 2361 & 1358$\dagger$ & 858$\dagger$ & 922$\dagger$ & 13364, 10402$\dagger$\\
    NGC~628-E & 2361 & 1119 & 4720$\dagger$ & 965 & 1560$\dagger$ & 13364, 9796$\dagger$\\
    NGC~1433 & 2376 & 1116 & 962 & 1140 & 986 & 13364\\
    NGC~1566 & 2382 & 1119 & 965 & 1143 & 989 & 13364\\
    NGC~3351-N & 2190 & 1110 & 1050 & 670 & 830 & 15654\\ 
    NGC~3351-S & 2361 & 1062 & 908 & 1062 & 908 & 13364\\ 
    \hline    
    \end{tabular}
    \label{TAB:exptime}
	\end{center}
    \begin{tablenotes}
	\small
\item[$a$] Note: $HST$ imaging exposure times for the galaxies and data sets analysed in this paper. Data for NGC~1433, NGC~1566, and NGC~3351 originate completely from {\em WFC3/UVIS}, whereas NGC~628 observations also include imaging taken with {\em ACS/WFC} (as indicated with the $^{\dagger}$ symbol).  Accordingly, the $B$-band data for NGC~628 (both fields) actually comes from the $ACS$/$F435W$ filter not~$WFC3$/$F438W$. $HST$ program ID (PID) is given in the final column.  Note that we combined the two fields of NGC~3351 into a single drizzled mosaic, whereas (owing to the non-negligible difference in filters between {\em WFC3/UVIS} and {\em ACS/WFC}) the NGC~628 data are kept separate for each field (e.g.\ NGC~628-C and NGC~628-E) though there is some spatial overlap.
    \end{tablenotes}
\end{table*}

\begin{figure*}
   \includegraphics[width=0.24\textwidth]{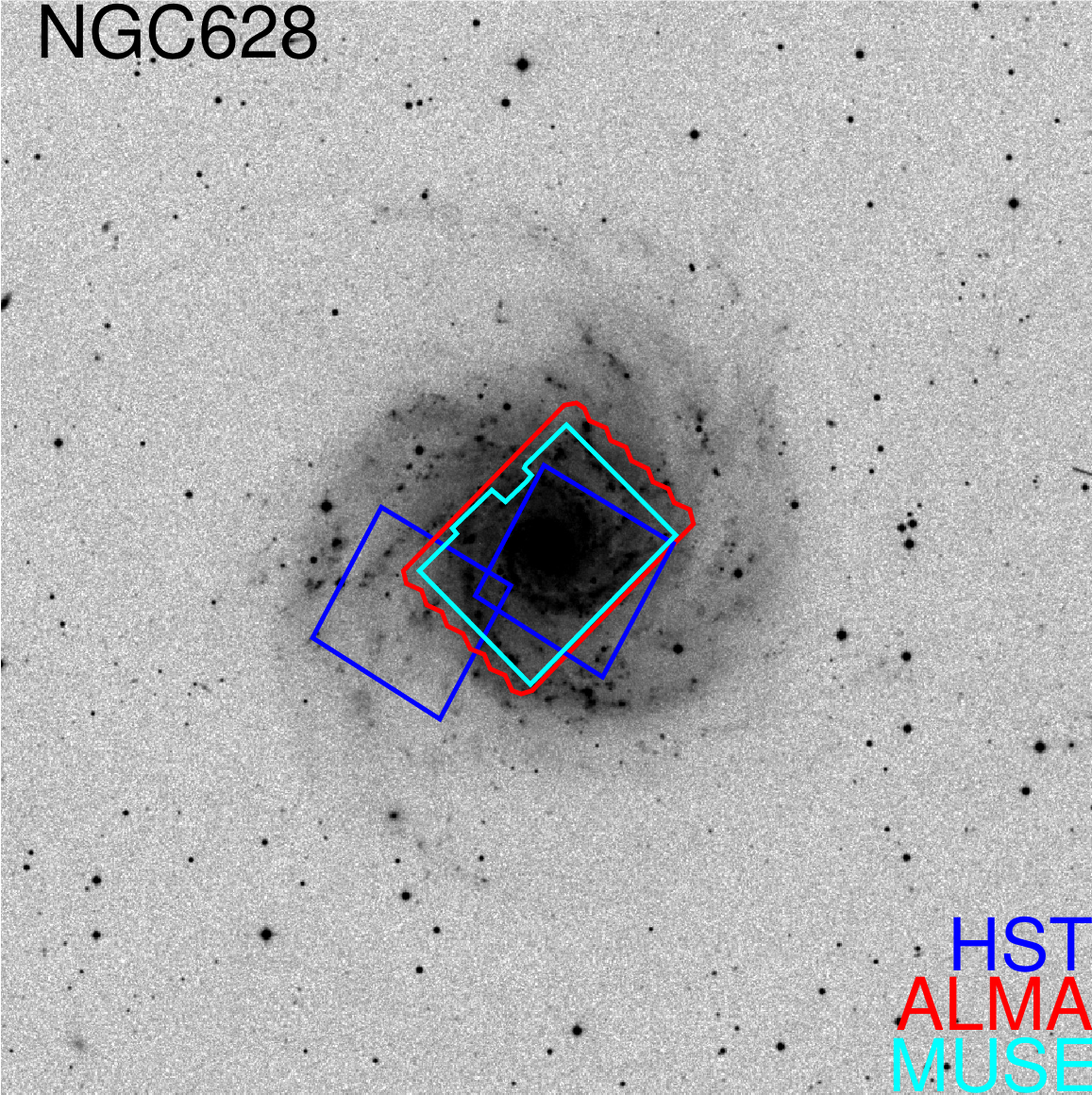}
   \includegraphics[width=0.24\textwidth]{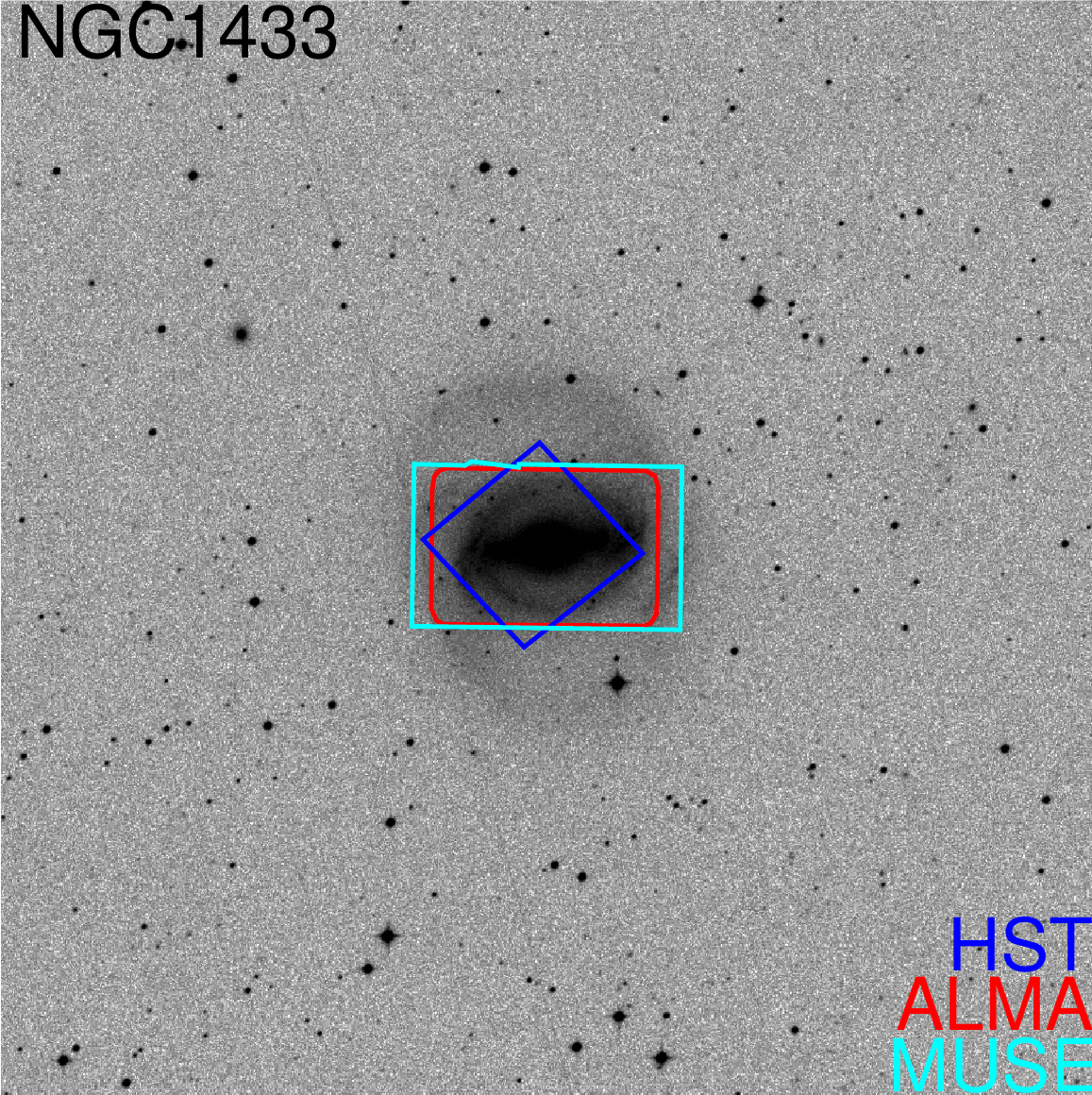}
   \includegraphics[width=0.24\textwidth]{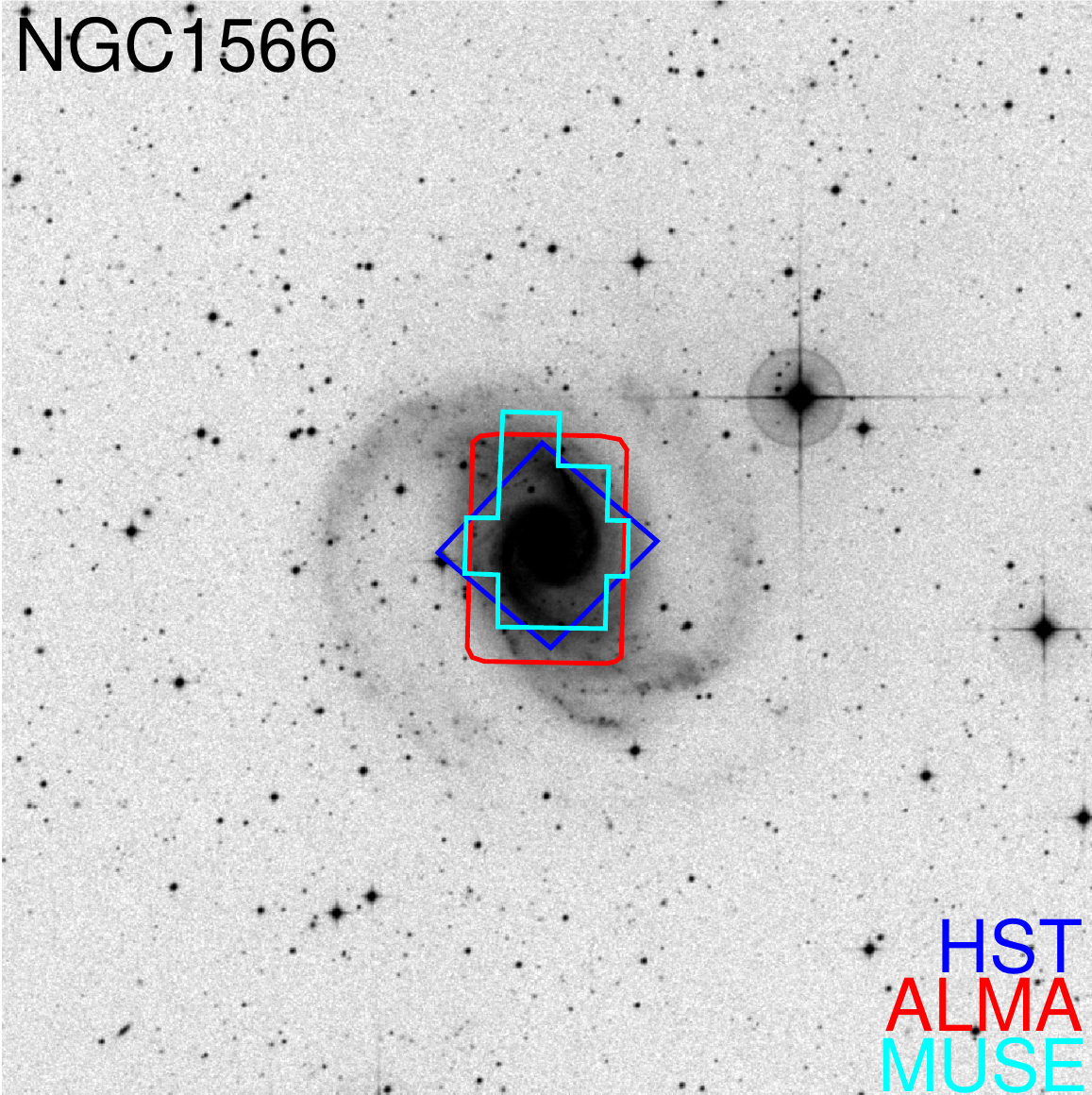}
   \includegraphics[width=0.24\textwidth]{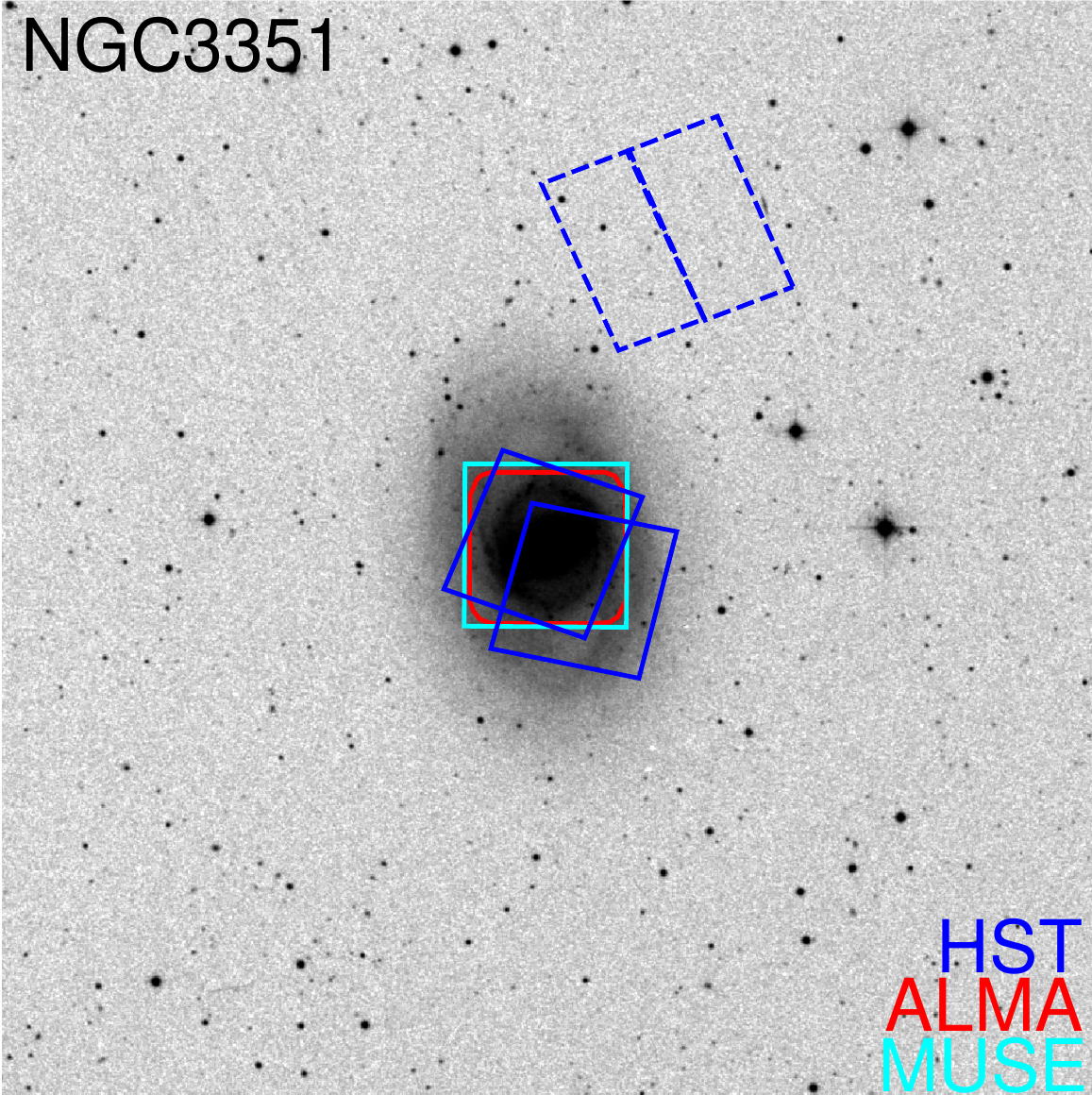}
   \caption{Footprints of the $HST$ (blue, only {\em WFC3/UVIS} shown), ALMA (red), and MUSE (cyan) coverage for the galaxies analysed in this paper overlaid on Digital Sky Survey (DSS) imaging.  The PHANGS-HST imaging for NGC~3351 also included {\em ACS} $V$- and $I$-band parallel observations (footprint shown as dotted blue box) to enable distance measurements \citep[][]{Anand2021}. Images are oriented with North up and East left.  Thus NGC~628-E is the field on left, and NGC~628-C is the field centred on the galaxy.  For, NGC~3351 the PHANGS-HST {\em WFC3/UVIS} data is centred on the galaxy, and LEGUS imaging is offset to the SW.  For reference of scale, the blue {\em WFC3/UVIS} footprint is 7.8, 6.5, 14.1, 7.9 kpc across, respectively for NGC~628, NGC~1433, NGC~1566, and NGC~3351.}
    \label{FIG:footprints}
\end{figure*}

\section{Application to PHANGS-HST galaxies}
\label{sec:results}

In this Section we present the results of our cluster detection and classification methods. We analysed the four PHANGS-HST galaxies for which LEGUS also produced a publicly available cluster catalogue, enabling direct comparison between the surveys.  Tables~\ref{TAB:gal_properties} and~\ref{TAB:exptime} provide relevant information on each of the galaxies and the $HST$ observations, respectively.  Figure~\ref{FIG:footprints} displays the footprint of the $HST$ images (for primary observations we only show {\em WFC3/UVIS} coverage) and the ALMA CO, MUSE IFU (Integral Field Unit) spectroscopy sky coverage from the broader PHANGS project.  Complete details of the observations and data reduction are given in \citet{PHANGSHSTsurvey}. 

We begin here with summary statistics of the cluster candidate sample, then describe candidate classification results, in both the human inspection and machine learning contexts. Finally, we present a detailed comparison of the Class~1 and~2 cluster populations identified by PHANGS-HST and by LEGUS. 

\subsection{Candidate statistics}
\label{sec:Candidateresults}

In Table~\ref{tab:candidate_stats}, for each analysed galaxy, we summarise information about the objects which are detected by our method and pass our selection criteria.  As described in Table~\ref{TAB:exptime} we process NGC~628 as two separate fields without mosaicking them together due to camera/\linebreak[0]{}filter differences.  Furthermore, we present the statistics for NGC~3351 broken into two columns because additional observations were obtained as part of PHANGS-HST, increasing spatial coverage to better match the ALMA observations, thereby increasing the exposure time in the region of overlap between LEGUS and PHANGS-HST imaging.  Specifically, we analyse: (i) a LEGUS-only data set (NGC~3351-S in Tab.~\ref{TAB:exptime}) and (ii) a mosaic constructed from all available images for NGC~3351. 

\clearpage
\onecolumn
\begin{landscape}
\begin{table*}
\begin{threeparttable}
 \caption{Detection and candidate statistics}
 \label{tab:candidate_stats}
 \begin{tabular}{llrrrrrr}
  \hline
Row & Quantity & NGC~628-C & NGC~628-E & NGC~1433 & NGC~1566 & NGC~3351\tnote{$e$} & NGC~3351 \\
(\#)&   &         &         &         &         & (LEGUS-only) & (all data) \\
  \hline

(1)&All-source catalogue entries, Sec.~\ref{sec:dolphot}      &     646263 & 761914 & 266386& 377133& 220905 & 452247 \\
(2)&General criteria satisfied, Sec.~\ref{sec:generalcriteria}    &    38776 & 19425 & 7992 & 34588 & 8079 & 16487 \\
(3)&Morphology (MCI) criteria satisfied\tnote{$a$}, Secs.~\ref{sec:modelregions} and~\ref{sec:EmpiricalSelectionReg}     &  10482 & 3450 & 3180 & 13455 & 3784 & 5415 \\
(4)&Super-HD (bright object) criteria satisfied\tnote{$a$}, Sec.~\ref{sec:maglimits}        &   57 & 25 & 12 & 436 & 99 & 89 \\
(5)&General + super-HD criteria only               & 12 & 4 & 6 & 130 & 26 & 29 \\
(6)&Human sample size (semi-empirical selection region or super-HD) \tnote{$b$}                       &   5695 & 1366 & 1230 & 3757 & 1705 & 2366 \\
(7)&Machine Learning sample size (model cluster region or super-HD)\tnote{$c$}               &     7679 & 2117 & 1989 & 8822 & 2398 & 3389 \\
(8)&LEGUS automatic candidates       & 3080\tnote{$d$} & 593\tnote{$d$} & 1099 & 2752 & 1389 & \nodata \\
(9)&LEGUS classified candidates (Classes 1, 2, 3, and 4)       & 1559 & 381 & 306  & 1061 &  618 & \nodata \\
  \hline
\end{tabular}
\begin{tablenotes}
\small
\item[$a$]{For Rows~(3) and~(4), we check the named criteria without enforcing the general conditions of Row~(2).}
\item[$b$]{The human sample is defined as the set of sources meeting general criteria, and either: (1) morphological criteria (e.g.\ MCI error, $V$-band $\mathrm{S/N}$) and lying in the semi-empirical selection region of the MCI plane (but outside the stellar exclusion region), or (2) meeting the super H-D criteria. In galaxies with very large source populations, only a bright subset of the human sample candidates are eventually classified interactively by a human.}
\item[$c$]{The Machine Learning sample is similar to the human sample, except that morphologically-selected sources are tested against the [broader] model cluster region rather the semi-empirical selection region.} 
\item[$d$]{Owing to their condition of having four good photometric bands in order for selection, the footprint of the LEGUS census is significantly reduced in these two targets.  See the notes of Table~\ref{tab:classification_stats} for more detail.  Because of this the ratio of Row~(7) to Row~(8) is inflated in NGC~628-C and NGC~628-E.}
\item[$e$]{This column pertains to analysis of the archival LEGUS data for NGC~3351 (NGC~3351-S in Table~\ref{TAB:exptime}) without any use of the new NGC~3351-N data obtained by PHANGS-HST, even in the area of overlap.}
\end{tablenotes}
\end{threeparttable}
\end{table*}
\end{landscape}
\twocolumn
\clearpage

Table~\ref{tab:candidate_stats} principally gives counts pertaining to our method, but we also include the total number of sources in the LEGUS `automatic' catalogue and the subset which were classified by the LEGUS team.   This allows an initial assessment regarding the efficacy of our new techniques.

Row~(1), giving the combined number of \codename{DOLPHOT} and \codename{DAOStarFinder} sources, shows that the initial set of all-source detections coming from \codename{DOLPHOT} is overwhelmingly large compared to the actual number of reliably identifiable stellar clusters.  At least for the distance range we probe, this all-source census is dominated by stars, including some of marginal significance (as is customary for PSF-fitting resolved stellar photometry prior to selection of `good stars' via quality assurance metrics).  
Row~(2), listing counts after application of general photometric selection criteria (first paragraph of Sec.~\ref{sec:generalcriteria} and also Sec.~\ref{sec:doubles}), demonstrates the drastic reduction in the tally of potential candidates due to our limits on photometric error (in multiple bands).  Row~ (3) and~(4) show the number of objects from Row~(1) meeting our separated conditions of morphological appearance and luminosity, described in Secs.~\ref{sec:MCI} and \ref{sec:maglimits}, respectively.  In this table, the general conditions of Row~(2) are not enforced for tabulating Rows (3) and~(4), indicating the small fraction of the initial all-source catalogue having suitably accurate MCI measurements, and the minimal population of objects brighter than the H-D limit.  Row~(5) gives the number of sources meeting our general selection criteria and being selected as a candidate {\em only} due to being very luminous.  

Our eventual selection, summarised in Fig.~\ref{fig:flowchart}, is the result of demanding the set of general criteria and {\em either} the morphological {\em or} luminosity based criteria.  We report the final number of cluster candidates so selected in Row~7 (ML~sample) whereas Row~6 (human sample) gives the subset distinguished as occupying the semi-empirical selection region or being in the super H-D tail.   
Though not broken down in the table, most ($\sim50{-}70\%$) candidates originate either within the semi-empirical selection region or the highest resolution model cluster region (0.01 binning), or their common zone.  Candidates added due to their MCI plane location in the 0.02 or 0.04 bin-size model cluster regions are typically comparable in number to each other, each additional layer of area (at lower resolution) adding between $20{-}30$\% to the candidate sample.  Our most distant target, NGC~1566, is different, with the candidates originating more evenly across the MCI plane.

Rows~(8) and~(9) give the LEGUS candidate counts in total, and those with LEGUS classifications, respectively.  Comparison of our ML sample size (Row~7) with that of the LEGUS automatic catalogue (Row~8) shows that despite cuts (e.g.\ Rows 2, 3,~4) our tally of candidate clusters is about a factor 2$\times$ larger than LEGUS.  This is probably due to a combination of our allowance of more compact cluster morphologies as candidates ($\sim$4\% with CI  < LEGUS limit) and the substantial difference in source detection methods, since \codename{DOLPHOT} deblends neighbouring sources more effectively than \codename{SExtractor}.  We find that candidates allowed into the sample only by virtue of their absolute magnitude (being too bright for a single star) are rare (Row~5), with generally between several to $\sim$30 in each target. NGC~1566 is an exception, with 130 super H-D candidates not otherwise qualified based on morphological metrics, presumably due to a combination of being much further away than the other galaxies (18 Mpc versus $\lessapprox$~10 Mpc) and having nearly triple the HST-footprint-integrated SFR of any other target.

\subsection{Results from human classification}
\label{sec:Humanresults}

\begin{figure*}
 \includegraphics[width=1.3in]{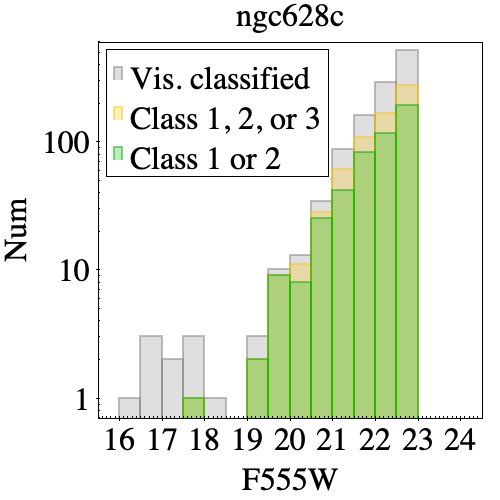}
 \includegraphics[width=1.3in]{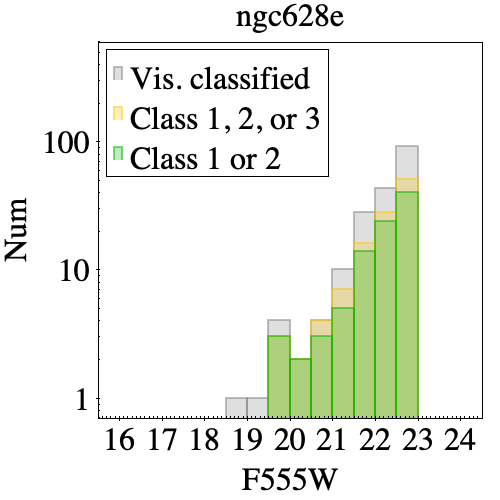}
 \includegraphics[width=1.3in]{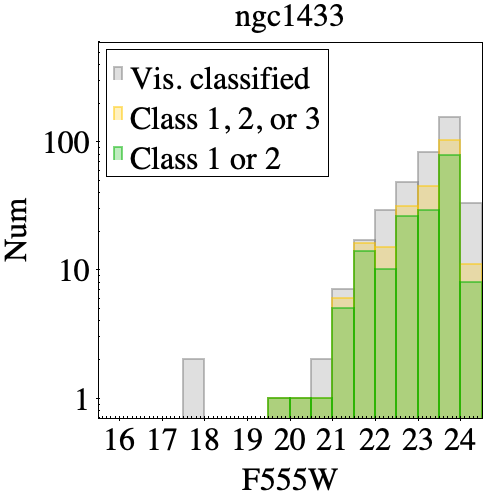}
 \includegraphics[width=1.3in]{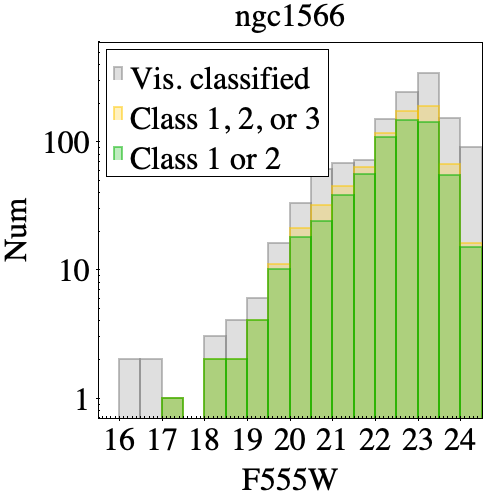}
 \includegraphics[width=1.3in]{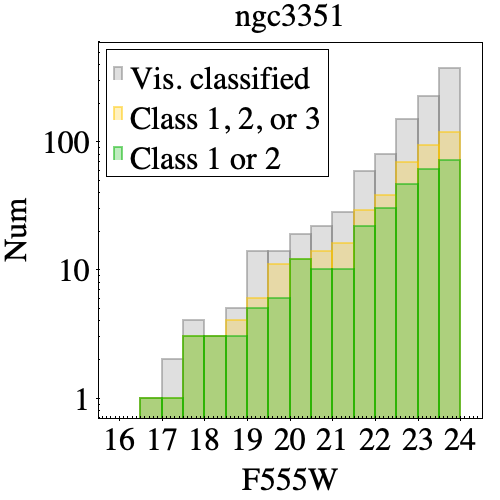}
 \includegraphics[width=1.3in]{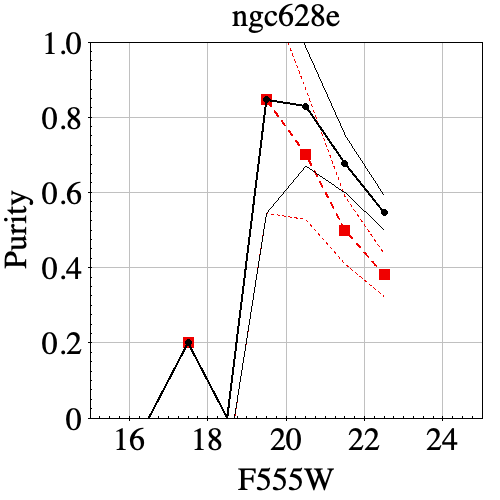}
 \includegraphics[width=1.3in]{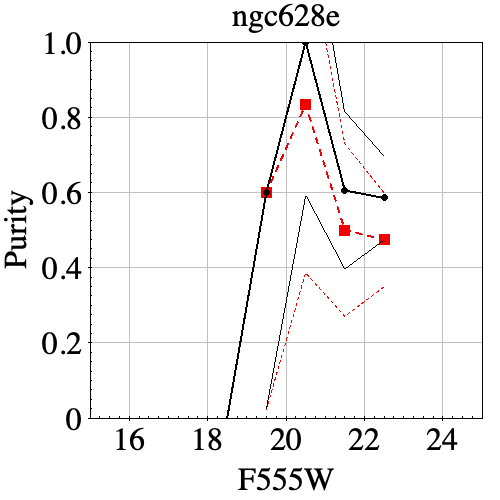}
 \includegraphics[width=1.3in]{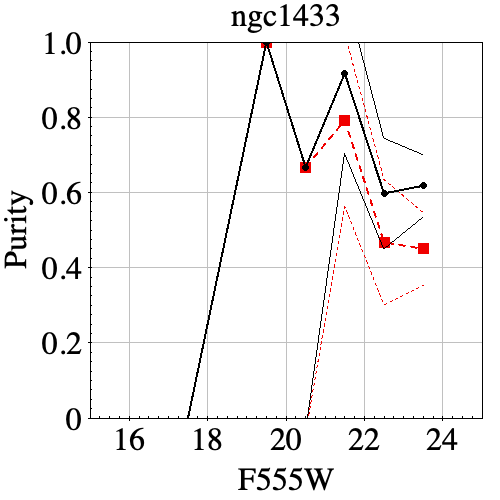}
 \includegraphics[width=1.3in]{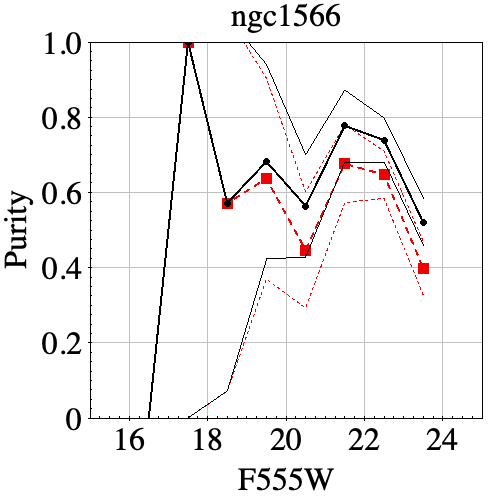}
 \includegraphics[width=1.3in]{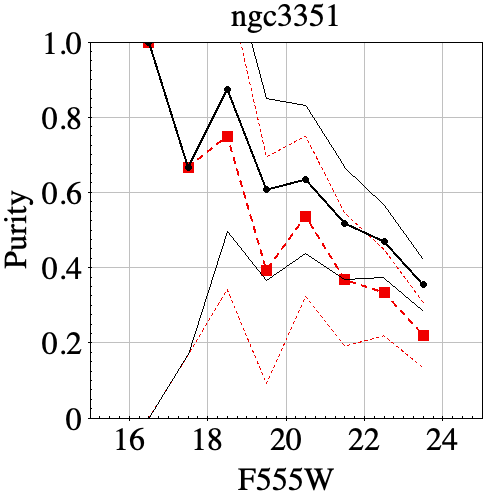}
 \caption{(Top row) Selection purity for human classified clusters assigned as candidates due to the semi-empirical selection region, or else being brighter than the H-D limit. Each histogram shows the object counts from a different target, plotting the overall set with available BCW classifications in grey, the BCW classified Class~1, 2, or~3 clusters in yellow, and BCW Class~1 or~2 clusters in green. (Bottom row) The same information but with the purity expressed as a fractional measure, with Class~1 or~2 results as red squares, and Class~1, 2,~3 as black dots. The $1 \sigma$ Poisson uncertainty is indicated with thinner red dashed and black solid lines.}
 \label{fig:Purity}
\end{figure*}

Drawing cluster candidates from the `human sample' (see Table~\ref{tab:candidate_stats}), one co-author (BCW) assessed source morphology and assigned classes following the method in Sec.~\ref{sec:humaninspection}.  The depth of the human classifications varied with the specific target, going deeper in cases with few candidates and shallower in highly populous targets.  We are able to use these classification data to assess the effectiveness of our selection criteria, and (in the future, with human classifications aggregated across all PHANGS-HST targets) to improve the training of ML networks.   Because human classifications are confined to the portion of the MCI plane we call the semi-empirical selection region, with the exception of rare super H-D sources, we assess compact cluster selection purity ($\equiv$~\#~Class~1+2 / \#~Candidates) and Class~3 contamination ($\equiv$~\#~Class~3 / \#~Candidates) only within this area, leaving the outer model cluster regions beyond this for a later study.

Figure~\ref{fig:Purity} shows the Class~1+2 selection purity attained in all targets, as a function of $V$-band magnitude.  We present the information both as a computed purity with Poisson error (bottom row, red markings), and as the associated counts versus magnitude (top row, green and grey histograms).  Purity generally increases from $\sim35{-}50$\% at 23~mag to $\sim70{-}85$\% brighter than 21~mag.  In NGC~1566 the data suggest a flatter relationship.  Note that our ability to accurately constrain the purity worsens as the population becomes brighter due to reduced counting statistics.  Nevertheless, the overall trend across the sample is clear and the typical purity within the semi-empirical selection region is in excess of the comparable statistic for LEGUS (44\%, 45\%, 37\%, 44\%, 32\% for NGC~628-C, NGC~628-E, NGC~1433, NGC~1566, and NGC~3351 (LEGUS-only), respectively). In Fig.~\ref{fig:Purity}, we also present a purity assessment for Class~1+2+3 (black markings in bottom row), finding that this metric including compact associations alongside compact clusters is typically $10{-}15$\% higher. 

Our method does not recover a majority of the Class~3 objects identified by LEGUS, but this is by design since we are treating these multi-peak compact associations with the dedicated method of \citet{Larson2021}.  Specifically, our selection areas in the MCI plane were determined on the basis of only Class~1 and~2 single-peak objects, both synthetic and observed.  Class~3 objects generally lie below (more negative MCI$_{\mathrm{out}}$ at fixed MCI$_{\mathrm{in}}$) Class~1 and~2 objects.  Some compact associations (Class~3) do survive our selection as the selection areas for clusters (Class~1 and~2) had to be made broad enough to retain outlying Class~1 and~2 objects.  Such surviving Class~3 objects should be considered contaminants, and we only keep them for completeness and comparison to prior work (LEGUS). We strongly advocate using the multi-scale associations of \citet{Larson2021} instead.  This is particularly important if the emphasis is on studying young star formation products, as our Class~3 objects would be highly incomplete in such a context.  


We achieved our goal of minimising Class~3 contamination (\#~Class~3 / \#~Candidates) amongst the objects considered as possible clusters.  For the depth attained by PHANGS-HST human classification, such contamination ranged from 8 to 16\% across the galaxies in our study.  We do not compare to similar metrics for LEGUS, as they aimed to recover Class~3 compact associations in addition to Class~1 and~2 clusters (incidentally, this is a likely reason for their usage of SExtractor as a detection algorithm).   

\subsection{Results from ML classification}
\label{sec:MLresults}

\begin{table*}
\begin{threeparttable}
 \caption{Classification outcomes for candidates in the ML sample, and comparison to BCW human\tnote{$a,b,c,d$} and LEGUS human\tnote{$e$} classification. Totals are given following the multiple strategies we considered for interpretation of the ML results.  For comparison with human classification of LEGUS we provide cluster counts for the subset brighter than $M_{V} = -6$ in parentheses.  Note that the depth reached for BCW classifications conducted as part of PHANGS-HST is variable, sometimes not reaching as faint as $M_{V} = -6$.}
 \label{tab:classification_stats}
 \begin{tabular}{llrrrrrr}
  \hline
Quantity & Option & NGC~628-C & NGC~628-E & NGC~1433 & NGC~1566 & NGC~3351\tnote{$f$} & NGC~3351 \\
&   &         &         &         &         & (LEGUS-only) & (all data) \\
  \hline
ML sample size  &                    &     7679 ( 5286 ) & 2117 ( 982 ) & 1989 ( 464 ) & 8822 ( 8768 ) & 2398 ( 1264 ) & 3950 ( 1887 )\\
  \hline
\multirow{3}{*}{ML Class~1}& VGG, preferred network     & 1218 ( 871 ) & 317 ( 161 ) & 210 ( 44 ) & 796 ( 796 ) & 320 ( 170 ) &  440 ( 221 )\\
 & ResNet                                               & 1141 ( 849 ) & 313 ( 169 ) & 286 ( 55 ) & 996 ( 971 ) & 359 ( 184 ) &  458 ( 239 )\\
  & Combined network                                    & 1156 ( 847 ) & 310 ( 156 ) & 225 ( 47 ) & 778 ( 778 ) & 323 ( 172 ) &  429 ( 219 )\\
  \hline
\multirow{3}{*}{ML Class~2}& VGG, preferred network     & 377 ( 251 ) & 157 ( 71 ) & 132 ( 39 ) & 398 ( 398 ) & 206 ( 98 ) & 247 ( 128 )\\
 & ResNet                                               & 306 ( 183 ) & 186 ( 50 ) & 214 ( 36 ) & 534 ( 533 ) & 285 ( 110 ) & 399 ( 137 )\\
 & Combined network                                     & 301 ( 197 ) & 150 ( 60 ) & 149 ( 31 ) & 392 ( 391 ) & 214 ( 94 ) & 272 ( 125 )\\
 \hline
\multirow{3}{*}{ML Class~3}& VGG, preferred network     & 385 ( 327 ) & 115 ( 72 ) & 206 ( 58 ) & 1143 ( 1143 ) & 202 ( 119 ) & 337 ( 201 )\\
 & ResNet                                               & 392 ( 309 ) & 110 ( 54 ) & 206 ( 55 ) & 950 ( 950 ) & 219 ( 125 ) & 240 ( 150 )\\
 & Combined network                                     & 344 ( 294 ) & 98 ( 58 ) & 202 ( 53 ) & 977 ( 977 ) & 200 ( 117 ) & 282 ( 175 )\\
 \hline
\multirow{4}{*}{ML Class~1 or 2}& VGG, preferred network& 1595 ( 1122 ) & 474 ( 232 ) & 342 ( 83 ) & 1194 ( 1194 ) & 526 ( 268 ) & 687 ( 349 )\\
 & ResNet                                               & 1447 ( 1032 ) & 499 ( 219 ) & 500 ( 91 ) & 1530 ( 1504 ) & 644 ( 294 ) & 857 ( 376 )\\
 & Mode consensus                                       & 1218 ( 929 ) & 417 ( 203 ) & 329 ( 82 ) & 1068 ( 1068 ) & 478 ( 250 ) & 629 ( 312 )\\
 & Combined network                                     & 1457 ( 1044 ) & 460 ( 216 ) & 374 ( 78 ) & 1170 ( 1169 ) & 537 ( 266 ) & 701 ( 344 )\\
\hline
\hline
Human Class~1 && 260 ( 260 )\tnote{$a$} & 51 ( 51 )\tnote{$a$} & 87 ( 51 )\tnote{$b$} & 377 ( 377 )\tnote{$c$} & 113 ( 113 )\tnote{$d$} & 136 ( 136 )\tnote{$d$} \\
Human Class~2 && 211 ( 211 )\tnote{$a$} & 40 ( 40 )\tnote{$a$} & 90 ( 50 )\tnote{$b$} & 257 ( 257 )\tnote{$c$} & 92 ( 91 )

\tnote{$d$} & 162 ( 162 )\tnote{$d$} \\
Human Class~3 && 167 ( 167 )\tnote{$a$} & 14 ( 14 )\tnote{$a$} & 52 ( 28 )\tnote{$b$} & 114 ( 114 )\tnote{$c$} & 68 ( 67 )\tnote{$d$} & 124 ( 124 )\tnote{$d$} \\
Human Class~1 or 2 && 471 ( 471 )\tnote{$a$} & 91 ( 91 )\tnote{$a$} & 177 ( 101 )\tnote{$b$} & 634 ( 634 )\tnote{$c$} & 205 ( 204 )\tnote{$d$} & 298 ( 298 )\tnote{$d$}\\
Human Classified && 1039 (1039)\tnote{$a$} & 174 (174)\tnote{$a$} & 436 (251)\tnote{$b$} & 1442 (1442)\tnote{$c$} &  569 (564)\tnote{$d$} & 1063 (1063)\tnote{$d$} \\ 
\hline
\hline
LEGUS Class~1 && 334\tnote{$e$} & 92\tnote{$e$} & 51 & 258\tnote{$c$} & 118 & \nodata\\
LEGUS Class~2 && 357\tnote{$e$} & 80\tnote{$e$} & 61 & 214\tnote{$c$} & 80 & \nodata\\
LEGUS Class~3 && 326\tnote{$e$} & 87\tnote{$e$} & 56 & 261\tnote{$c$} & 94 & \nodata\\
LEGUS Class~1 or 2 && 691\tnote{$e$} & 172\tnote{$e$} & 112 & 472\tnote{$c$} & 198 & \nodata\\
  \hline
   \end{tabular}
  \begin{tablenotes}
\small
\item[$^a$] {PHANGS-HST human classification (by BCW) for NGC~628-C and NGC~628-E reached mag 23.0 uniformly.}
\item[$^b$] {PHANGS-HST human classification (by BCW) for NGC~1433 reached mag 24.1 uniformly.}
\item[$^c$] {PHANGS-HST human classification (by BCW) for NGC~1566 reaches mag 23.5 for the complete field and is supplemented with spot check regions (one running E/W across centre, another covering the S~corner of field) inspected to 24.34.  LEGUS classification availability for NGC~1566 is complicated as it was a mix of pure human and ML (Human: 258 C1, 214 C2, 261 C3, 328 C4; Total: 478 C1, 404 C2, 691 C3, 868 C4). Here we use the human classifications.}
\item[$^d$] {PHANGS-HST human classification (by BCW) for NGC~3351 reached mag 24.0 uniformly.}
\item[$^e$] {The LEGUS human classification effort (\citealt{Adamo2017}) for NGC~628-C and NGC~628-E was confined to a region somewhat smaller than the complete field that we analysed.  Specifically, the LEGUS team required cluster detection in at least four bands to be considered for inspection, and in these targets the survey included archival data with differing coverage.  As a result, only clusters in the overlap region between {\em WFC3/UVIS} and {\em ACS/WFC} imaging were classified.  For NGC~628-C, this omits a peripheral area surrounding the {\em WFC3/UVIS} coverage.  For NGC~628-E, this omits approximately the easternmost quarter of the {\em WFC3/UVIS} field and about half of the WNW portion of the {\em ACS/WFC} coverage.  Imposing the same sky area cut to PHANGS-HST yields the following counts for `ML Class~1 or~2' in NGC~628-C, VGG 999(758), ResNet 1056(760), consensus 856(670), combined 952(722); and in NGC~628-E, VGG 392(199), ResNet 411(187), consensus 344(177), combined 375(183).  The numbers just given in parentheses have sky and magnitude limit cuts making them directly comparable to the NGC~628 values in the last line of the table (LEGUS Class~1 or~2).}
\item[$f$]{This column pertains to analysis of the archival LEGUS data for NGC~3351 (NGC~3351-S in Table~\ref{TAB:exptime}) without any use of the new NGC~3351-N data obtained by PHANGS-HST, even in the area of overlap.}\end{tablenotes}
 \end{threeparttable}
\end{table*}

As currently implemented in our pipeline, we conduct ML classification using two different neural network architectures (ResNet and VGG), each pre-trained as described in \citet{Wei2020}.  For each network we evaluate the ensemble of class~predictions ten times, allowing us to determine a mode cluster classification for each source.  Just as with human classifiers, these modes do not always agree between networks or  between networks and the BCW human classification.
We are then faced with deciding how to best use the ML results, in the context of our goal to provide cluster classifications with the highest completeness (fraction of true Class~1 or~2 clusters classified as such) and lowest contamination (population of non-clusters classified as clusters).  We have several options:
\begin{itemize}
    \item [(1)] Identify the network which agrees with human classification of Class~1 or~2 most frequently.  That is, select a `preferred network' (ResNet or VGG) and adopt its output as the final classification.
    \item [(2)] Combine the modes of each network (per candidate) into a `mode consensus classification', by accepting a candidate as a compact cluster (either Class~1 or~2, but unspecified) if either network predicts that and the other network does not predict Class~4 (star or artefact). 
    \item [(3)] Evaluate the mode of all 20 (ten ResNet and ten VGG) predictions in a unified sense, to yield a `combined network classification'.
\end{itemize}

We now examine the outcomes associated with these options, and then use the difference between resulting compact cluster counts to: quantify the uncertainty of our ML classification methods, and to bracket the actual number of detectable Class~1 and~2 clusters in our data set.


Option~1 (`preferred network') is the most straightforward way to proceed, though it should be expected to provide lower accuracy than other choices -- since the confusion matrices of \citet{Wei2020} suggest the accuracy of each network is approximately on par with the consistency achieved by a single human if classifications are blindly redone after a period of years.  There is no redundancy against misclassification associated with specific downfalls of a network.  However, it has the benefit of simplicity and retains distinction between Class~1 and~2 (the `mode consensus classification' would not).  In order to decide which network is preferred we compared the classifications for PHANGS-HST human Class~1 and~2 objects (independently per target) to ResNet and VGG classifications.  We find that in 3 of 5 targets the number of identically classified sources is highest for the VGG network.  For NGC~1433, ResNet is more frequently in agreement with human classification.  The two networks are about equivalent versus Human classification for NGC~3351.  Relaxing the set of PHANGS-HST Human classified sources to also include Class~3 (so 1, 2, or~3), we find that VGG classifications are more closely aligned with the Human assessment in 4 of 5 targets, and VGG and ResNet are about equivalent for NGC~1433.  Therefore, we take VGG as our `preferred network'.  \citet{Whitmore2021} independently come to the same conclusion, using different methods. Comparison of ResNet and VGG mode predictions for human Class~4 sources, shows that ResNet is very slightly more likely to agree with human classification that a contaminant is not a cluster.  Table~\ref{tab:classification_stats} lists the number of compact clusters (Classes~1 and~2) and compact associations (Class~3) identified in each of our targets (plus NGC~3351 LEGUS-only) via the preferred network,~VGG.  For completeness we also give the tallies according to ResNet.

Option~2 (`mode consensus classification') is an attempt to synthesise the classification output of both trained networks, and, in doing so, provide mitigation against misclassification by VGG or ResNet operating alone.  As defined above, this option can be thought of as demanding that a source is agreed upon to not be a star or artefact (by both network architectures), and has been classified as~1 or~2 by at least one of them.  As such, it is a balance between completeness (inclusivity) and contamination.  Mode consensus classification results for all five targets are given in Table~\ref{tab:classification_stats}.  The compact cluster counts are lower than for the preferred network option, but presumably benefit from reduced contamination.  

Option~3 (`combined network'), in which the mode is evaluated directly over set of 10+10 (ResNet+VGG) models, is also a means of synthesising all available runs of ML.  It has the advantage over option~2 of retaining the discrimination between classifications 1, 2, 3, and~4.  However, it is unclear if the possible drawback of adding information resulting from a non-preferred network (here ResNet) is overcome by the increase in statistical significance due to more ML classification votes.  We include counts from the `combined network' option to inform future work. In general, combined network cluster counts are between those of VGG and ResNet, or slightly lower than either.  We note this `combined network' option may have more meaning in a situation where networks (possibly even the same architecture) are trained according to different strategies, such as training with observed clusters \citep[e.g.][]{Wei2020} versus training with synthetic clusters and artificial stars.  Discussion of other varied training strategies is given in \citet{Whitmore2021}.

The choice of which option to use will ultimately be driven by the specific science use-case.  If high completeness or retention of information regarding Class~1 versus~2 is important then we advise adopting the straight VGG mode classification.  When forming distribution functions of cluster properties from such VGG classified objects, one could opt to weight by the fraction of VGG votes agreeing with the VGG mode, effectively placing emphasis on the most certain classifications.  For those studies that require low contamination and can accept the inability to distinguish Class~1 from~2, we suggest use of the `mode consensus classification'.  We emphasise that the census of clusters obtained in this manner will still be more complete than what is provided via PHANGS-HST human classification.  Presently, we discourage use of the `combined network' classification option until such time as diverse training strategies have been implemented.

Table~\ref{tab:classification_stats} indicates that the ratio of Class 1 to Class 2 for ML (here VGG) is typically a factor of 2 to 4, but this same ratio is around 1 to 1.5 for human based tallies.  The main reason for this is the brighter magnitude limit for human classification and does not imply systematic differences in how candidates are classified (between ML and human). For instance, human Class 1 clusters are generally brighter than ML Class 1 because of the limits imposed by inspection.

What can we say regarding the uncertainty of Class~1 and~2 cluster counts in our analysis? For each galaxy/field we interpret the `mode consensus classification' tally as a lower limit on the number of Class~1+2 sources, and the maximum of VGG and ResNet Class~1+2 cluster counts (VGG and ReSNet individually) as an upper limit.  Adopting the mean of these three tallies as a representative count, we can express the bracketed range as a percent uncertainty with respect to the mean.  We find $\pm$13\%, $\pm$9\%, $\pm$22\%, $\pm$18\%, $\pm$16\% respectively for NGC~628-C, NGC~628-E, NGC~1433, NGC~1566, and NGC~3351.  
Unfortunately, analogous percentage uncertainties from LEGUS analysis are not available for comparison. 



\subsection{Comparison of class~1 and~2 clusters in PHANGS-HST versus LEGUS}
\label{sec:PHANGSLEGUScomparison}

\begin{figure*}
 \includegraphics[width=6.5in]{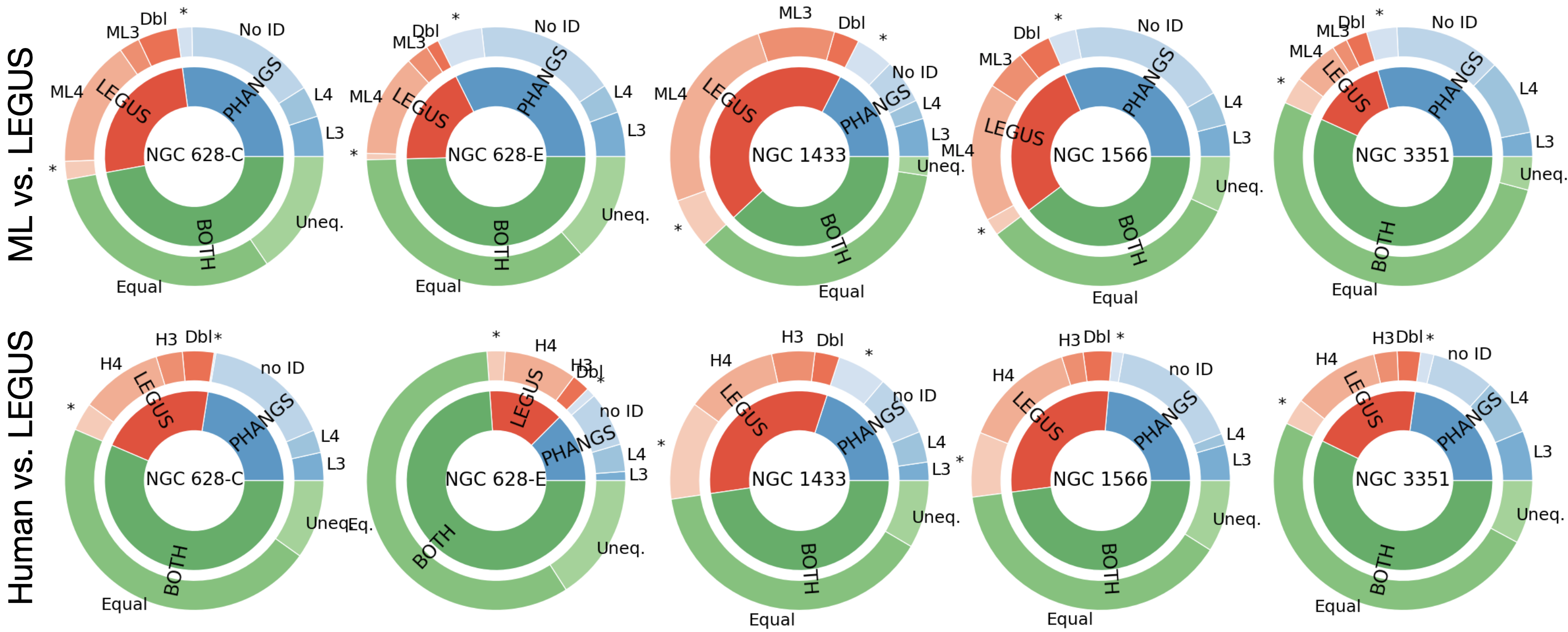}
 \caption{Class~1 and~2 cluster population comparison between PHANGS-HST and LEGUS. The inner ring of each doughnut plot illustrates the fractional division of the comparison set into: objects belonging to the Class~1+2 survey intersection, (green, 'BOTH'); objects identified as Class~1 or~2 only by PHANGS-HST (blue, 'PHANGS'); and objects identified as Class~1 or~2 only by LEGUS (red, 'LEGUS').  The outer ring breaks these populations down further, showing the cause of the inconsistency, whether due to differing classification (labels ending in 3 or~4, e.g. 'H3', 'H4', 'ML3', 'ML4', 'L3', 'L4' -- where, for instance, 'H3' indicates the source was classified by a human as Class 3 in the survey for which it did not survive as Class 1 or 2), non-detection as a candidate (`No ID'), flagging as a double ('Dbl'), or some other unspecified reason ('*').  For objects that do appear as either Class 1 or 2 in both surveys, the green outer wedge indicates the fraction having equal Class (1 and 1, or 2 and 2, labelled 'Equal') or swapped class (1 and 2, labelled 'Unequal'). See the text for detailed description of each subclass.  (Top row) Comparison outcome based on PHANGS-HST machine learning classifications.  (Bottom row) Comparison outcome based on human classification.  Note that the comparison set of objects is determined independently per object and also for ML and human classifications, with the ML outcomes ultimately based on a larger number of objects extending to fainter magnitudes.  See the text for details.}
 \label{fig:Doughnuts}
\end{figure*}

\begin{table*}
\begin{threeparttable}
 \caption{Comparison between the Class~1 and Class~2 cluster populations of PHANGS-HST and LEGUS}
 \label{tab:PHANGSLEGUScomparison}
 \begin{tabular}{lrrrrr}
  \hline
Quantity & NGC~628-C & NGC~628-E & NGC~1433 & NGC~1566 & NGC~3351\tnote{$a$} \\
   &         &         &         &         & (LEGUS-only) \\
  \hline
ML comparison set & 929 & 250 & 126 & 1270 & 264 \\
ML set: $M_{V}$ limit  & -6.15 & -6.0 & -6.15 & -7.0 & -6.2 \\
ML set: PHANGS    & 690 (74\%) & 205 (82\%) & 70 (55\%) & 906 (71\%) & 228 (86\%) \\
ML set: LEGUS     & 677 (74\%) & 169 (68\%) & 104 (83\%) & 869 (68\%) & 186 (70\%) \\
{\bf ML set: Both\tnote{$b$} } & {\bf 438 (47\%)} & {\bf 124 (50\%)} & {\bf 48 (38\%)} & {\bf 505 (40\%)} & {\bf 150 (57\%)} \\
ML set: PHANGS not LEGUS & 252 (27\%) & 81 (32\%) & 22 (17\%) & 401 (31\%) & 78 (29\%) \\
ML set: LEGUS not PHANGS & 239 (26\%) & 45 (18\%) & 56 (44\%) & 364 (29\%) & 36 (14\%) \\
  \hline
Human comparison set & 457 & 88 & 130 & 780 & 241 \\
Human set: $M_{V}$ limit  & -7.0 & -7.0 & -6.15 & -7.8 & -6.2 \\
Human set: PHANGS    & 361 (79\%) & 76 (86\%) & 88 (68\%) & 558 (71\%) & 193 (80\%) \\
Human set: LEGUS     & 354 (77\%) & 77 (87\%) & 88 (68\%) & 596 (76\%) & 186 (77\%) \\
{\bf Human set: Both\tnote{$b$} } & {\bf 258 (56\%)} & {\bf 65 (74\%)} & {\bf 62 (48\%)} & {\bf 374 (48\%)} & {\bf 138 (57\%)} \\
Human set: PHANGS not LEGUS & 103 (22\%) & 11 (13\%) & 26 (20\%) & 184 (23\%) & 55 (23\%) \\
Human set: LEGUS not PHANGS & 96 (21\%) & 12 (14\%) & 42 (32\%) & 222 (28\%) & 48 (20\%) \\

  \hline
   \end{tabular}
  \begin{tablenotes}
\small
\item[$a$]{This column pertains to analysis of the archival LEGUS data for NGC~3351 (NGC~3351-S in Table~\ref{TAB:exptime}) without any use of the new NGC~3351-N data obtained by PHANGS-HST, even in the area of overlap.}
\item[$b$]{Lines in boldface show that if one considers the union of Class 1 and 2 clusters in both surveys to be the true population, PHANGS-HST and LEGUS agree at the 48--74\% (56\% median) level for human classified objects brighter than $m(V)\sim23.5$ and 38--57\% (47\% median) level for PHANGS-HST ML classified objects brighter than $m(V)\sim24$.}
\end{tablenotes}
 \end{threeparttable}
\end{table*}

\begin{figure*}
 \includegraphics[width=6.5in]{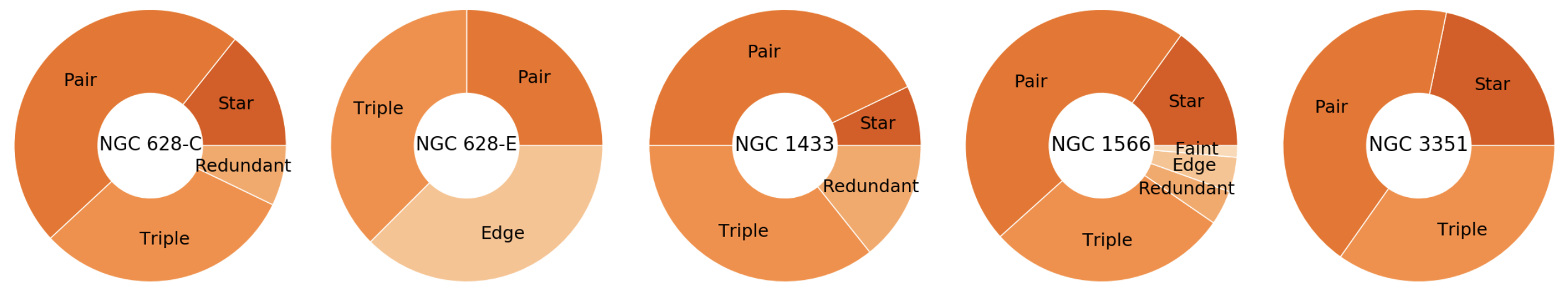}
 \caption{Human classification results for the comparison subset of LEGUS Class~1 or~2 objects labelled as Class~4 (contaminant) by PHANGS-HST. See the text for details of each subclass. Low number statistics are a concern for NGC~628-E and NGC~1433, so the most attention should be focused on the other targets (having more sources represented) in these plots.}
 \label{fig:DoughnutsH4subclass}
\end{figure*}

\begin{figure}
\includegraphics[width=\columnwidth]{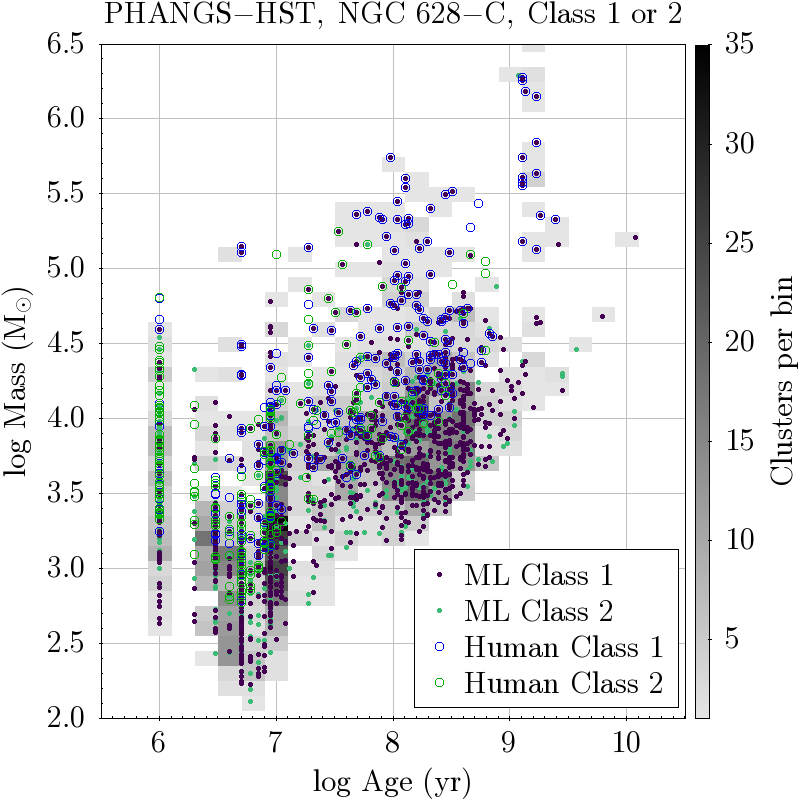}
\caption{Cluster mass versus age for Class~1 or~2 clusters in NGC~628-C. With open circles we plot the clusters recovered via human classification.  Dots represent the ML classified population.  Green colour is used for Class~2 (asymmetric) clusters, whereas blue circles and black dots represent Class~1 objects.  Note the marked extension of the ML population to lower masses and higher ages.  The greyscale histogram behind the plotted points shows the number density of ML clusters (Class~1 and~2 together). }
\label{fig:MassAgeplot}
\end{figure}

\begin{figure*}
\includegraphics[width=3.25in]{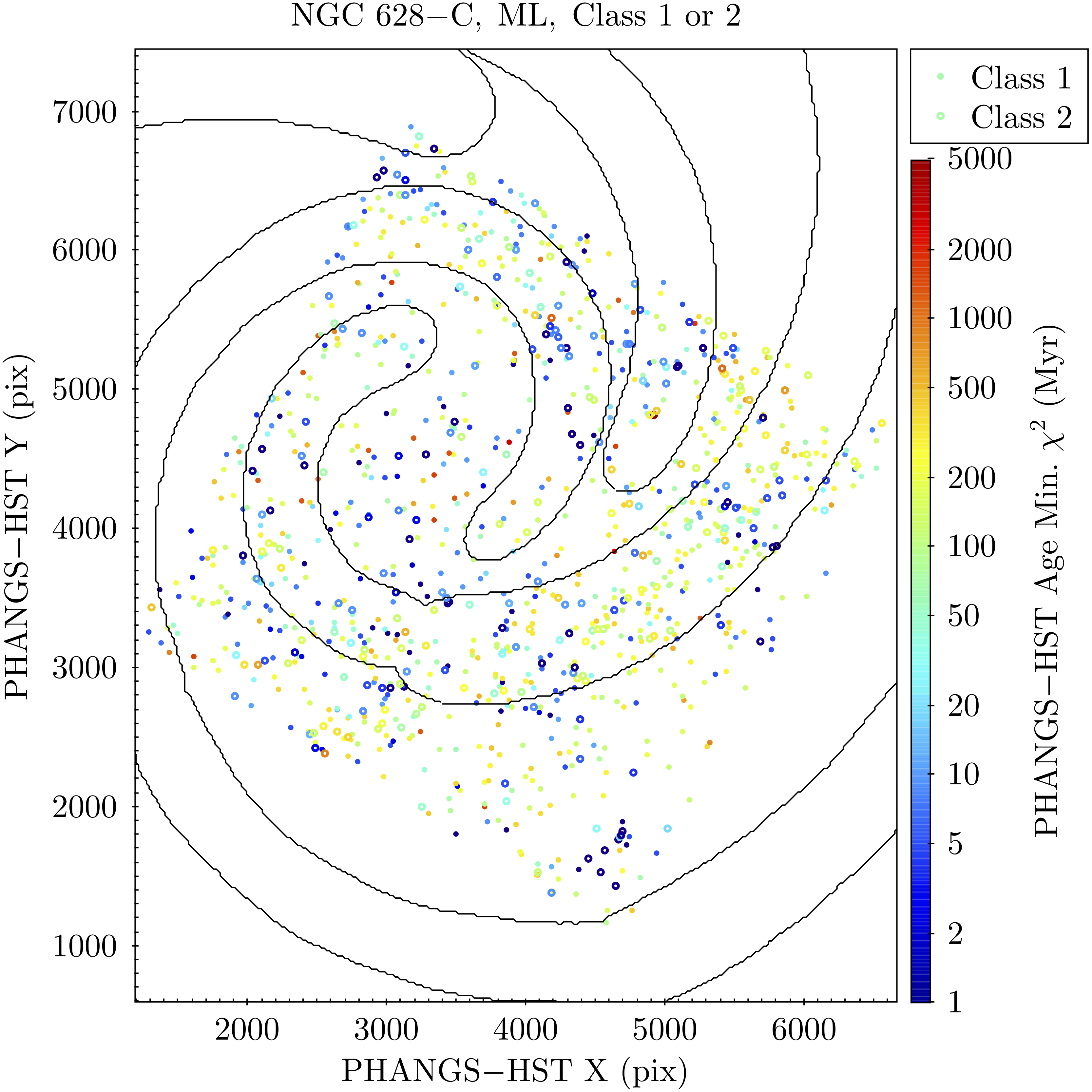}
\includegraphics[width=3.25in]{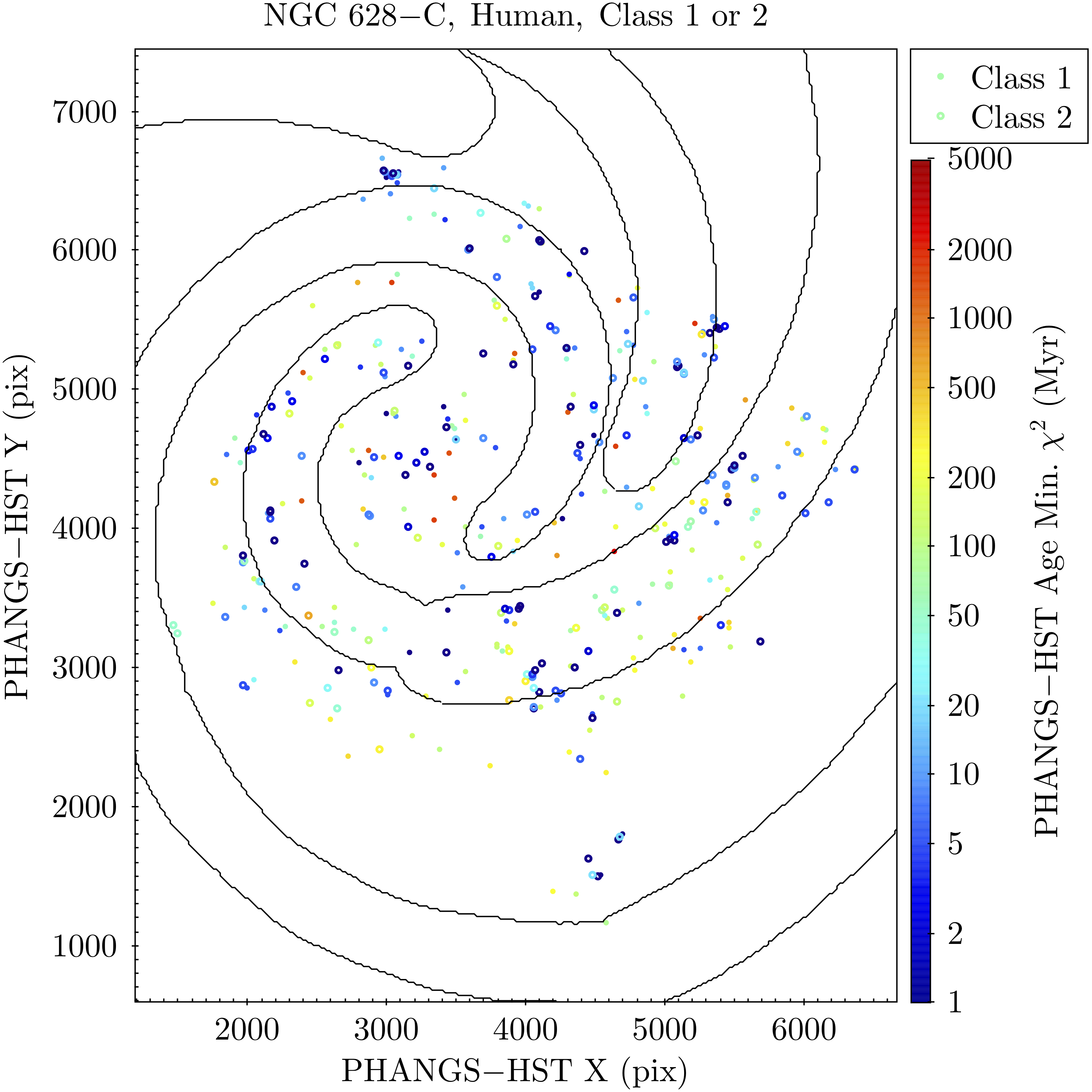}
\caption{Spatial distribution of PHANGS-HST Class~1 and~2 clusters identified in NGC~628-C, restricted to the area of {\em WFC3/UVIS}+{\em ACS/WFC} overlap, as the accuracy of our age determination is negatively impacted by the loss of $NUV$- and $U$-band photometry outside the (relatively smaller) {\em WFC3/UVIS} footprint.  (Left) The distribution of recovered Class~1 or~2 clusters according to the ML catalogue.  (Right) The equivalent figure using human classifications.  Note how the increased number of clusters in the ML catalogue makes it substantially easier to see spatial trends in age, such as those associated to spiral structure (the arms defined by \citealt{Querejeta2021} are outlined). Note that much of the very young star-forming population is excluded here, as we do not show Class~3 objects \citep[see multi-scale associations of][instead]{Larson2021}. This higher recovery rate of clusters will directly improve cross-correlation analyses with other objects such as CO clouds.}
\label{fig:spatialage}
\end{figure*}
It is vital to develop an understanding for potential differences between the Class~1 and~2 cluster populations identified by our new method and the techniques used by others.  As noted earlier in the text, we chose to compare against the LEGUS survey, as several galaxies are in common between the projects and, even for targets that differ, equivalent five band $NUV$-$U$-$B$-$V$-$I$ $HST$ data are employed.  We anticipated both random and systematic differences in the resulting catalogues, with the former associated primarily to classification uncertainty and the later linked to changes in source detection / candidate selection procedures.  We did not undertake a head-to-head comparison of the Class~3 objects (compact associations) resulting from our cluster identification pipeline versus LEGUS, as our methods were tuned to preferentially identify single-peaked cluster-like objects.

We first generate a set of sources to guarantee a fair comparison between the surveys.  We account for differences in photometry and coverage. LEGUS classified cluster candidates down to $M_{V} = -6$, effectively setting the faint end for comparison.  Not surprisingly, slight differences in aperture corrections, photometric scatter (from minimal shifting of candidate centres), and occasional changes in aperture size suggest that in some cases we should back off from the $-6$ limit to ensure that objects brighter than $M_{V} = -6$ in PHANGS are also above this limit in LEGUS.
The magnitude limits were allowed to vary by target and for machine learning classification-based comparisons versus human classification.  The adopted values are given in Table~\ref{tab:PHANGSLEGUScomparison}.  For our ML-based comparison the limit is generally up to $0.2$~mag brighter than $M_{V} = -6$, though LEGUS only classified sources down to $M_V=-7$ in NGC~1566.  For our human classification-based analysis, the limit is even brighter for NGC~628-C, NGC~628-E, and NGC~1566, as these galaxies have very rich cluster populations and it was not practical for our expert human classifier (BCW) to inspect the entire population.
Finally, of importance when setting the scope of comparison, for NGC~628-C and NGC~628-E the observational data were a mix of {\em WFC3/UVIS} and {\em ACS/WFC} imaging.  These cameras have a different size on the sky and were not forced to have identical orientation.  LEGUS only searched for clusters in the intersection of {\em WFC3/UVIS} and {\em ACS/WFC} coverage, whereas we allow cluster candidates anywhere with coverage in at least three bands, including~$V$ ($F555W$).  Accordingly we cut back the footprint of the survey comparison to the region with LEGUS classifications.  Our eventual `comparison sets' are defined as the union of Class~1 or~2 clusters from either PHANGS-HST or LEGUS meeting the magnitude and sky area criteria described above. Table~\ref{tab:PHANGSLEGUScomparison} gives the number of sources in the comparison sets as the first row in each section of the table.

Within this comparison set for each target we tallied the number of Class~1 or~2 clusters: (a) found by PHANGS-HST, (b) found by LEGUS, (c) found by both surveys, (d) found by PHANGS-HST but not LEGUS, and (e) found by LEGUS but not PHANGS-HST.  These cluster counts are listed in Table~\ref{tab:PHANGSLEGUScomparison}, along with their associated percentage of the entire comparison set.  The percentages are also presented graphically in Fig.~\ref{fig:Doughnuts}.  In a panel devoted to each target and the type of PHANGS-HST classification (ML or human), the {\bf inner ring} of each doughnut plot shows basic results of the survey comparison.  Relative to the entire comparison set (think of this as anything that could possibly be a Class~1 or~2 cluster),
the surveys agree at a significant percentage (case~(c) above), approximately 50\% for ML (median 47\%) and slightly higher for human classifications (median 57\%).  Agreement is weakest for NGC~1433, with 38\% for ML and 40\% for human classification.  Survey agreement also falls for NGC~1566, which we believe is a consequence of its substantially larger distance.  This also makes sense in terms of NGC~1433, which \citet{Anand2021} have shown is actually at a distance of 18.6~Mpc rather than the 8.3~Mpc we assumed (in this paper only) for consistency with LEGUS.  The percentages discussed above should not be interpreted in terms of completeness, since some objects will be identified as clusters by one survey but not by the other, 
and because the source list itself includes at least some non-clusters from both surveys.
Rather, this $\sim50{-}60$\% agreement reflects that both LEGUS and PHANGS-HST are likely returning Class~1 and~2 cluster samples with at least this level of accuracy.

At face value, the higher and generally rather similar percentages of the comparison set reported for PHANGS-HST (case~(a) -- blue+green in Fig.~\ref{fig:Doughnuts}) and LEGUS (case~(b) -- red+green) suggest that recovery of the true Class~1 or~2 population is indeed higher than the intersection statistics above require as a minimum.  The median percentage recovered by the surveys for ML is 73\% and for human classification is 77\%, calculated by integrating over both surveys without preference.  Of course, both surveys are likely subject to contamination and incompleteness.  Contamination in particular will make the case~(a) and~(b) percentages skewed misleadingly high for the survey with the issue and low for the other survey.  Without a more detailed look at the objects seemingly missed by either survey it is difficult to interpret any further than to say that both surveys most likely recover on the order of 75\% of true Class~1 or~2 clusters meeting the magnitude/footprint constraints, and if systematic identification/\linebreak[0]{}misclassification exists then this figure will drop or rise accordingly depending on which survey has the problem.  

Blue and red shaded portions of the {\bf outer rings} of the doughnut plots in Fig.~\ref{fig:Doughnuts} aim to better understand the types of sources for which the surveys find different classifications.
{\em We begin with objects identified as Class~1 or~2 by PHANGS-HST but not by LEGUS (shown in blue).}  The wedges labelled L3 and L4 were
classified as Class~3 and Class~4, respectively, by LEGUS (but Class 1 or~2 by PHANGS-HST).  Approximately one quarter to one third of the PHANGS-HST-only Class~1 or~2 population falls into this category.
We note that this happens more often in NGC~3351 than for other targets (42\% for ML, 58\% for human).  In the case of human classification, this may be due to the greater exposure depth (which was not used in ML), in the overlapping LEGUS and PHANGS-HST fields.
An even larger contribution ($\sim40{-}70$\%) to this misclassification category comes from sources which were somehow not even considered by LEGUS (labelled `no ID').
These result from intrinsic differences in the types of sources detected by SExtractor versus \codename{DOLPHOT}, or due to automated cuts made by LEGUS after the detection stage (CI, number of bands with good photometry).  
A minority of sources, belonging to doughnut wedges labelled with asterisks, were at least considered as LEGUS candidates and seem to have met the conditions for LEGUS classification given by \citet{Adamo2017} but did not receive an eventual class for some reason we have not been able to determine.
{\em Next, we examine the sources deemed Class~1 or~2 by LEGUS but not by PHANGS-HST (red region in the outer doughnut).}  A first category is comprised of sources we flag as `doubles' (see Sec.~\ref{sec:doubles}), where two objects are detected within 2.5~pixels of one another.
Extended sources at such small separations, even if real and not a redundant \codename{DOLPHOT} detection of the wings of the brighter extended object, would be challenging to classify accurately.  We find that $10{-}20$\% of the LEGUS Class~1 or~2 objects not included by PHANGS-HST fall into this category.  \citet{Whitmore2021} also comment on the preferential inclusion of `doubles' by LEGUS, and find they are typically assigned to Class~2.  Differences in classification between surveys can also be seen as many PHANGS-HST Class~1 or~2's are called Class~3 or~4 by LEGUS.  In terms of Class~3 (see ML3, H3 doughnut wedges), they amount to $\sim$15\%.  One comment applicable to both survey classification efforts is the difficulty in classifying objects in crowded environments -- this is particularly relevant to the distinction between Class~3 and Class~1+2.  For example, imagine taking a confident Class~1 or~2 cluster and placing it in a region with many point sources or even other clusters.  Based on the relative location of objects, it is possible that such a Class~1 or~2 is frequently judged by the classifier (either ML or human) as an apparent Class~3 compact association because the only evidence remaining of cluster-nature is the very inner profile of the source, which can be overridden or overlooked due to environment.  We have no reason to believe this happens more often in either survey, and the ML3/H3 fractions are indeed similar to H3 in Fig.~\ref{fig:Doughnuts}.  A far more significant (and systematic) classification trend between LEGUS and PHANGS-HST is represented by the ML4/H4 wedges which dominate the LEGUS-only Class~1 or~2 population.  In most targets more than 50\% (and up to $>70$\%) of LEGUS Class~1 or~2 sources not called 1 or~2 by PHANGS-HST are instead classified as contaminants by PHANGS-HST.  We return to this point below.  Finally, a small percentage of LEGUS Class~1 or~2 clusters are omitted from the PHANGS-HST Class~1 or~2 census for reasons we did not track down (red outer wedge labelled with asterisk), but possibly these sources failed cuts on MCI error. 

In the case of the human comparison set, we have more detailed information on the appearance of sources LEGUS called Class~1 or~2 and PHANGS-HST classified as contaminants (that is, Class~4 in the original LEGUS system, and now subdivided by BCW into Classes 4.1, 4.2, \ldots, 4.12 in order to specify the variety of contaminant cluster candidate, e.g. single star, stellar pair, stellar triple, saturated star, diffraction spike, galaxy nucleus, background galaxy, etc. -- see \citet{Whitmore2021} for details).  In Fig.~\ref{fig:DoughnutsH4subclass}, we show doughnut charts for these human Class 4 (`H4') sources, splitting them into up to six BCW subclasses (others of the 12 were not encountered).  Specifically, the H4 sources are classified as a single star, pair or triple stars, redundant (another accepted Class~1, 2,~3 within 5 pixels), edge artefact, or too faint to classify.  This categorical assessment of contributors to the H4 LEGUS-only population is most informative/\linebreak[0]{}reliable in cases having many sources.  For this reason NGC~628-E and NGC~1433 should be viewed cautiously, as they each have $\le 15$ H4 sources (only~8 in NGC~628-E).  

The first observation to be made from Fig.~\ref{fig:DoughnutsH4subclass} is that sources appearing to BCW as single stars are a minority in this population, amounting to less than 25\% in all cases.  The bulk of H4 sources are thought to be pairs or triples of stars according to the PHANGS-HST human classification.  In general pairs constitute $\sim45\%$, and triples $\sim33\%$, of the H4 LEGUS-only population.  {\em Combined with the already flagged `doubles' from the analysis above, we conclude that the majority of objects called Class~1 or~2 by LEGUS and not by PHANGS-HST are actually multiple stellar sources not meeting the source surface density needed to qualify as Class~3 associations}. \citet{Whitmore2021} describes how this distinction is set by counting peaks within a 5 pixel radius.  This effect is probably accentuated by situations in which SExtractor does not deblend such multiples and places a source between the actual peaks.  This is a particular strength of using the PSF-fitting source detection algorithm (\codename{DOLPHOT}) for our work, though it comes with the price of spurious detections in the wings of bright objects (largely eliminated by our doubles check, Sec.~\ref{sec:doubles}) and the need to explicitly eliminate redundant counting of associations due to proper deblending of subclumps (last paragraph of Sec.~\ref{sec:doubles}).
The subclasses not yet described (redundant, edge, too faint) remain in the minority across the entire H4 population, with faint objects only cropping up in NGC~1566 and edge artefacts in NGC~1566 and NGC~628-E.  The slightly more numerous `redundant' population can be attributed to differences in source detection between surveys.  

Lastly, we emphasise one final point from Fig.~\ref{fig:Doughnuts} concerning exact classification agreement.  In the outer ring of the doughnut plots, in two shades of green, we indicate the fraction of Class~1 or~2 clusters in common between surveys that were given the same Class~(e.g.\ 1~and~1; 2~and~2) with the darker wedge, and those that disagreed (e.g.\ 1~and~2; 2~and~1) with the lighter colour.  Quite commonly the survey classification agreed on specific Class for this subset of the the overall comparison set.  We direct the reader to \citet{Whitmore2021} for a detailed, extensive discussion of confusion between classes, including a discussion of confusion matrices for cross-survey (PHANGS-HST versus LEGUS) and internal (VGG versus ResNet, human versus VGG, etc.) comparisons.

Complementing the limited PHANGS-HST-ML to LEGUS-human classification results (top row of Fig.~\ref{fig:Doughnuts}) of this Section, we direct the reader to the more extensive analysis of similar issues by \citet{Whitmore2021}.  In particular, \citet{Whitmore2021} introduce use of the colour-colour diagram as a 'figure of merit' to further test the relative performances of the different methods.

\subsection{Resulting PHANGS-HST cluster catalogues}

This paper is not intended to provide astrophysical analysis of the cluster populations recovered by our methods, but rather to focus on the adopted procedures.  Nevertheless, without presentation of some scientifically relevant plots, it would be impossible to fully appreciate the improvement we have attained relative to previous works, except in the sense of the comparison described in Sec.~\ref{sec:PHANGSLEGUScomparison}. In Fig.~\ref{fig:MassAgeplot}, we show the PHANGS-HST Class~1+2 clusters identified by human and ML classification for NGC~628-C. Fig.~\ref{fig:spatialage} shows the spatial distribution of clusters for the same field (trimmed to the {\em WFC3/UVIS}+{\em ACS/WFC} footprint) colour-coded by age in both contexts.  These figures displaying our results for one target are both internally instructive, emphasising the more expansive cluster population recovered using ML classification relative to human-based efforts, and also can be compared versus the results of \citet{Adamo2017} who used the same data.  Appendix A contains equivalent figures for all the fields analysed in our paper.

\section{Discussion}
\label{sec:discussion}

\subsection{General comments on our overall strategy}

A limiting step in most studies of extragalactic star clusters is the manual effort needed to remove artefacts from candidate lists. It is difficult to automatically identify candidate star clusters with a high degree of certainty due to environmental complexity (effects such as crowding and high background), in addition to resolution limits.  The MCI approach we introduced was originally designed to reduce the number of candidate star clusters that would need to be manually examined to remove artefacts such as pairs of stars, background galaxies, edge effects, etc. The basic strategy was to leverage photometry from several apertures rather than just the two used to determine the concentration index (CI), which has been the standard way to filter out artefacts. In this way, selection criteria could be enforced that insured that the profile followed a typical profile of a cluster rather than a star, pair of star, or other more complicated morphologies. Original trials using this approach were largely successful, reducing the size of the candidate list by about~50\%.

During implementation, several other considerations reduced this percentage somewhat, while enabling other improvements to the overall strategy. One example is that the selection routine for both stars and cluster candidates has been unified to use a single primary method (\codename{DOLPHOT}), rather than using two completely different methods employed by many other projects. This eliminates the necessity for subsequently cross matching the stellar and cluster catalogues after totally independent selection/\linebreak[0]{}classification decisions have been made, which would otherwise be a significant source of noise and complexity at a later stage of a project. However, a side-effect of the use of \codename{DOLPHOT} is that it introduces a much larger initial candidate list since \codename{DOLPHOT}'s selection routine uses an iterative method to optimally deblend neighbouring sources and find fainter, inconspicuous sources in the wings of bright objects. This is the main source of the growth of our candidate cluster lists. Another similar, but less important effect from a growth standpoint is the development of a more sophisticated method of removing point-like objects, allowing us to include proportionately more (of particularly compact) cluster candidates than possible using only the concentration index, hence improving the completeness levels. 

The MCI approach also enables other potential improvements, many of which will be more evident in later stages of the PHANGS-HST project. For example, the development of cluster models needed to define selection regions for the MCI values will be useful in the analysis of various physical characteristics of the clusters, such as whether the objects are likely to be bound or unbound, whether certain kinds of potential profiles (e.g.\ very diffuse) are not actually found in nature, and how the profiles of clusters change as a function of age.  

In summary, while some of the initial goals of the MCI method have not been fully realised (i.e.\ the reduction in the candidate list), the overall improvements to the project is evident and is expected to have long range benefits relative to the PHANGS-HST science goals and cluster studies in general.

\subsection{Realised scientific advantages of our method}

Our fully integrated pipeline for cluster detection, identification, and ML classification produces reliable catalogues of clusters with minimal human effort.  These catalogues have clear advantages over prior efforts, even using the same observational data and neglecting the relative ease with which they can be generated.  This subsection is devoted to highlighting such scientific benefits in broad terms. 

The advantage of reproducibility, and presumed replicability, should not be understated. Even within LEGUS, the premier cluster study prior to PHANGS-HST, attaining reproducibility was challenging \citep[][]{Perez2021}.  The burden of human classification relative to the limited pool of trained team member classifiers ($\sim 12$) led to compromises in the uniformity of the adopted procedure.  For instance, in general for each galaxy only a small subset (3~people) of classifiers examined the candidates \citep[][]{Adamo2017}.  This could impart systematic differences in the effective classification scheme between targets, if individual volunteers gave more weight to different aspects of the classification procedure. In cases where the number of candidates per target were very high, the classification task was divided across the galaxy, assigning stripes of coverage to different subsets of the classification group.  It should be said that membership in each subset was shuffled, and the mode was adopted as final judgement amongst the classification votes per candidate, in hopes of minimising systematic effects. The adopted LEGUS method represents the best that can be achieved with a small group of human volunteers.  \footnote{Experimentation with citizen science based classification was undertaken by LEGUS as well, to potentially overcome the limits of person power, but the lack of interactive inspection tools (image contrast, radial profiles) on current platforms led to the conclusion this was not yet feasible. We note that in the more resolved case of the Magellanic Clouds, and M31 / M33 (with $HST$), citizen science cluster identification has proven rather successful \citep[e.g.][]{Johnson2012}.}  

For PHANGS-HST, at no stage in the ML classification branch of our pipeline does human judgement enter the procedure in real time, so in one sense our analysis is reproducible.  However, the training of the ML network embodies the non-reproducible\footnote{\citet{Whitmore2021} highlight the fact that even an expert reviewer, conducting classifications of the same objects separated by a period of several years, will assign a small fraction of clusters to differing classes.} expert knowledge of cluster classification of co-author BCW, as well as the collective efforts of LEGUS classifiers (used for training by \citealt{Wei2020} alongside BCW-only classifications).  Because there are humans involved in the construction of the training sets,  the outcome of the procedure may change if the training set is constructed by a different human[s], though \citet{Wei2020} conclude that LEGUS and BCW training sets seem to produce comparable accuracies. We expect ML classification accuracy to improve with further training method experimentation and as the database of human classified objects grows and becomes fully representative, both factors described by \citet{Whitmore2021}.  Systematics can still exist in our ML classification, dependent upon the overall suitability of the training sets to the actual objects being classified, and possibly on the degree to which the training sets are balanced among classes (clusters versus stars versus artefacts) or even among cluster morphologies within a class (physically large clusters versus compact).  In regard to overall suitability, the majority of galaxies in the full PHANGS-HST sample are more distant than the LEGUS targets that were used by \citet{Wei2020}.  This is one of the main reasons that we still undertake human classification in our project, with the goal of accumulating a training sample that probes a more fully representative range of distance spanning the Local Volume (e.g.\ LEGUS) out to objects beyond the Virgo cluster (some of the PHANGS-HST targets).  In essence, the construction of a training set should be seen as an integral part of the ML procedure, not as something that precedes it. To summarise, we hope that full adoption of ML classification by the star cluster community (see also important work by \citealt{Perez2021} and \citealt{Bialopetravicius2019}) will lead to replicability amongst groups that often have somewhat different data sets, and such replicability can build consensus regarding our understanding of cluster formation / evolution / destruction. 

Secondly, and in a more practical sense, another methodological advantage is that by relying on ML classification we enable complete evaluation of all candidates down to our detection (selection) limit.  Naturally, classifications of fainter objects can be expected to individually be less confident (distinguishing between Class 1 and Class 2 for example) than for bright analogues in the same environment (see \citealt{Whitmore2021}).  However, integrated over an ensemble of many objects the cluster / not-a-cluster ML evaluation (that is, Class~1 or~2, or not) is still reliable near the faint limit of our candidate selection ($V$-band $\mathrm{S/N} \sim 10$), based on spot checks of deeper human classification versus the ML classification.  We emphasise further that by classifying our complete list of candidates we are able to include the often substantial, easy to classify population of faint candidates occupying isolated, low surface brightness, low confusion environments.  These typically get missed by human classification efforts, unless specifically remedied post facto (e.g. over 100 faint diffuse sources were added manually to the NGC 4449 sample in \citealt{Whitmore2020}).  

Unfortunately, improved recovery of faint clusters enabled by ML is counteracted by losses in our ability to determine accurate photometric ages for young star clusters with mass $\lessapprox10^{4}$ M$_\odot$ due to stochasticity \citep[e.g.][]{Fouesneau2014,Krumholz2015,Hannon2019,Whitmore2020}. In some science use cases requiring single object characterisation, this may limit the usefulness of clusters at faint apparent magnitude, particularly as galaxy distance decreases.

Even so, ML classification for complete candidate samples dramatically increases the number of recovered clusters, translating into statistical improvement in the age--mass diagram, reaching significantly lower mass at fixed age, or alternatively greater age at fixed mass.  In Fig.~\ref{fig:MassAgeplot}, we show the PHANGS-HST Class~1+2 clusters identified by human (open circles) and ML classification (dots).  Class~2 clusters are colour-coded green. Notice how the ML population reaches masses about 0.5~dex ($\sim3\times$) lower than the human classification.  In the alternative perspective, at fixed mass (at least for intermediate age clusters), the ML catalogue probes up to a factor ten older in age.  Appendix~A includes age--mass diagrams for all the fields analysed in this paper.  The degree of human-to-ML improvement varies within the five fields, owing to differences in depth of the human classifications.  

There are numerous trickle-down scientific gains to expect coming from expansion of the identified cluster population.  For example, improvements in the age--mass diagram will impact studies of the cluster mass function by allowing the bins in age that are typically used to each extend to lower mass (e.g.\ R. Chandar et al. in prep, note that the intermediate age clusters are often examined separately as they have survived removal from their natal environment). This will generally boost the statistical significance of mass function fit results through increased numbers, and may perhaps allow sensitivity to detect broken power law mass functions.  Far better populated age--mass diagrams will yield improvement in modelling of cluster disruption (mass dependent versus mass independent) as it provides additional statistics for low mass  clusters in an evolved, intermediate age state.   We anticipate $\Gamma$ \citep[the fraction of star formation occurring in bound clusters, e.g.][]{Bastian2008,Kruijssen2012} determinations will increase, given that some of the newly recovered clusters are of Class 1 and have ages significantly larger than their crossing time (thus are likely bound at high confidence).

Another trickle-down scientific gain from the increased number of clusters is improvement of studies that use clusters as `clocks' to estimate timescales for physical processes via cross-correlation analyses (such as cloud disruption, via comparison to the ALMA CO catalogues unique to PHANGS-HST).  This is due to the increased surface density of the recovered cluster population in the ML census versus human classification outcome, as best appreciated in Fig.~\ref{fig:spatialage}. Note the improved clarity of patterns in age in the ML cluster distribution.  Appendix~A presents the full set of ML classified catalogue spatial distributions. In the timescale `clock' application, having a more complete cluster catalogue means that the power of apparent physical association between cluster and cloud will contribute to the cross-correlation function at the `correct' length scale more frequently than otherwise in the limited human catalogue.  Viewed in another way, thinking in terms of nearest neighbour analysis, incompleteness in a cluster catalogue might lead to the `wrong' neighbour being selected.  This being said, measurements \citep[e.g.][]{Cook2019} of $\Gamma$ in main sequence galaxies like those in the PHANGS-HST sample are generally $\lessapprox30$\%, and will almost certainly remain less than 50\% even with expanded cluster catalogues.  Therefore, in order to capture the majority of the star formation activity, we stress the importance of using other localized tracers such as the most compact multi-scale associations (e.g. \citealt{Larson2021}) as well for timescale 'clock' applications.  

\begin{figure*}
\includegraphics[width=6.5in]{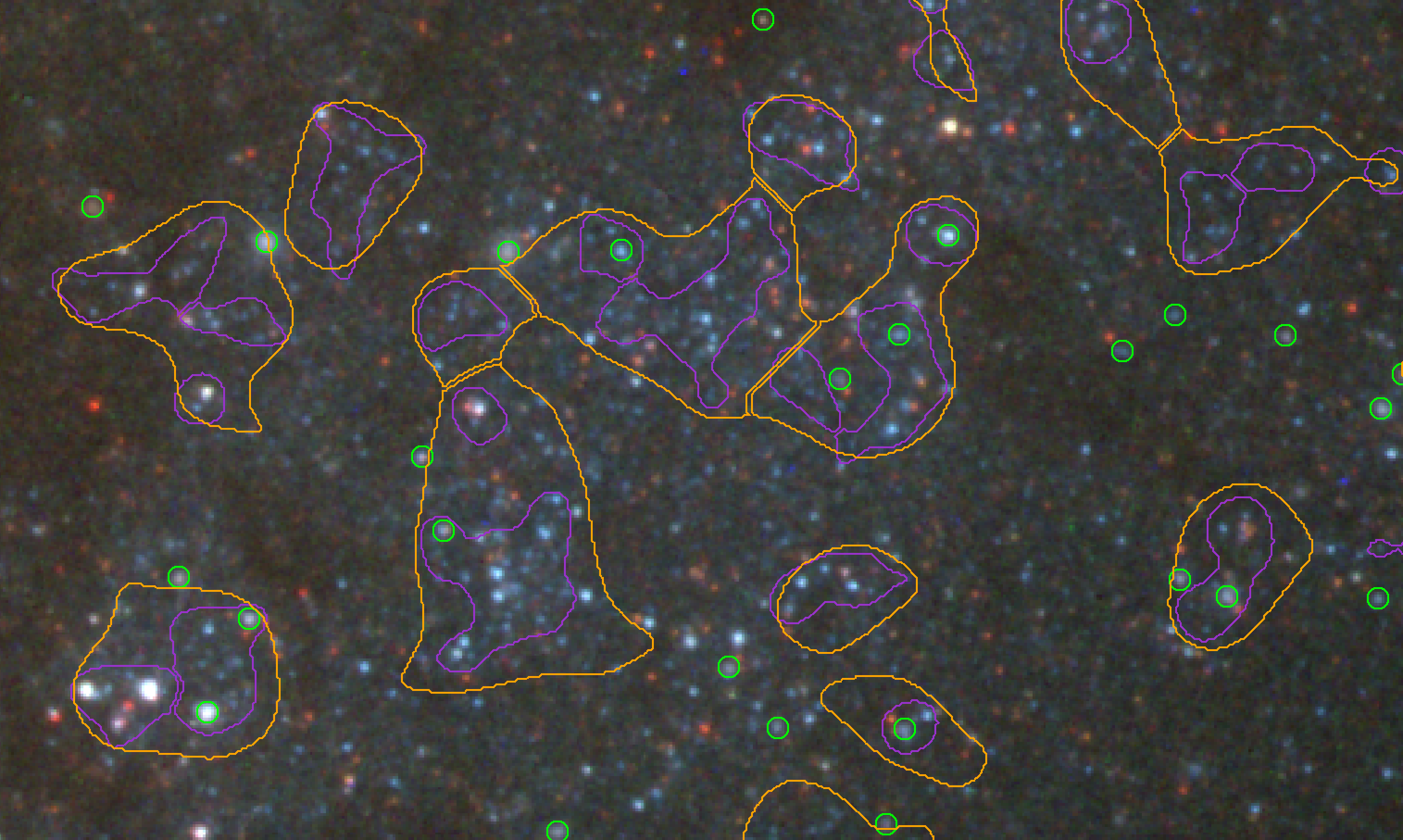}
\caption{For a small portion of the NGC~628-C field, located in a spiral arm west of the galaxy center, we show PHANGS-HST Class~1 and~2 ML clusters (as green circles) and the hierarchical associations of \citet{Larson2021} determined on 32 and 64 pc scales using NUV point-sources (as purple and orange polygons, respectively).  The HST image is a colour composite constructed from $I,V,B$ as red, green, and blue. } 
\label{fig:AssocCluster}
\end{figure*}

Lastly, although discussed at length in the previous Section, we end the presentation of scientific benefits associated to our methods by reiterating the improvement we achieved with respect to excluding (often random) pairs and triples of point-like sources from the Class~1 and~2 census.  This reduction in contamination is partially due to our specific attention to this issue but also due to our adoption of \codename{DOLPHOT} rather than SExtractor as source detection code.  Our choice to use \codename{DOLPHOT} as a starting point also strategically facilitates a self-consistent inventory of point sources and clusters, and enables determination of multi-scale stellar associations \citep[e.g.][]{Larson2021}.  Fig.~\ref{fig:AssocCluster} illustrates the combined census of star formation products attained by PHANGS-HST by displaying the associations and clusters of NGC~628-C together.  It is abundantly clear that a complete view of recent (few 10$^7$ yr) star formation requires methods that capture structures spanning from very likely bound ($\Pi >$ many), possibly but indeterminately bound ($\Pi \lessapprox$ few), to unbound and already dissolved/disrupted.  Another illustration of the innovative PHANGS-HST approach to combined recovery of clusters and associations can be found in Fig.~9 of \citet{PHANGSHSTsurvey}, highlighting a region of NGC~3351 with all components of our multi-observatory PHANGS program displayed.  With the union of PHANGS cluster, association, CO cloud (from ALMA), and {\sc H\,ii} region (from MUSE and traditional narrowband imaging) catalogues we will be able to place unprecedented constraints on the cyclic process of star formation in nearby galaxies.

\section{Concluding remarks}
\label{sec:conclusions}

In this paper we have described a new analysis pipeline developed for the PHANGS-HST survey, designed to detect, measure, optimally select, and morphologically classify star cluster candidates.

Our pipeline employs innovative strategies such as: (1) source detection and deblending using PSF photometry code \codename{DOLPHOT}, leading to self-consistency between cluster and association catalogues, (2) cluster candidate selection based on multiple concentration indices (MCI$_{\mathrm{in}}$ and MCI$_{\mathrm{out}}$) that more fully characterise the source morphology than possible with a single CI, (3) selection criteria optimised with guidance from synthetic clusters and previously confirmed clusters, and with special allowance for potential candidates more luminous than the Humphreys-Davidson (H-D) limit, and lastly (4) fully-pipelined ML classification of cluster candidates.


To test the performance of our method, we apply the pipeline to four galaxies: NGC~628, NGC~1433, NGC~1566, and NGC~3351, with assumed distances between 8.3 and 18 Mpc. Across the four galaxies, the median candidate sample size selected for ML (human) classification is 3389 (2366) with range of 1989--8822 (1230--5695).  The median number of candidates in each galaxy classified by a human (limited by practicality) is 1039 with range of 174--1442;
the median number of these sources which are verified to be single-peaked star clusters (Class 1 or 2) is 298 with range of 91--634. 
For NGC~1566, at 18 Mpc which is close to the median distance of the PHANGS-HST sample, for such human verified clusters, the median stellar mass is $3.1\times10^4~M_\odot$ with a 25--75 percentile interval of $\sim1.5{-}7.0\times10^4~M_\odot$.  These values drop by about $3{-}5\times$ for the other targets in the current paper (at 8--10 Mpc).  All candidates have been classified using ML \citep[deep transfer learning CNN models of][]{Wei2020}, yielding Class 1 or 2 cluster classifications for 4292 objects (for VGG, 17\% of the ML sample).  The ML-classified Class 1 or 2 samples reach $\sim$1 magnitude fainter, $\sim2\times$ lower mass, and are $2{-}5\times$ larger in number, than the human classified samples. The resulting age-mass diagrams and spatial distributions are provided in Appendix~A.

These four galaxies were selected for study because they have star cluster catalogues which have been independently produced by the LEGUS program, using a standard approach for selecting candidates.  Comparison between the LEGUS catalogues and the ones resulting from our PHANGS-HST pipeline allow for one test of the effectiveness of our new approach.  Based on the union of the two catalogues, 48--74\% (56\% median, see lower portion of Fig.~\ref{fig:Doughnuts} and note ($b$) of Tab.~\ref{tab:PHANGSLEGUScomparison}) of human verified Class 1 and 2 star clusters appear in both catalogues.  The remainder, clusters which are unique to one of the catalogues, are discovered in roughly equal proportion by the two methods.  Interestingly, we find that the majority of clusters unique to LEGUS are classified as contaminants in the PHANGS-HST catalogues, and the majority of those unique to PHANGS-HST appear to be likely clusters not captured by the LEGUS pipeline.  

Overall, we find that our new selection method is effective, and furthermore, yield a substantially larger cluster population than previously identified in each of our target galaxies.  These cluster catalogues, and the forthcoming ones from the other 34 galaxies in the PHANGS-HST program, will transform our understanding of the link between cluster populations and environmental properties of the ISM (CO measured with ALMA; {\sc H\,i} with the JVLA, Karl G. Jansky Very Large Array; and optical nebular emission lines with MUSE IFU spectroscopy), especially when combined with our upcoming JWST observations recovering embedded clusters while also clarifying PAH and small dust grain properties.  The complete potential of our PHANGS-HST data set would not have been accessible without successful pipeline implementation of ML classification, owing to the very large number of cluster candidates.  This seamless integration of automated classification has not been achieved, and deployed to an entire galaxy sample, previously in the field of star clusters.  As noted in Section~\ref{sec:MLclassification} and \citet{Whitmore2021}, we expect that the accuracy of ML classification will further improve as more extensive experimentation with varied training strategies is undertaken.  Even so, the current accuracy is on par with an expert human classifier.

Our analysis of synthetic clusters provides a database with measured properties for a rather diverse set of parametric Moffat cluster morphologies.  We indicated such linkage between cluster morphology and metric values could be used to further simulate the distribution of observed characteristics for a mock cluster population.  Folding in completeness estimates as a function of cluster properties would allow us to interpret the ensemble measured properties of a detected cluster population in terms of assumed (cluster formation rate, destruction $d^2N/dM\,dt$, and the distribution of $A_V$ and morphology).  This forward-modelling approach will be explored in future work.  Aspects of this idea, omitting information on cluster morphology, have been published in the context of hierarchical Bayesian analysis by \cite{KrumholzLEGUS2019}.  By incorporating cluster structural measurements, either MCI-based or via Moffat fitting, and ideally also the population of OB associations that were once clusters, we can expect the utility of such work to grow considerably.




\section*{Acknowledgements}

Based on observations made with the NASA/ESA Hubble Space Telescope, obtained from the data archive at the Space Telescope Science Institute. STScI is operated by the Association of Universities for Research in Astronomy, Inc. under NASA contract NAS 5-26555.  Support for Program number 15654 was provided through a grant from the STScI under NASA contract NAS5-26555.

Most of the plots in this paper were generated with \codename{TOPCAT} \citep{TOPCAT} and/or its sister command-line package \codename{STILTS} \citep{STILTS}, both developed and generously released/maintained for public use by Mark Taylor.  A significant amount of interactive data exploration and testing was conducted using \codename{TOPCAT}.

Our pipeline makes extensive use of the following software packages: \codename{DOLPHOT, photutils, astropy, matplotlib, numpy, IMFIT, pytorch, CIGALE}.  We extend our appreciation to their respective developers.

This research has made use of the NASA/IPAC Extragalactic Database (NED) which is operated by the Jet Propulsion Laboratory, California Institute of Technology, under contract with NASA. 

JMDK gratefully acknowledges funding from the Deutsche Forschungsgemeinschaft (DFG, German Research Foundation) through an Emmy Noether Research Group (grant number KR4801/1-1) and the DFG Sachbeihilfe (grant number KR4801/2-1), as well as from the European Research Council (ERC) under the European Union's Horizon 2020 research and innovation programme via the ERC Starting Grant MUSTANG (grant agreement number 714907).

SCOG and RSK acknowledge support from the DFG via SFB 881 “The Milky Way System” (sub-projects A1, B1, B2 and B8) and from the Heidelberg cluster of excellence EXC 2181-390900948 “STRUCTURES: A unifying approach to emergent phenomena in the physical world, mathematics, and complex data”, funded by the German Excellence Strategy. They also acknowledge funding from the European Research Council via the ERC Synergy Grant “ECOGAL” (grant 855130).

TGW acknowledges funding from the European Research Council (ERC) under the European Union’s Horizon 2020 research and innovation programme (grant agreement No. 694343).

\section{Data Availability}

The imaging observations underlying this article can be retrieved from the Mikulski Archive for Space Telescopes at \url{https://archive.stsci.edu/hst/search_retrieve.html} under proposal GO-15654. High level science products, including science ready mosaicked imaging, associated with HST GO-15654 are provided at \url{https://archive.stsci.edu/hlsp/phangs-hst} with digital object identifier \doi{10.17909/t9-r08f-dq31}



\bibliographystyle{mnras}
\bibliography{phangs_clusteridentification}


\appendix
\section{Summary Figures}

This appendix presents the age--mass and age-coded spatial distribution plots for all the fields we processed as part of this paper.  
%

\begin{figure*}
\includegraphics[width=4.5in]{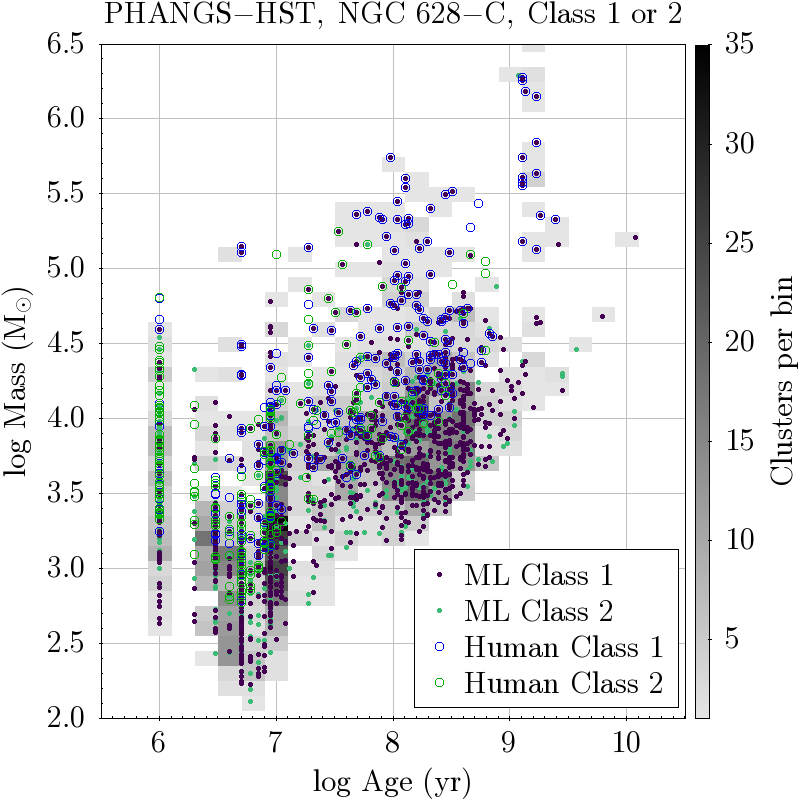}
\includegraphics[width=4.5in]{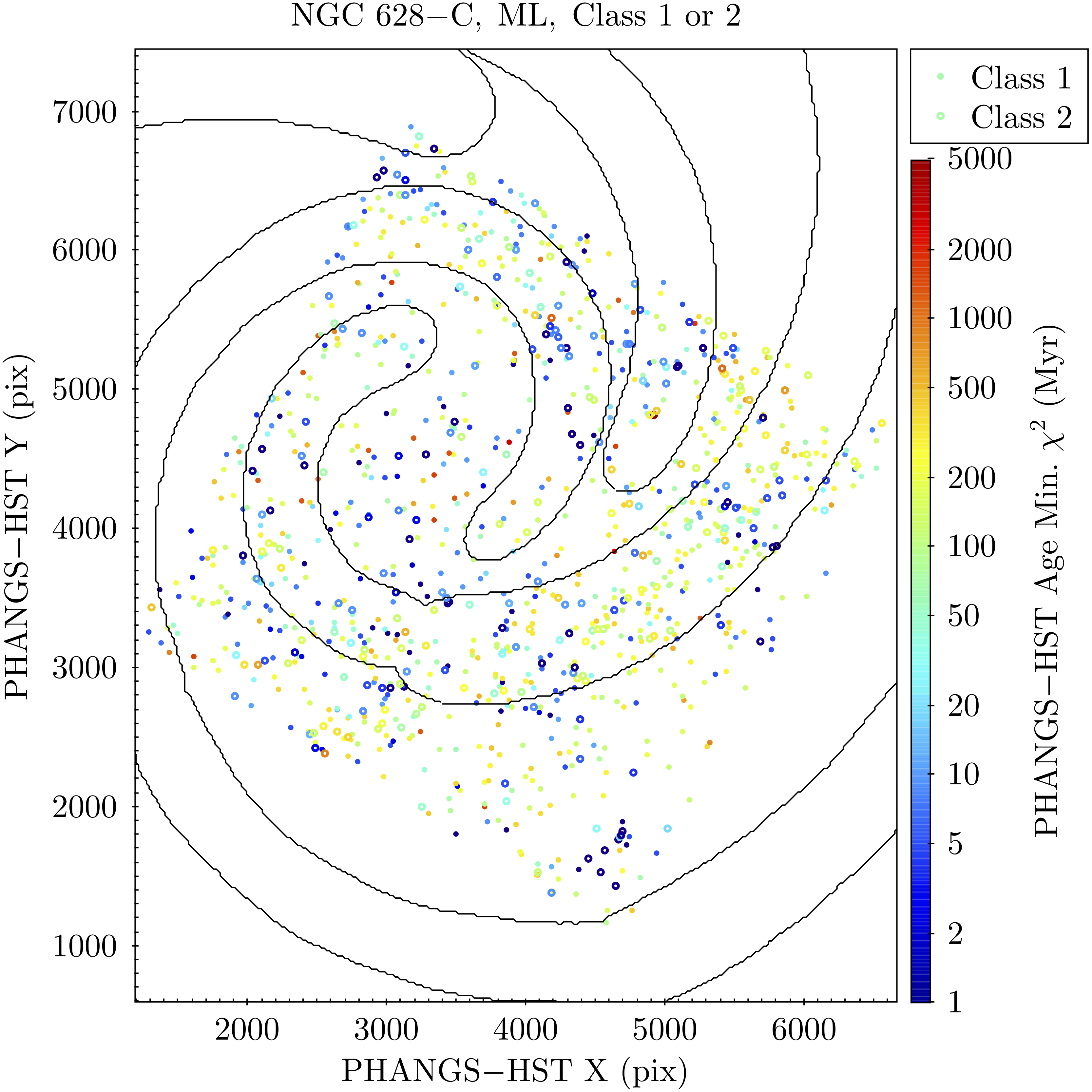}
\caption{The cluster population of the NGC~628-C field as displayed in Fig.~\ref{fig:MassAgeplot} and Fig.~\ref{fig:spatialage}, duplicated here for easy comparison to Figs.~\ref{fig:Summary_ngc628e}--\ref{fig:Summary_ngc3351}.}
\label{fig:Summary_ngc628c}
\end{figure*}

\begin{figure*}
\includegraphics[width=4.5in]{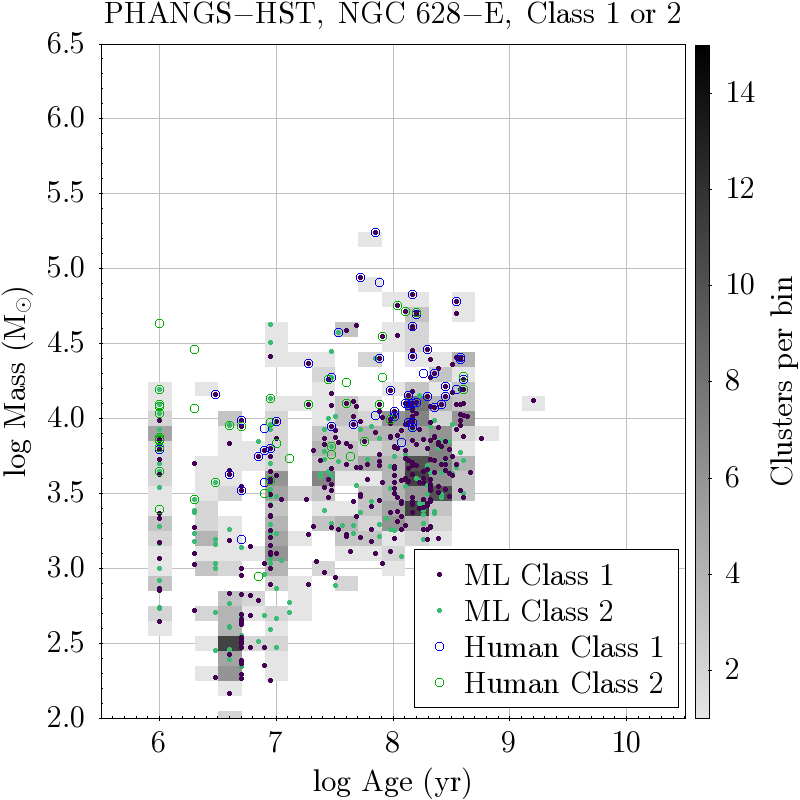}
\includegraphics[width=4.5in]{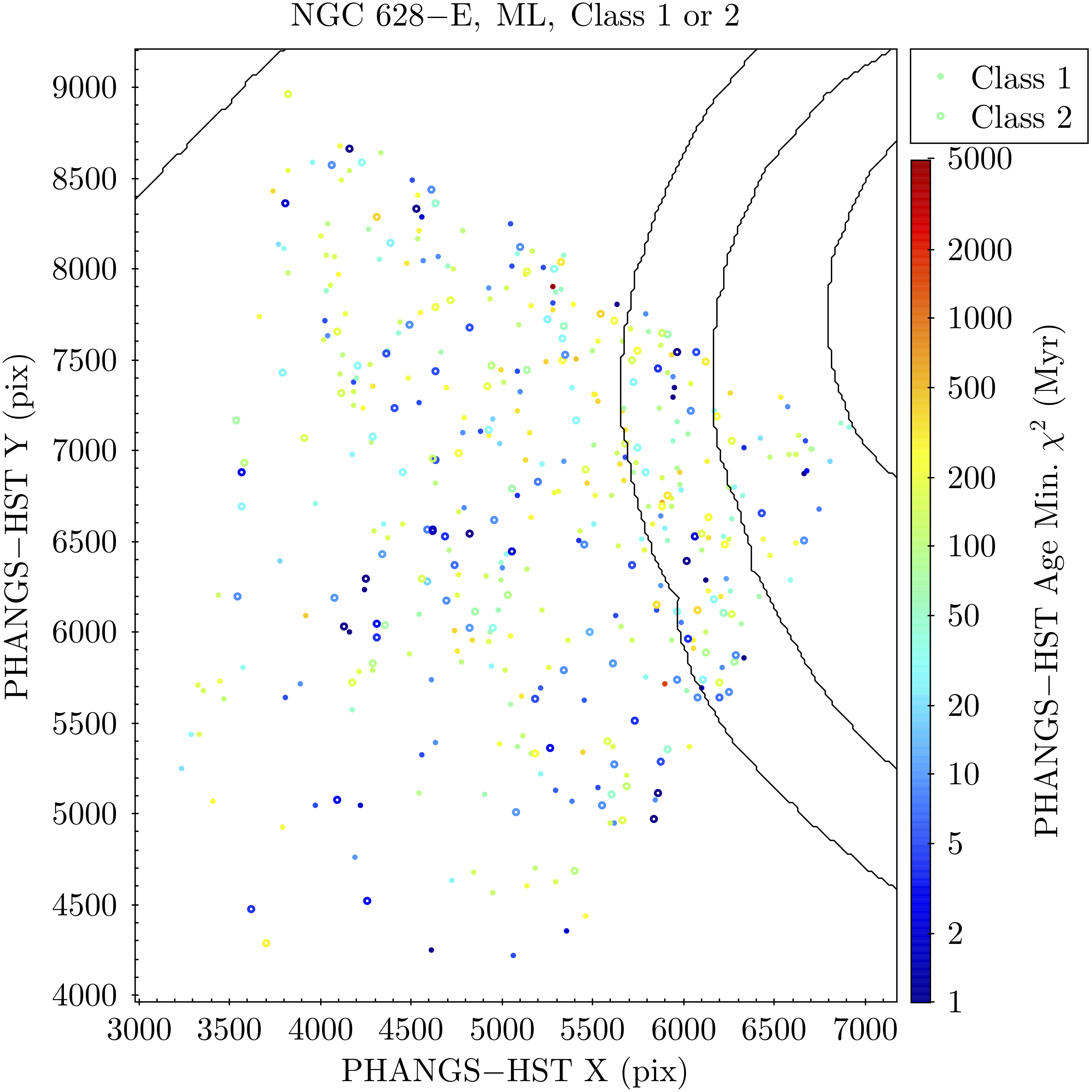}
\caption{As in Fig.~\ref{fig:Summary_ngc628c}, except for NGC~628-E.  In the lower panel, clusters are only plotted if they have {\em WFC3/UVIS}+{\em ACS/WFC} coverage. Spiral arms features extending outward from the centre of the galaxy (lying off plot to the right) are marked.}
\label{fig:Summary_ngc628e}
\end{figure*}

\begin{figure*}
\includegraphics[width=4.5in]{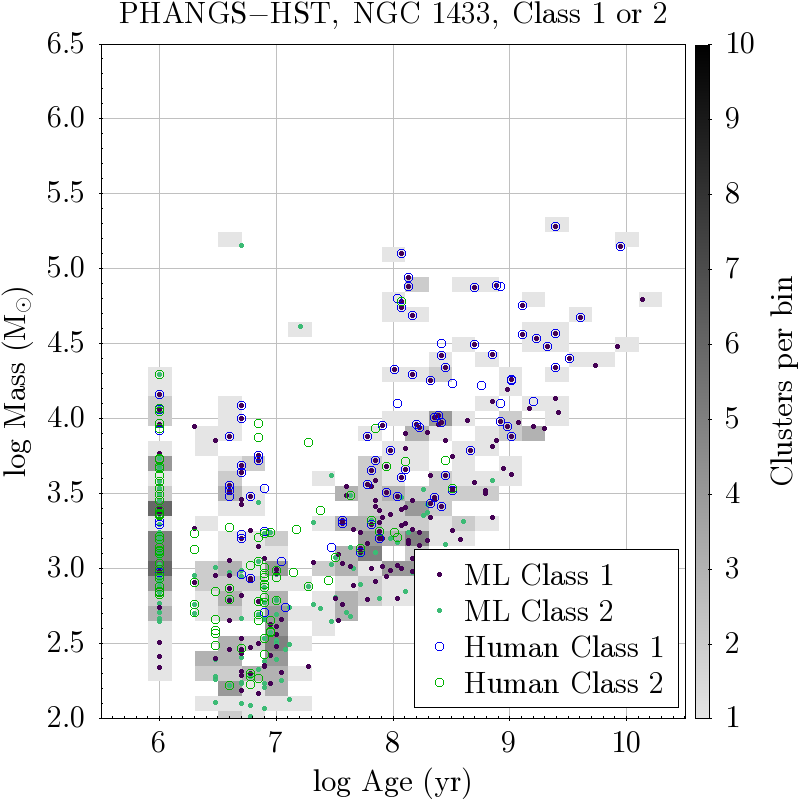}
\includegraphics[width=4.5in]{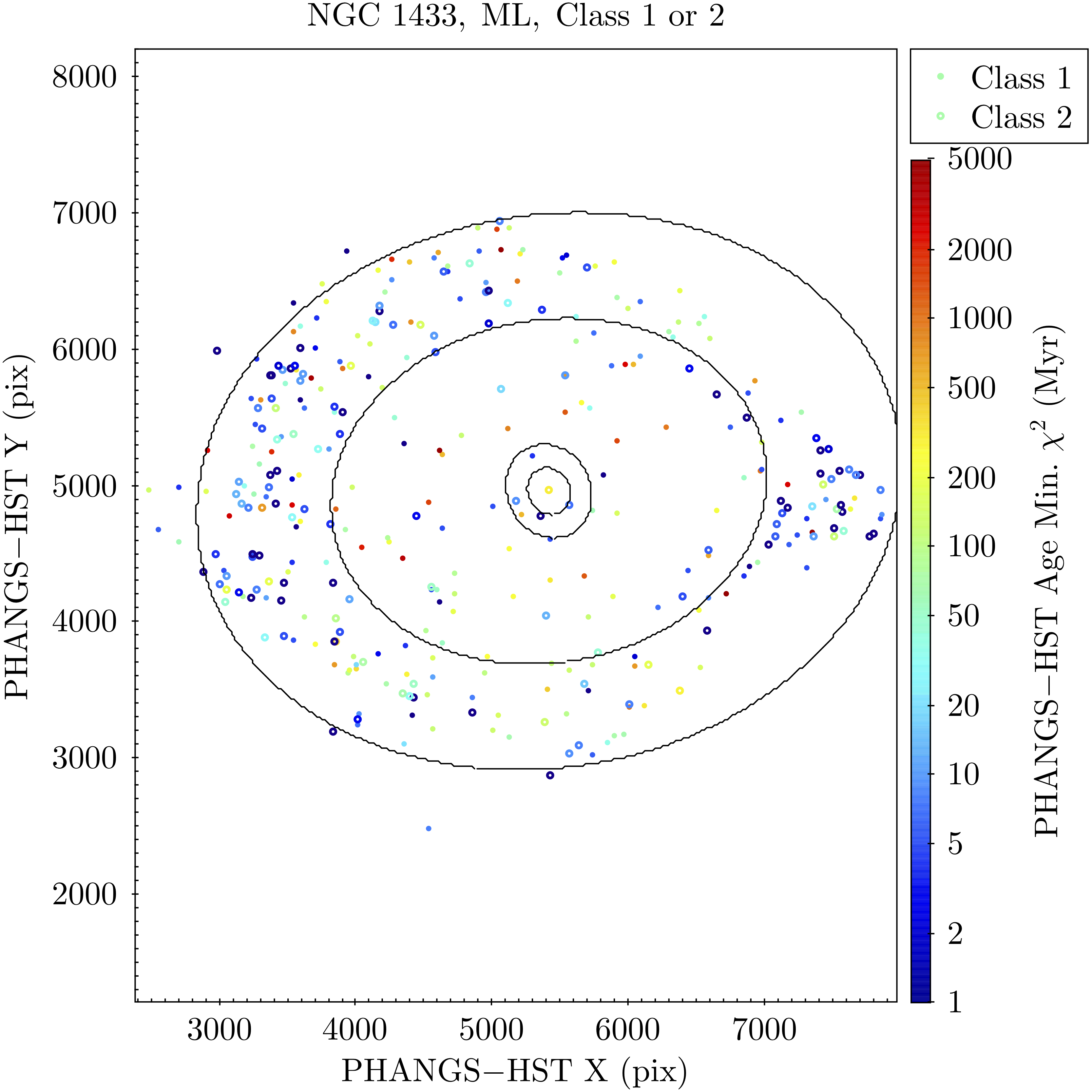}
\caption{As in Fig.~\ref{fig:Summary_ngc628c}, except for NGC~1433.  We show the ring-like structural features identified by the environmental masks of \citet{Querejeta2021}.  Note that the $HST$ field does not fully cover the NW and SW portions of the outer ring.}
\label{fig:Summary_ngc1433}
\end{figure*}

\begin{figure*}
\includegraphics[width=4.5in]{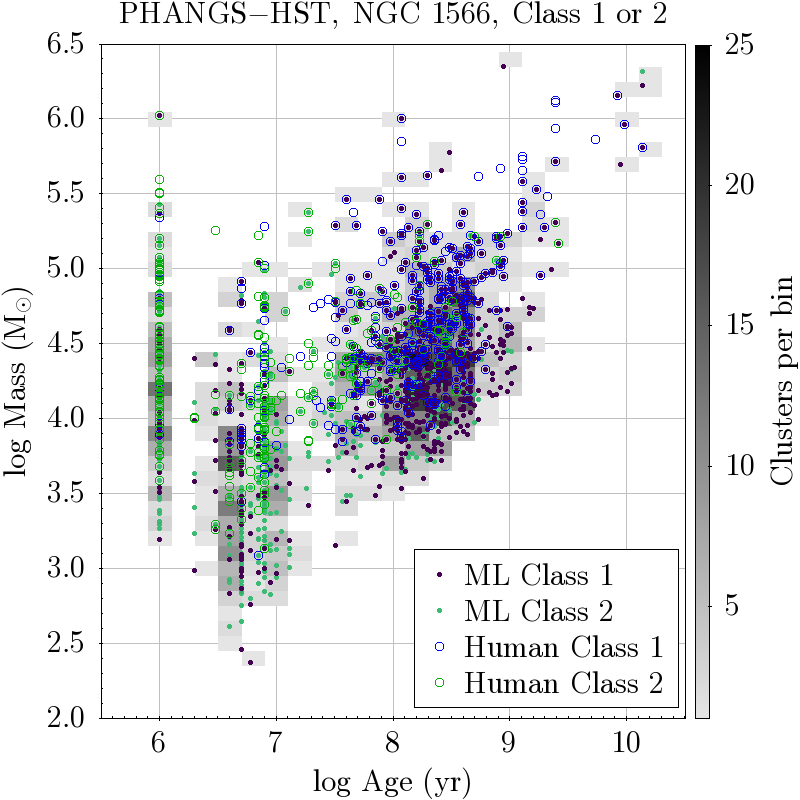}
\includegraphics[width=4.5in]{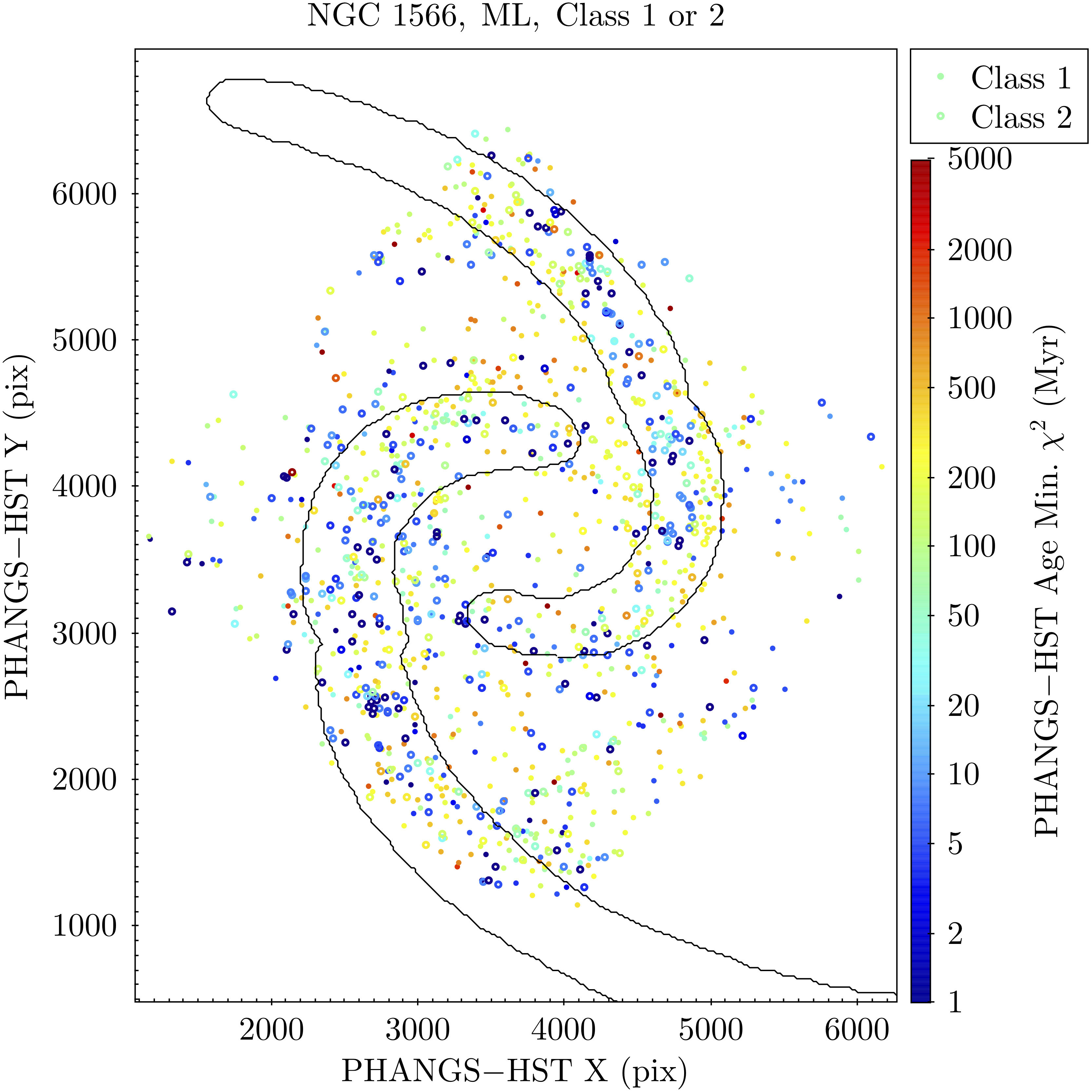}
\caption{As in Fig.~\ref{fig:Summary_ngc628c}, except for NGC~1566. The spiral arms of \citet{Querejeta2021} are marked.}
\label{fig:Summary_ngc1566}
\end{figure*}

\begin{figure*}
\includegraphics[width=4.5in]{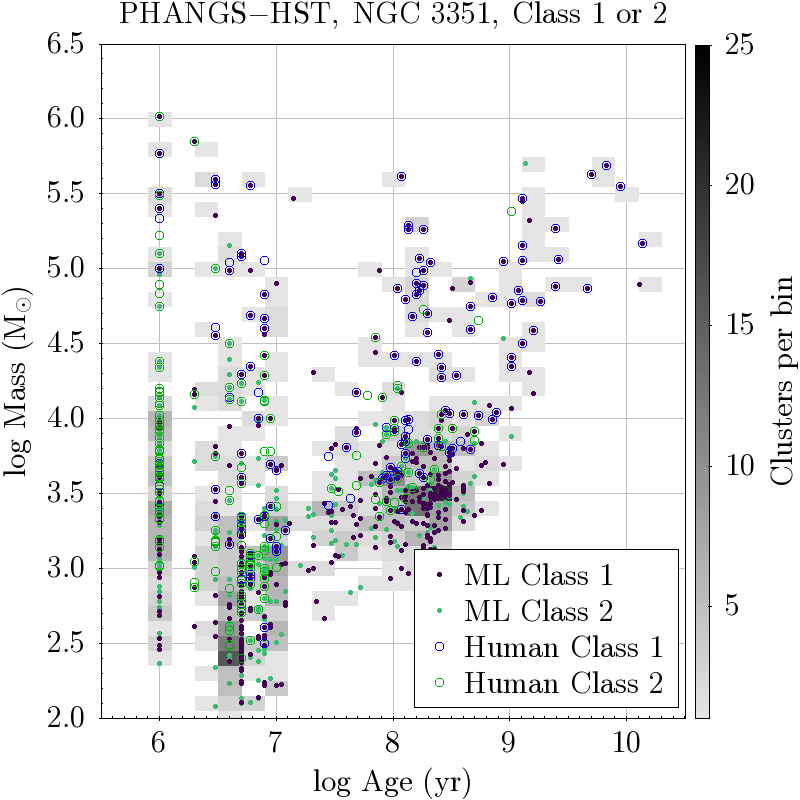}
\includegraphics[width=4.5in]{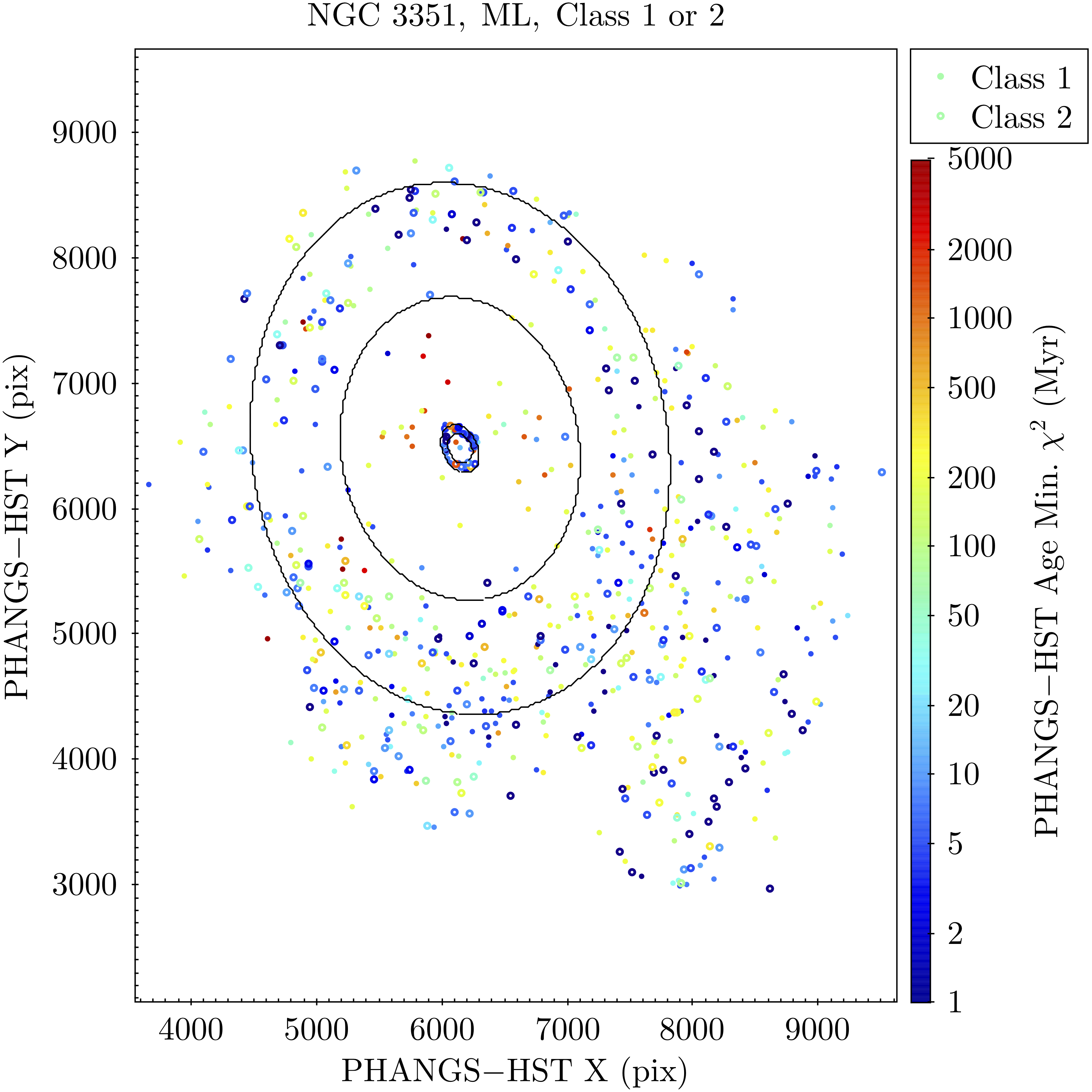}
\caption{As in Fig.~\ref{fig:Summary_ngc628c}, except for NGC~3351. \citet{Querejeta2021} ring-like structural features are drawn.}
\label{fig:Summary_ngc3351}
\end{figure*}

\section{\codename{DOLPHOT} parameters}

\codename{DOLPHOT} serves as the primary source detection algorithm for PHANGS-HST, feeding both the cluster identification pipeline and the multi-scale association finder of \citet{Larson2021}.  As such, even though many choices are similar to usage of \codename{DOLPHOT} by prior studies \citep[e.g.\ LEGUS,][]{Sabbi2018}, PHAT \citep[][]{Williams2014,Williams2021}, we provide a listing of relevant parameters here so as to allow for reproducible research.

We note that some parameters may be further tweaked before the final PHANGS-HST data products are released.  In particular, we anticipate reducing SecondPass in order to help mitigate `doubles' (redundant detections of extended objects) in the cluster context and over-decomposition of diffuse emission into point sources in the association context.  If this, and any other changes are made, they will be reported in the catalogue release papers.

\begin{table*}
 \caption{\codename{DOLPHOT} parameters adopted for PHANGS-HST}
 \label{tab:dolphotparam}
 \begin{tabular}{ll}
  \hline
Parameter & Meaning\\
  \hline
\multicolumn{2}{c}{Exposure-specific parameters:}\\ 
\hline
img*\_RAper = 8.0         &photometry aperture size (float) \\
img*\_RChi = 3.0          &aperture size for determining Chi (float); if <=0 use RAper \\
img*\_RSky = 15 35        &radii defining sky annulus (float>=RAper+0.5) \\
img*\_RSky2 = 4 10        &radii defining sky annulus (for FitSky=2 option) \\
img*\_RPSF = 13           &PSF size (integer>0) \\
img*\_aprad = 10          &radius for aperture correction $HST$-DISABLED (0.5" for {\em WFC3/UVIS} and {\em ACS/WFC})\\ 
img*\_apsky = 15 25       &sky annulus for aperture correction \\
\hline
\multicolumn{2}{c}{The following parameters affect the finding and measurement of stars:}\\ 
\hline
RCentroid = 2           &centroid box size (integer>0) \\
SigFind = 3.0           &sigma detection threshold (float)\\ 
SigFindMult = 0.85      &Multiple for quick-and-dirty photometry (float>0)\\ 
SigFinal = 3.5          &sigma output threshold (float) \\
MaxIT = 25              &maximum iterations (integer>0) \\
PSFPhot = 1             &photometry type (integer/0=aper,1=psf,2=wtd-psf) \\
PSFPhotIt = 3           &number of iterations in PSF-fitting photometry (integer>=0)\\ 
FitSky = 3              &fit sky? (integer/0=no,1=yes,2=small,3=with-phot) \\
SkipSky = 2             &spacing for sky measurement (integer>0) \\
SkySig = 2.25           &sigma clipping for sky (float>=1) \\
NegSky = 1              &allow negative sky values? (0=no,1=yes)\\ 
NoiseMult = 0.10        &noise multiple in imgadd (float) \\
FSat = 0.999            &fraction of saturate limit (float) \\
PosStep = 0.1           &search step for position iterations (float)\\ 
dPosMax = 2.5           &maximum single-step in position iterations (float)\\ 
RCombine = 1.5          &minimum separation for two stars for cleaning (float)\\ 
SigPSF = 5.0            &min $\mathrm{S/N}$ for psf parameter fits (float) \\
\hline
\multicolumn{2}{c}{Settings to enable/disable features:}\\
\hline
UseWCS = 1              &use WCS header info in alignment (integer 0=no, 1=use to est. shift/rotate/scale, 2=use to est. full distortion sol.)\\ 
Align = 2               &align images? (integer 0=no,1=const(x/yoff),2=lin(x/yoff+scale),3=cube(x/yoff+distortion,4=full3rdorderpolynom) \\
AlignIter = 2           &number of iterations on alignment? (integer>0) \\
AlignTol = 0            &number of pixels to search in preliminary alignment (float>=0)\\ 
AlignStep = 1.0         &stepsize for preliminary alignment search (float>0) \\
AlignOnly = 0           &exit after alignment \\
Rotate = 1              &allow cross terms in alignment? (integer 0=no, 1=yes)\\ 
SubResRef = 1           &subpixel resolution for reference image (integer>0) \\
SecondPass = 5          &second pass finding stars (integer 0=no,1=yes,>1=multiple passes)\\ 
SearchMode = 1          &algorithm for astrometry (0=max SNR/chi, 1=max SNR) \\
Force1 = 1              &force type 1/2 (stars)? (integer 0=no,1=yes) \\
PSFres = 1              &make PSF residual image? (integer 0=no,1=yes) \\
psfstars =              &Coordinates of PSF stars \\
psfoff = 0.0            &coordinate offset (PSF system - \codename{DOLPHOT} system)\\ 
ApCor = 1               &find/make aperture corrections? (integer 0=no,1=yes) \\
VerboseData = 1         &to write all displayed numbers to a .data file\\ 
\hline
\multicolumn{2}{c}{Flags for $HST$ modes:}\\
\hline
ForceSameMag = 0        &force same count rate in images with same filter? (integer 0=no, 1=yes) \\
FlagMask = 4            &photometry quality flags to reject when combining magnitudes \\
CombineChi = 0          &combined magnitude weights uses chi? (integer 0=no, 1=yes) \\
ACSuseCTE = 0           &apply CTE corrections on {\em ACS} data? (integer 0=no, 1=yes) \\
WFC3useCTE = 0          &apply CTE corrections on {\em WFC3} data? (integer 0=no, 1=yes) \\
ACSpsfType = 0          &use Anderson PSF cores? (integer 0=no, 1=yes) \\
WFC3UVISpsfType = 0     &use Anderson PSF cores? (integer 0=no, 1=yes) \\
WFC3IRpsfType = 0       &use Anderson PSF cores? (integer 0=no, 1=yes) \\
InterpPSFlib = 1        &interpolate PSF library spatially \\
  \hline
 \end{tabular}
\end{table*}


\bsp	
\label{lastpage}
\end{document}